\documentclass[lettersize,journal]{IEEEtran}
\usepackage{amsmath,amsfonts}
\usepackage{amssymb}
\usepackage{algorithmic}
\usepackage{algorithm}
\usepackage{array}
\usepackage[caption=false,font=normalsize,labelfont=sf,textfont=sf,parskip=0pt]{subfig}
\usepackage{textcomp}
\usepackage{stfloats}
\usepackage{url}
\usepackage{verbatim}
\usepackage{graphicx}
\usepackage{cite}
\usepackage{float}
\usepackage{xcolor}
\usepackage{upquote}
\usepackage{mathtools}
 \DeclareUnicodeCharacter{221E}{\ensuremath{\infty}}
\begin{document}

\title{Non-Minimum-Phase Resonant Controller \\ for Active Damping Control: Application to Piezo-Actuated Nanopositioning System}

\author{Aditya M. Natu,~\IEEEmembership{Student Member,~IEEE}, S. Hassan HosseinNia,~\IEEEmembership{Senior Member,~IEEE}

\thanks{This work was financed by Physik Instrumente (PI) SE \& Co. KG and co-financed by Holland High Tech with PPS Project supplement for research and development in the field of High Tech Systems and Materials.}
\thanks{Aditya M. Natu, and S. Hassan HosseinNia are with the Department of Precision and Microsystems Engineering, Delft University of Technology, Mekelweg 2, 2628 CD Delft, The Netherlands (e-mail:  a.m.natu@tudelft.nl; s.h.hosseinniakani@tudelft.nl)}}

\markboth{Non-Minimum-Phase Resonant Controller (NRC)}%
{Shell \MakeLowercase{\textit{et al.}}: A Sample Article Using IEEEtran.cls for IEEE Journals}

\IEEEpubid{}

\maketitle

\begin{abstract}
Nanopositioning systems frequently encounter limitations in control bandwidth due to their lightly damped resonance behavior. This paper presents a novel Non-Minimum-Phase Resonant Controller (NRC) aimed at active damping control within dual closed-loop architectures, specifically applied to piezo-actuated nanopositioning systems. The control strategy is structured around formulated objectives for shaping sensitivity functions to meet predetermined system performance criteria. Leveraging non-minimum-phase characteristics, the proposed NRC accomplishes complete damping and the bifurcation of double resonant poles at the primary resonance peak through a constant-gain design accompanied by tunable phase variation. The NRC demonstrates robustness against frequency variations of the resonance arising from load changes and is also capable of damping higher-order flexural modes simultaneously. Furthermore, by establishing high gains at low frequencies within the inner closed-loop and integrating it with a conventional PI tracking controller, the NRC achieves substantial dual closed-loop bandwidths that can exceed the first resonance frequency. Moreover, the NRC significantly diminishes the effect of low-frequency reference signals on real feedback errors while effectively rejecting disturbances proximate to the resonance frequency. All contributions are thoroughly formulated and exemplified mathematically, with the controller's performance confirmed through an experimental setup utilizing an industrial nanopositioning system. The experimental results indicate dual closed-loop bandwidths of 895 Hz and 845 Hz, characterized by $\pm3$ dB and $\pm1$ dB bounds, respectively, that surpass the resonance frequency of 739 Hz.
\end{abstract}


\section{Introduction}
\IEEEPARstart{N}{anopositioning} stages are employed for high-resolution positioning tasks, ranging from subnanometers to a few hundred micrometers \cite{fleming2014design}. These systems find applications in various fields, including scanning probe microscopy (SPM) \cite{tuma2013four,clayton2009conditions,shan2013design}, imaging using atomic force microscopy (AFM) \cite{rana2014performance,maroufi2014high,yong2012design}, wafer and mask alignment in lithography \cite{lan2007review,li2006sub}, and even micro/nano-manipulation in biological processes such as DNA sequencing \cite{hyun2013threading}. 

Depending on the specific task within an application, nanopositioning systems are required to track various types of references, such as periodic or arbitrary signals. However, all require a fast response to the controlled inputs. To ensure accurate reference tracking, these systems typically employ sensor-based feedback control architectures that mitigate errors arising from excited system dynamics or external disturbances \cite{devasia2007survey}. In response to the growing demands for higher throughput and resolution, often exceeding the sub-nanometer level, considerable emphasis has been placed over the past two decades on designing optimized systems and closed-loop control architectures to maximize control bandwidth \cite{huang2024design,yong2012invited,gu2014modeling,chen2021damping}.

Section \ref{SystemArchitecture} provides an overview of the typical system design and the associated challenges. Section \ref{TraditionalControl} offers a concise overview of the various control approaches developed in the state-of-the-art to address these challenges. Finally, Section \ref{Contributions} outlines the contributions made in this work, which will form the foundation of this paper. 

\subsection{System Design Architecture and Dynamics}
\label{SystemArchitecture}
\begin{figure}[t]
    \centering
    \includegraphics[width=0.8\columnwidth]{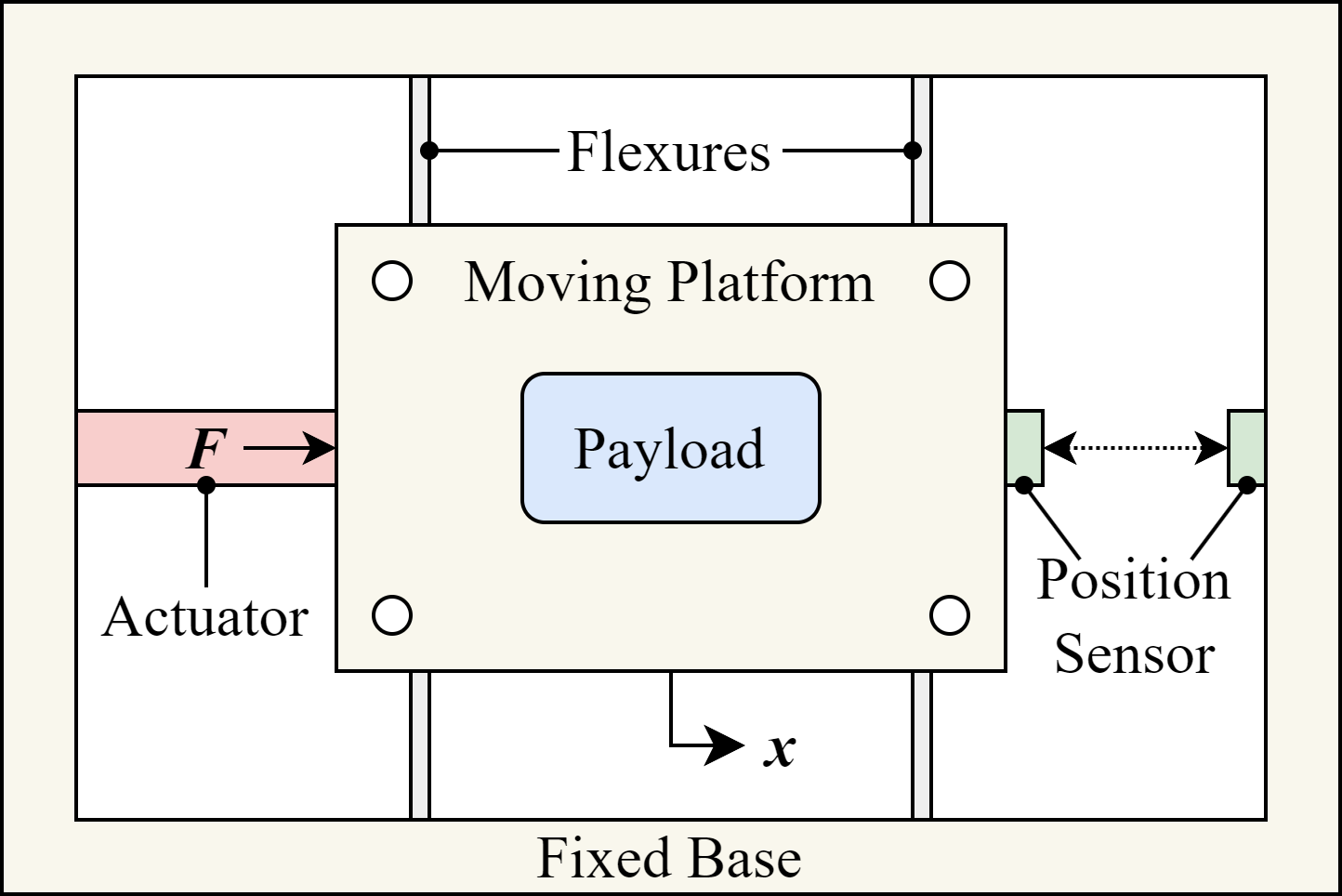}
    \caption{A system design schematic of piezo-actuated nanopositioners.}
    \label{fig:SystemDesign}
\end{figure}

As illustrated in Fig. \ref{fig:SystemDesign}, typical single-degree-of-freedom nanopositioning stages comprise a moving platform displaced laterally by the force generated by the actuator. Most of these stages employ piezoelectric stack actuators because of their advantageous properties, such as generating large forces, high stiffness, bandwidth, and resolution. The design integrates parallel flexures as a guiding mechanism for the platform, providing several benefits, including zero backlash and frictionless operation. To measure platform displacement, position sensors, such as dual-plate capacitive sensors, are employed, where the capacitive change due to the platform's movement produces an output voltage, which for a few micrometers the range of motion is highly linear \cite{fleming2014design}. 

To enable high control bandwidths, the design specification of nanopositioning stages is to have a very high first resonance frequency ($\omega_n$) in the actuation direction, which can be obtained by highly stiff stage-guiding flexures ($k_s$). However, this inevitably results in a much shorter range of motion ($r \propto 1/k_s$) \cite{yong2012design}. Thus, this often leads to a trade-off in which the stiffness of the guiding flexures, made of materials with an inherent low structural damping coefficient ($\zeta_n \approx 0.01$), is compromised to produce a sufficient travel range. The typical mass-spring-damper configuration of the stages is represented by second-order frequency dynamics:

\begin{equation}
\label{GeneralSecondOrderPlant}
G(s)=\frac{g \omega_n^2}{s^2+ \eta_n s+\omega_n^2},
\end{equation}
where, $s=i\omega$ denotes the Laplace variable, with $i$ and $\omega$ being the imaginary number and frequency, respectively, and $\eta_n = 2 \zeta_n \omega_n$.

The flexures also exhibit additional bending modes, observable by the sensor, which leads to the appearance of higher-order modes in the system frequency response, as illustrated in Fig. \ref{fig:GeneralFreqResp}, which, if excited by the high-frequency components of the reference signal, will further deteriorate the positioning accuracy. Additionally, voltage amplifiers are often employed in conjunction with the stage to drive the actuators with high voltages, resulting in a low-pass filtering effect due to the series configuration of the piezo-actuator's capacitance with the amplifier's input resistance. In addition, the need for high-resolution
analog-to-digital conversion (A/D) and filtering for capacitive sensor signals to achieve sub-nano resolution digital displacement signals inevitably results in a significant delay \cite{san2015modified}. Consequently, the system exhibits substantial phase lag within the frequency range of interest ($\omega<\omega_n$). Therefore, a general representation of the system dynamics is represented by:
\begin{figure}[t]
    \centering
    \includegraphics[width=\columnwidth]{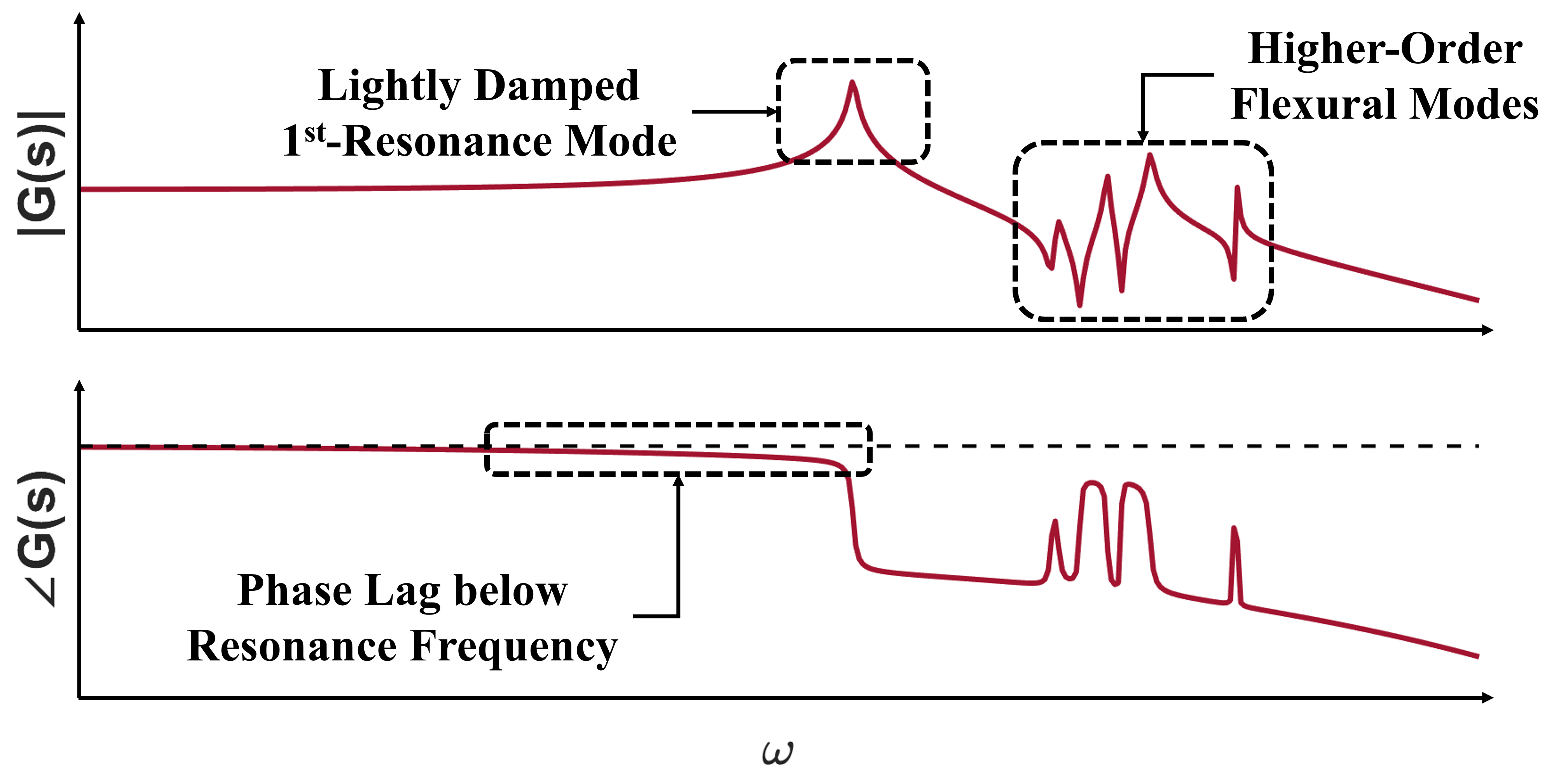}
    \caption{Schematic of the typical frequency response of a piezo-actuated nanopositioning system.}
    \label{fig:GeneralFreqResp}
\end{figure}
\begin{equation}
\resizebox{1\hsize}{!}{$G(s)=\left(\underbrace{\frac{\omega_n^2}{s^2+\eta_n s+\omega_n^2}}_{\substack{\text{Dominant} \\ \text{ Resonant Dynamics}}}+\underbrace{\sum_{m=2}^N \frac{\omega_m^2}{s^2+ \eta_m s+\omega_m^2}}_{\text{Higher-Order} \\ \text{ Mode Dynamics}}\right) \underbrace{\frac{g\cdot\omega_a}{s+\omega_a}}_{\substack{\text{Actuator} \\ \text{Amplifier} \\ \text{Dynamics}}} \underbrace{e^{-\tau s}}_{\text{Delay}}$},
\end{equation}
where, $m$ represents the higher-order modes, $\omega_a = R\cdot C$, with $R$ being the amplifier input impedance and $C$ being actuator capacitance, $g$ is the amplifier gain, and $\tau$ is the time delay. 
\subsection{State-of-the-Art Feedback-Control Methods}
\label{TraditionalControl}
\begin{figure}[!t]
\captionsetup[subfloat]{farskip=0pt,captionskip=0pt}
\centering
\subfloat[]{\includegraphics[width=0.9\columnwidth]{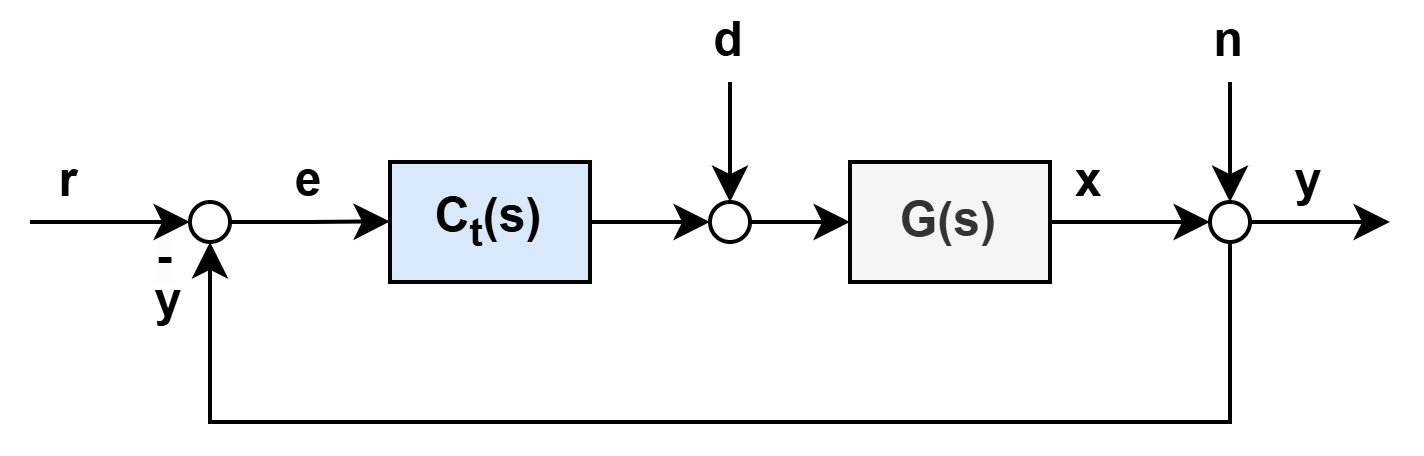}
\label{fig:TraditionalArchitecture}}
\hfil
\subfloat[]{\includegraphics[width=1\columnwidth]{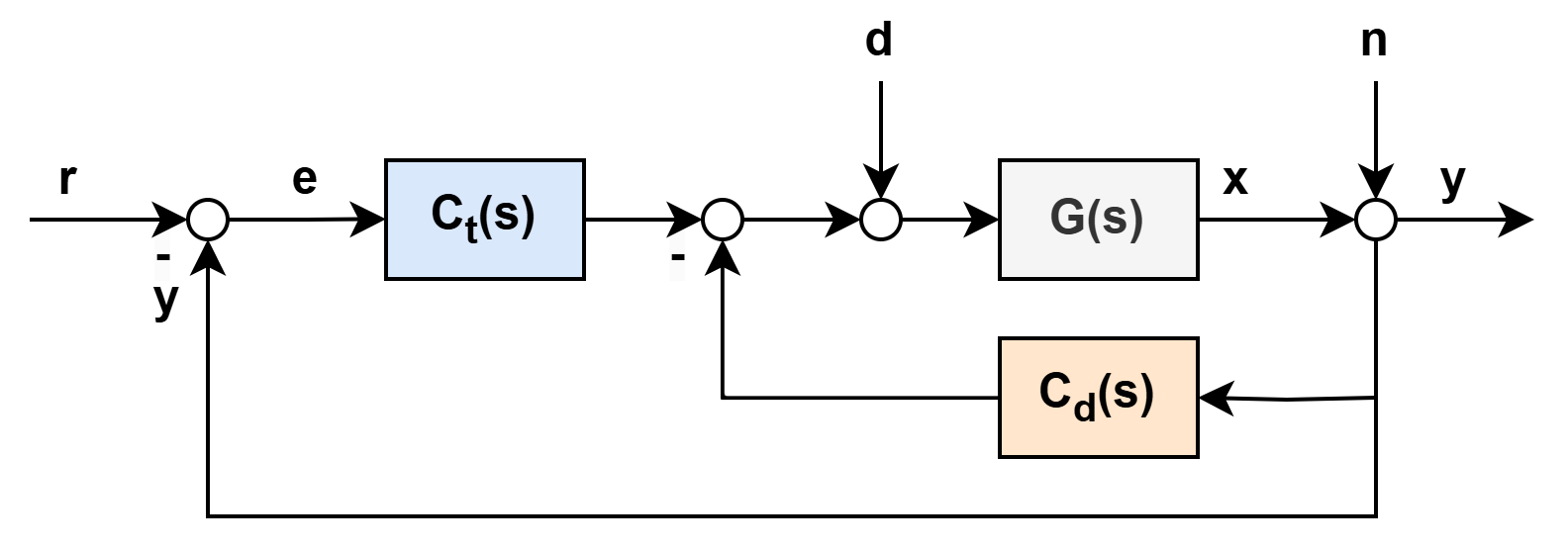}
\label{fig:DualLoopArchitecture}}
\caption{(a) Conventional feedback control architecture, (b) Dual closed-loop control architecture incorporating active damping control.}
\label{fig:ControlArchitectures}
\end{figure}

Conventionally, simple linear proportional-integral (PI) controllers have been primarily utilized to track references in a sensor-based feedback control architecture, as presented in Fig. \ref{fig:ControlArchitectures}(a). However, the bandwidth that such controllers can achieve is severely limited to less than 2\% of the dominant resonance frequency of the system due to the highly low-damped nature of the system resonance peak \cite{fleming2009nanopositioning}. To overcome this limitation, inversion techniques, such as notch filters, are often combined with tracking controllers to suppress the resonant dynamics, allowing for a higher bandwidth \cite{abramovitch2008semi,feng2017high}. However, implementing such filters requires a very accurate representation of the system model and proves to be highly sensitive to variations in the system dynamics due to changing payload mass.

Alternatively, active damping control has been researched and developed, where damping control is implemented in an inner feedback loop within a dual-loop architecture, as illustrated in Fig. \ref{fig:ControlArchitectures}(b). Different techniques, including positive position feedback (PPF) \cite{moon2017selection,li2015positive}, integral resonant control (IRC) \cite{fleming2009new,al2013integral}, resonant control (RC) \cite{ling2019robust,das2014mimo}, positive velocity and
position feedback (PVPF) \cite{russell2015simultaneous}, positive acceleration, velocity, and position feedback (PAVPF) \cite{li2017positive,babarinde2019simultaneous}, integral force feedback (IFF) \cite{fleming2009nanopositioning,teo2014optimal}, etc., have been shown to provide good damping performance with modest insensitivity to variations in resonance frequency. However, for instance, in PPF, an arbitrary configuration of the resonant poles cannot be realized in the s-plane while ensuring stability. IFF requires the addition of force sensors, etc. \cite{feng2022high}. In IRC, if designed to consider model uncertainty in the control design, the system robustness needs to be improved \cite{feng2023high}. Additionally, when the system, damped with most of these methods, is included in an integral tracking loop to obtain high gains for reference tracking, the system is still limited by a low gain margin \cite{fleming2009nanopositioning}.

\subsection{Research Contributions}
\label{Contributions}
The paper initially formulates objectives and guidelines to shape desired sensitivity functions in dual closed-loop control architectures. In line with these, this paper presents a novel non-minimum phase resonant controller (NRC) for active damping control with application to a piezo-actuated nanopositioning system. The main contributions of this work are as follows.
\begin{enumerate}
    \item The controller leverages the characteristics of a non-minimum phase system to enable a constant-gain controller with tunable phase variation to dampen the first resonance peak completely. Additionally, the controller not only dampens the double resonant poles but also effectively splits them in the frequency domain. These aspects are discussed in detail in \ref{NMPC_INTRO} and \ref{ADC_NMPC}.

    \item The controller's damping performance, tuned for the unloaded system, is robust to variations in the system's resonance frequency due to load variations and ensures complete damping of resonant poles. This is demonstrated in \ref{Robustness_LoadVariations}.

    \item The controller, aimed at damping the dominant resonance peak, is also capable of sufficiently damping the higher-order flexural modes of the system. This is formulated in \ref{DampingMultipleModes}.

    \item The controller facilitates the creation of low-frequency high gains in the inner closed loop. In conjunction with a standard PI tracking controller, it enables the achievement of high closed-loop bandwidths, characterized by both $\pm$1 dB and $\pm$3 dB crossings, which can also exceed the system's first resonance frequency. This aspect is detailed in \ref{ClosedLoopDynamics}.

    \item  The proposed NRC, in dual closed-loop control architecture, reduces the contribution of the low-frequency reference to the real feedback errors accumulated in the system while effectively rejecting disturbances around the resonance frequency. This is also discussed in \ref{ClosedLoopDynamics}.
\end{enumerate}

The rest of the paper is outlined as follows: Section \ref{Dual Closed Loop Architecture} presents a perspective on sensitivity shaping for general dual closed-loop control architectures, Section \ref{NMPC_INTRO} presents the proposed NRC for active damping control with tuning guidelines, robustness to load variations, multimode damping and high-frequency taming. Section \ref{Dual Loop Control based on NMPRC} presents the combination of NRC with a conventional PI tracking controller and its tuning for desired dual closed-loop performance requirements. Section \ref{ExperimentalSection} presents the experimental setup and results with the proposed method, which demonstrate compliance with the analytical study. Finally, Section \ref{Conclusions} concludes this paper.

\section{Dual Closed-Loop Control Architecture}
\label{Dual Closed Loop Architecture}
In conventional closed-loop structures, as illustrated in Fig. \ref{fig:ControlArchitectures}(a), the sensitivity of the system output has been extensively analyzed in the literature, demonstrating its correlation with closed-loop performance under various system inputs \cite{schmidt2020design}. Within the dual closed-loop control architecture that incorporates active damping, it is imperative to understand how the sensitivity functions evolve and can be delineated. In this section, the sensitivity functions are specified for a general dual closed-loop control architecture (Fig. \ref{fig:ControlArchitectures}(b)), relating the measured and actual positions (\ref{RedefinedSensitivities} and \ref{RealFeedbackErrorSensitivity}, respectively). Moreover, based on general dual closed-loop performance objectives, guidelines for shaping these sensitivities are provided in \ref{LoopShapingGuidelines}.

\subsection{Redefined Sensitivities}
\label{RedefinedSensitivities}
The primary variables that serve as input within the system include the reference $r$, the process disturbance $d$, and the output disturbance $n$. Sensor noise within the measurement system is incorporated into $n$, thus characterizing the system output $y$ as the measured output.

The damping controller $C_d(s)$ and the tracking controller $C_t(s)$ collaboratively enhance bandwidth in the dual closed-loop system. In Fig. \ref{fig:ControlArchitectures}(b), the dual closed-loop transfer function $T_{yr}(s)$, mapped from reference $r$ to position output $y$, is given by:
\begin{equation}
\label{CL_RT}
T_{yr}(s) = \frac{G(s) C_t(s)}{1+G(s)\left(C_t(s)+C_d(s)\right)}.
\end{equation}

The dual closed-loop transfer function $S_{yn}(s)$, widely known as the sensitivity function, mapping from the output disturbance $n$ to the measured output $y$ can be expressed as: 
\begin{equation}
\label{CL_S}
S_{yn}(s) = \frac{1}{1+G(s)\left(C_t(s)+C_d(s)\right)}.
\end{equation}

Subsequently, the  dual closed-loop transfer function $PS_{yd}(s)$, known as the process sensitivity function, mapping from the process disturbance $d$ to the measured output $y$ can be expressed as: 
\begin{equation}
\label{CL_PS}
PS_{yd}(s) = \frac{G(s)}{1+G(s)\left(C_t(s)+C_d(s)\right)}.
\end{equation}

Dual closed-loop functions in \eqref{CL_RT} and \eqref{CL_S} lack complementarity, unlike standard feedback architectures. By examining the interaction between the tracking and damping controllers, control strategies can be optimized for reference tracking, disturbance rejection, and noise attenuation.
\subsection{Real Feedback Error Sensitivity}
\label{RealFeedbackErrorSensitivity}
The three dual closed-loop functions defined in \ref{RedefinedSensitivities} depict the impact of any of the inputs on the measured output $y$. However, in reality, the actual positioning error $e_{r}=r-x$ is of greater interest and importance than the control error $e = r-y$, where $x$ is the actual or real position of the system \cite{schmidt2020design}. Thus, it is worthwhile to investigate how the real error $e_r$ maps to different inputs in the system to understand and compute the contribution of each of these inputs to the real error. Since $e_r$ is concerned with the $x$, the dual closed-loop transfer functions mapping from the inputs $r$, $d$, and $n$ to $x$ are first defined as follows:
\begin{equation}
\label{CL_RT_real}
T_{xr}(s) = \frac{G(s) C_t(s)}{1+G(s)\left(C_t(s)+C_d(s)\right)} = T_{yr}(s).
\end{equation}
\begin{equation}
\label{CL_PS_real}
PS_{xd}(s) = \frac{G(s)}{1+G(s)\left(C_t(s)+C_d(s)\right)} = PS_{yd}(s).
\end{equation}
\begin{equation}
\label{CL_S_real}
S_{xn}(s) = \frac{-G(s)\left(C_t(s)+C_d(s)\right)}{1+G(s)\left(C_t(s)+C_d(s)\right)} \neq S_{yn}(s).
\end{equation}

According to \eqref{CL_S} and \eqref{CL_S_real}, the primary distinction becomes evident in the dual closed-loop sensitivity function concerning input $n$. Consequently, $x$ can be articulated as:
\begin{equation}
    x = T_{xr}(s)r + PS_{xd}(s)d + S_{xn}(s)n.
\end{equation}

Using linear time-invariant system theory and applying a statistical addition assuming that the signals $r$, $d$, and $n$ are uncorrelated, $e_r$ can be expressed as:
\begin{equation}
\label{Eq_RealErrorSq}
\begin{aligned}
    e_{r}^2= &(r-x)^2 \\ = &\left(\frac{(r-x)}{r} r\right)^2+\left(\frac{(r-x)}{d} d\right)^2+\left(\frac{(r-x)}{n} n\right)^2\\
    = &\left(1-T_{xr}(s)\right)^2 r^2 + \left(PS_{xd}(s)\right)^2 d^2 + \left(S_{xn}(s)\right)^2 n^2   \\
    =&\left(\underbrace{\frac{1+G(s)C_d(s)}{1+G(s)\left(C_t(s)+C_d(s)\right)}}_{T'_{xr}(s)}\right)^2 r^2\\ &\left(\frac{G(s)}{1+G(s)\left(C_t(s)+C_d(s)\right)}\right)^2 d^2+ \\ &\left(\frac{-G(s)\left(C_t(s)+C_d(s)\right)}{1+G(s)\left(C_t(s)+C_d(s)\right)}\right)^2 n^2 \Longrightarrow \\
    e_{r}= & \sqrt{\left(T'_{xr}(s)r\right)^2 + \left(PS_{xd}(s)d\right)^2 + \left(S_{xn}(s)n\right)^2}.
\end{aligned}
\end{equation}
\textit{Note:} It should be emphasized that in such dual closed-loop control architectures, the pairs of transfer functions $T_{yr}(s)=T_{xr}(s)$, $T'_{xr}(s)$, and $S_{yn}(s)$, $S_{xn}(s)$ are complementary functions.

\subsection{Shaping Sensitivities for Dual Closed-Loop Control}
\label{LoopShapingGuidelines}
The objectives of the dual closed-loop control system can be formulated in the frequency domain by defining the desired shapes of the closed-loop and open-loop transfer functions. Although loop-shaping guidelines have been presented in the literature for independently tracking controllers and damping controllers \cite{wang2017tutorial,dastjerdi2018tuning,kaczmarek2023fractional}, we focus on the interaction of these two in a dual closed-loop architecture. As established in \ref{RealFeedbackErrorSensitivity}, what actually matters is to look at the real error feedback sensitivity functions, namely $T'_{xr}(s)$, $PS_{xd}(s)$, and $S_{xn}(s)$, and reduce their contributions to $e_r$ in the system. Thus, in this section, the objectives to shape the dual closed-loop transfer functions and, subsequently, the tracking ($C_t(s)$) and damping ($C_d(s)$) controllers will be presented.

\textit{Note:} Three notation to be used are defined as follows:
\begin{itemize}
    \item Dual Closed-Loop Control Bandwidth $\omega_c$
    \begin{equation}
        \omega_c:= \{ \omega \in \mathbb{R} \mid \omega \geq 0 \text{ and } |T_{xr}(s)|_{\omega > \omega_c} < 1 \}.  
    \end{equation}
    \item Tracking Controller Frequency $\omega_{C_t}$
    \begin{equation}
        \omega_{C_t}:= \{ \omega \in \mathbb{R} \mid \omega \geq 0 \text{ and } |C_{t}(s)|_{\omega < \omega_{C_t}} \gg 1 \} .
    \end{equation}
    \item Dual Loop Gain $L_D(s)$
    \begin{equation}
        L_D(s) := G(s) (C_t(s)+C_d(s)).
    \end{equation}
\end{itemize}

The dual closed-loop shaping objectives and the guidelines to achieve them are discussed below: 
\begin{figure}[t!]
    \centering
    \includegraphics[width=1\linewidth]{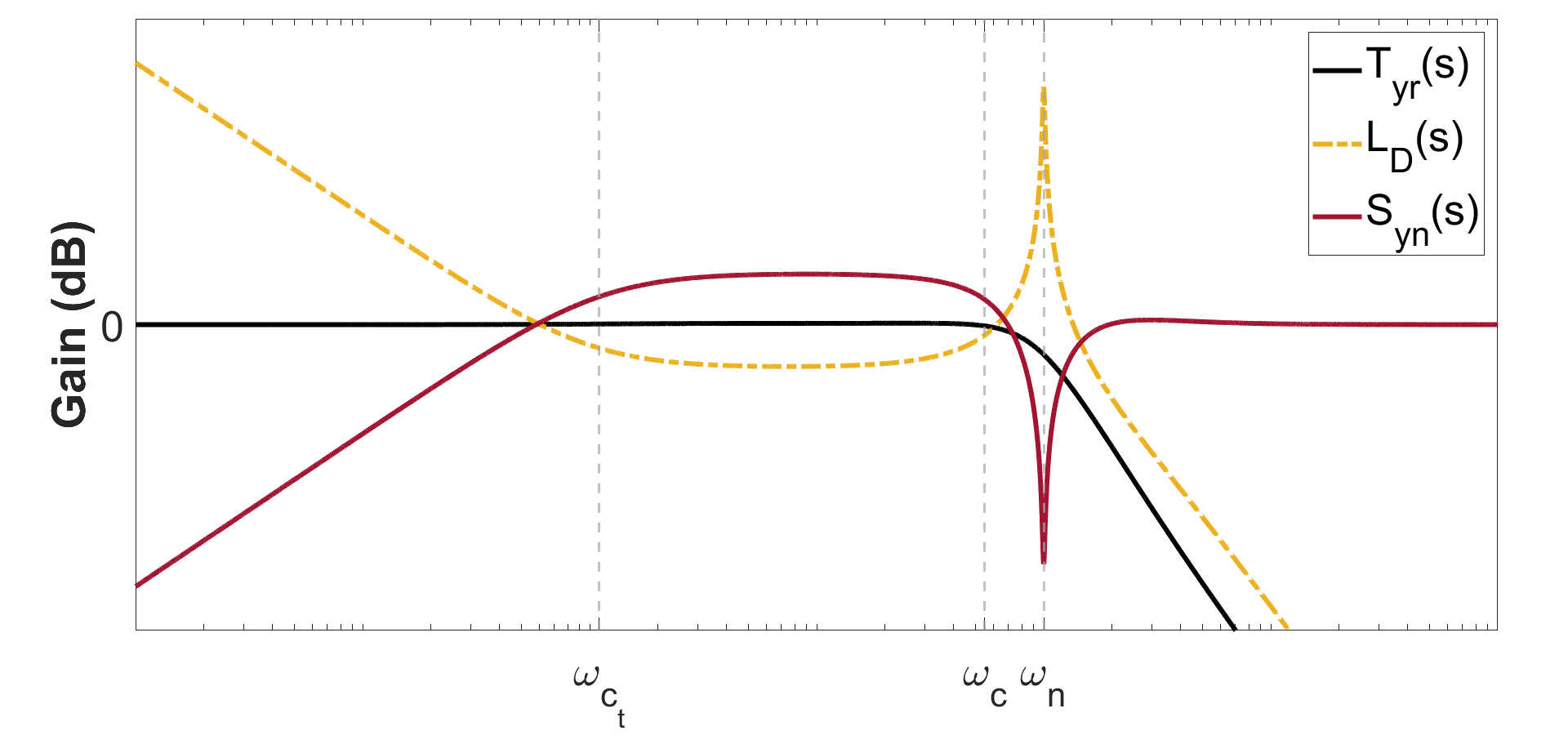}
    \caption{Illustration of the dual closed-loop shaping for a $2^\text{nd}$-order lightly-damped system.}
    \label{fig:DualLoopShaping}
\end{figure}   
\begin{enumerate}
    \item \textit{$O_1$: Maximizing Dual Closed-Loop Control Bandwidth} \\ The system should be able to track references ($x\approx r$) up to as high frequencies as possible. Due to the complementary nature of $T_{xr}(s)$ and $T'_{xr}(s)$, it also results in low real errors due to reference signals. The objective can thus be defined as:
    \begin{equation}
    \label{Eq_Objective1}
        \max \omega_c \big| \hspace{1mm} |T'_{xr}(s)| \approx 0 \Leftrightarrow |T_{xr}(s)| \approx 1 \hspace{1mm} \forall \hspace{1mm} \omega\leq\omega_c.
    \end{equation}
    This is achieved by ensuring $|G(s)C_d(s)|=1$ and $\angle G(s)C_d(s) = \pm \pi \hspace{1mm}\forall\hspace{1mm}\omega < \omega_c$ \eqref{Eq_RealErrorSq}. Thus, $|C_d(s)| \hspace{1mm} \forall \hspace{1mm}\omega < \omega_c$ should be a constant gain equal to $|G^{-1}(0)|$ as $|G(s)|=|G(0)| \hspace{1mm}\forall\hspace{1mm}\omega<\omega_n$.
    
    \item \textit{$O_2$: Maximizing Low-Frequency Disturbance Rejection} \\
    The system should be able to minimize the influence of low-frequency disturbances on the real error, implying a good disturbance rejection performance. The objective can thus be defined as:
    \begin{equation}
    \label{Eq_Objective2}
    \begin{aligned}
        \max \omega_{C_t} \big| \hspace{1mm} & |PS_{xd}(s)| \ll |G(s)| \hspace{1mm} \forall \hspace{1mm} \omega<\omega_{C_t} \\ & \Leftrightarrow |S_{yn}(s)| \ll 1 \hspace{1mm} \forall \hspace{1mm} \omega<\omega_{C_t}.
    \end{aligned}
    \end{equation}
    Taking into account $O_1$, this is achieved by ensuring $|C_{t}(s)|_{\omega < \omega_{C_t}} \gg 1$.
    
    \item \textit{$O_3$: Maximizing Active Damping Performance} \\
    The system resonance peak should be damped as much as possible to enable higher control bandwidths and to avoid excessive excitation of disturbances at that frequency. The objective can thus be defined as:
    \begin{equation}
    \label{Eq_Objective3}
    \begin{aligned}
        \max |L_D(s)|_{\omega=\omega_n} \big| \hspace{1mm} & |PS_{xd}(s)| \ll |G(s)| \text{ at } \omega=\omega_n \\ & \Leftrightarrow |S_{yn}(s)| \ll 1  \text{ at } \omega=\omega_n.
    \end{aligned}
    \end{equation}
    This is achieved by ensuring $|L_D(s)|_{\omega = \omega_n} \gg 1$ and preferably $\angle L_D(s)_{\omega = \omega_n} \approx \pm \pi$.
    
    \item \textit{$O_4$: Maximizing Noise Attenuation Performance} \\
     The system should be able to minimize the influence of high-frequency noise on real error, implying good noise attenuation performance. The objective can thus be defined as:
    \begin{equation}
    \label{Eq_Objective4}
    \begin{aligned}
        \min |L_D(s)|_{\omega\gg\omega_n} \big| \hspace{1mm} & |S_{xn}(s)| \approx 0 \hspace{1mm} \forall \hspace{1mm} \omega\gg\omega_n
        \\ & \Leftrightarrow |S_{yn}(s)| \approx 1  \hspace{1mm} \forall \hspace{1mm} \omega\gg\omega_n.
    \end{aligned}
    \end{equation}
    This is achieved by ensuring $|L_D(s)|_{\omega \gg \omega_n} \ll 1$. Thus, $|C_t(s)|\ll1$ and $|C_d(s)| \ll 1 \hspace{1mm} \forall \hspace{1mm}\omega \gg \omega_n$. 
\end{enumerate}

For dual closed-loop stability, there should be a sufficient phase margin $\phi_{m_i}$ at all three crossover frequencies $\omega_{cf_i}\hspace{1mm}\big|\hspace{1mm} |L_D(\omega_{cf_i})|=1;\hspace{1mm} i=\{1,2,3\}$.

However, there are some fundamental limitations with such a linear dual closed-loop control system, as discussed below.
\begin{enumerate}
    \item The frequencies ($\omega_{C_t}$) up to which the tracking controllers have high gains cannot be pushed to be arbitrarily large. From a stability perspective, the dual closed-loop control system should have sufficient stability margins to ensure robustness. Typically, high gains are obtained using integrators, where increasing integrator frequency ($\omega_i \approx \omega_{C_t}$) adversely affects stability margins. From a hardware perspective, the control inputs to the actuators cannot be arbitrarily high and fast due to the saturation and bandwidths of the actuators \cite{wang2017tutorial}.

    \item While maximizing the reference tracking performance by maximizing the dual closed-loop control bandwidth ($\omega_c$) ensures minimal real error accumulation due to references ($|T'_{xr}(s)|\approx0$) in the frequency regime of interest ($\omega\leq\omega_c$), it results in an inevitable trade-off of error accumulation due to noise ($|S_{xn}(s)|\approx1$). This fundamental limitation, known as the waterbed effect, transpires from Bode's integral theorem \cite{schmidt2020design}.
\end{enumerate}

Fig. \ref{fig:DualLoopShaping} elucidates the concept of dual closed-loop shaping objectives and limitations, as previously delineated. Although the state-of-the-art damping controllers introduced in \ref{TraditionalControl} adhere to this dual closed-loop shaping framework, their performance is restricted due to the interdependence of their gain and phase. Recognizing this, the present paper introduces and implements a novel active damping controller characterized by a constant-gain design and a tunable phase. This approach upholds the dual closed-loop shaping guidelines, exceeds the performance of existing state-of-the-art methods, and exemplifies the contributions presented in \ref{Contributions}.

\section{Non-Minimum-Phase Resonant Controller (NRC)}
\label{NMPC_INTRO}
\begin{figure}[t!]
    \centering
    \includegraphics[width=1\columnwidth]{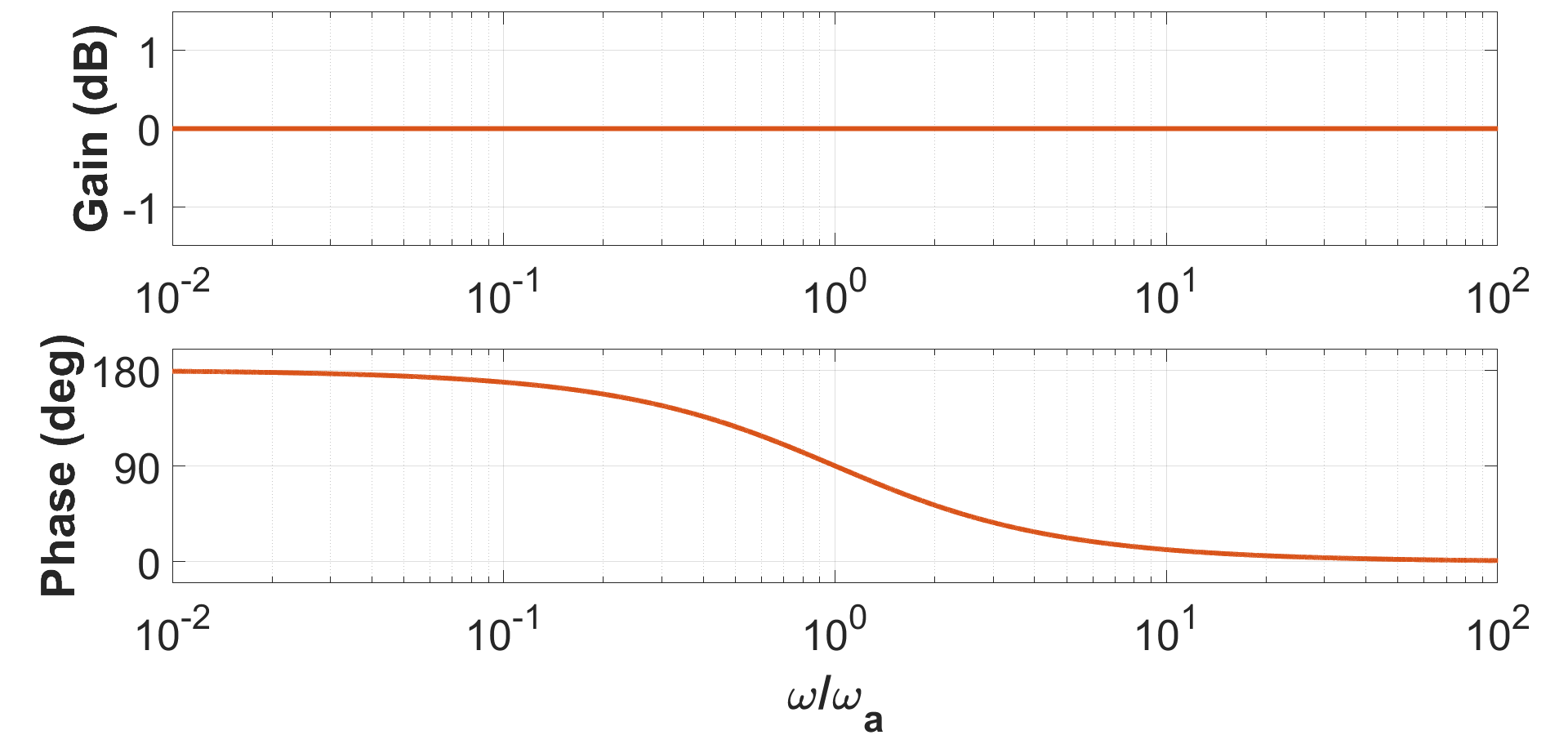}
    \caption{Frequency response of Non-Minimum Phase Resonant Controller $C_d(s)$ for $k=1$.}
    \label{fig:NMPC}
\end{figure}

In this section, we present a novel Non-Minimum-Phase Resonant Controller (NRC) for active damping control, described by the transfer function:
\begin{equation}
    C_d(s) = k\cdot\left(\frac{s-\omega_a}{s+\omega_a}\right),
\end{equation}
where $\omega_a$ signifies the tuned corner frequency of the controller, and \textit{k} represents the controller's gain. Fig. \ref{fig:NMPC} illustrates the frequency domain plot of the described controller. A distinctive characteristic of this controller is its constant gain across all frequencies, which is expressed as:
\begin{equation}
    \left| C_d(s) \right|= k \hspace{2mm} \forall \hspace{1mm} \omega\in [0, \infty).
\end{equation}

The phase response of the controller is defined as:
\begin{equation}
    \angle C_d(s) = \arctan \left(\frac{i\omega - \omega_a}{i\omega + \omega_a}\right).
\end{equation}

Notably, the phase of the controller transitions from \(\pm180^\circ\) to \(0^\circ\) as the frequency varies from \(0\) to \(2\omega_a\), and provides the \(-90^\circ\) phase at $\omega_a$. The controller consists of one pole in the left half-plane (LHP) at $s = -\omega_a$ and a zero in the right half-plane (RHP) at $s = \omega_a$, resulting in a non-minimum-phase characteristic.

The controller transfer function can also be interpreted in state-space representation by obtaining a minimal realization, where \( y\) and \( v\) are the input and output time signals to the controller. The state equation, with \( z \) as the state variable, can be written as:
\begin{equation}
\dot{z}(t) = -\omega_a z(t) + y(t),
\end{equation}
and the output equation is as follows:
\begin{equation}
v(t) = -2 \omega_a k z(t) + k y(t).
\end{equation}

Thus, the state-space matrices can be represented as \( A = [-\omega_a] \), \( B = [1] \), \( C = [-2\omega_ak] \), and \( D = [k] \).

\subsection{Active Damping Control with NRC}
\label{ADC_NMPC}
\begin{figure}[t!]
    \centering
    \includegraphics[width = 0.5\columnwidth]{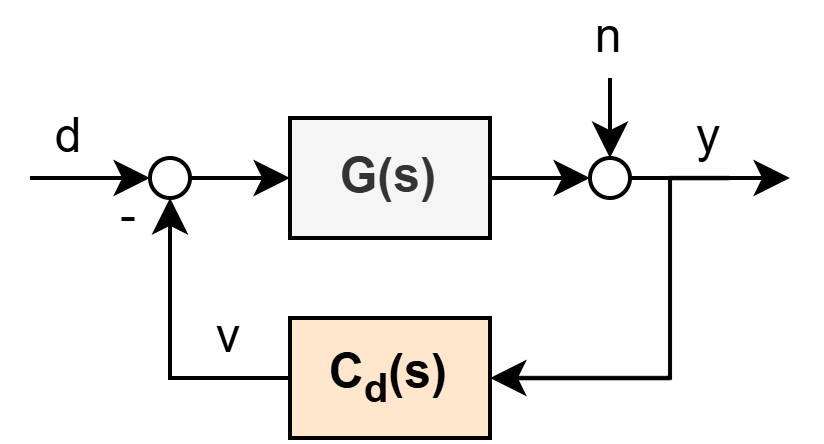}
    \caption{Inner closed-loop feedback architecture for Active Damping Control.}
    \label{fig:Inner_CL_Diagram}
\end{figure}

A negative feedback loop is implemented to enable active damping using the proposed NRC, as illustrated in Fig. \ref{fig:Inner_CL_Diagram}.
The tuning of the controller parameters is parametrized in terms of the general system parameters \eqref{GeneralSecondOrderPlant} to facilitate an easy tuning process. The controller's corner frequency is expressed as $\omega_a = n\omega_n$. Here, $n$ represents the normalized corner frequency with respect to the natural frequency of the plant. 
\begin{equation}
\label{NMPC_n_eq}
n = \frac{\omega_a}{\omega_n}.
\end{equation}

The controller gain $k$ is tuned as follows: 
\begin{equation}
\label{NMPC_Gain_Eq}
k = \gamma|G(s)|_{s=0}^{-1} =\gamma g^{-1} \hspace{1mm} \forall \hspace{1mm} \gamma\in(0,1],
\end{equation}
such that the loop's DC gain follows $0<|G(s)C_d(s)|_{s=0}\leq1$, ensuring closed-loop stability as detailed later.

The inner closed-loop transfer function from input disturbance $d$ to the system output $y$ can be formulated as: 
\begin{equation}
\label{ICL_Eqn}
\begin{aligned}
G_d(s) & =\frac{G(s)}{1+G(s)C_d(s)} \\
 &=\resizebox{0.85\hsize}{!}{$\frac{g \omega_n^2\left(s+n \omega_n\right)}{s^3+\left(n \omega_n+2 \zeta_n \omega_n\right) s^2+\left(2 \zeta_n n \omega_n^2+(1+\gamma) \omega_n^2\right) s+n(1-\gamma) \omega_n^3}$}.
\end{aligned}
\end{equation}

It can be seen that the inner closed-loop characteristic equation formulated comprises three poles ($p_{1,2,3}$) and one zero ($z_1$). For any $\omega_a>0$, $z_1$ lies in the LHP, ensuring a minimum phase behavior of the closed-loop function $G_d(s)$. 

The NRC parameters ($k,\omega_a$) influence the frequency response characteristics of the inner closed loop. The DC gain of $G_d(s)$ is given as: 
\begin{equation}
    |G_d(s)|_{s=0} = \frac{g}{1-\gamma}. 
\end{equation}

\subsubsection{Inner Closed-Loop Stability}
\label{StabilityAnalysis}
The characteristic equation for the presented inner closed-loop is given as: 
\begin{equation}
\label{CharEqn}
\resizebox{1\hsize}{!}{$s^3+\left(n \omega_n+2 \zeta_n \omega_n\right) s^2+\left(2 \zeta_n n \omega_n^2+(1+\gamma) \omega_n^2\right) s+n(1-\gamma) \omega_n^3 = 0$}.
\end{equation}

The stability conditions can be determined by applying the Routh-Hurwitz criterion to the characteristic equation \eqref{CharEqn}. The first two rows of the Routh–Hurwitz array are filled with these coefficients as follows:
\begin{equation}
\label{RouthArray}
\begin{array}{c|cc}
s^3 & 1 & 2 \zeta_n n \omega_n^2+(1+\gamma) \omega_n^2 \\
s^2 & n \omega_n+2 \zeta_n \omega_n & n(1-\gamma) \omega_n^3 \\
s^1 & \frac{\omega_n^2\left(2 \zeta_n n^2+2 n+4 \zeta_n^2 n+2 \zeta_n(1+\gamma)\right)}{n+2 \zeta_n} & 0 \\
s^0 & n(1-\gamma) \omega_n^3 & 0.
\end{array}
\end{equation}

To ensure stability, the first column of the Routh array must be positive \cite{aastrom2021feedback}. For $n,\omega_n,\zeta_n,\gamma > 0$, the first three terms are always positive based on \eqref{RouthArray}. Hence, the following condition must be satisfied for the fourth term:
\begin{equation}
\label{RouthCondition}
(1-\gamma) n \omega_n^3>0,
\end{equation}

Hence, \eqref{NMPC_Gain_Eq} and \eqref{RouthCondition} imply stability is preserved for $0<\gamma<1$.

\textit{Marginal Stability for $\gamma=1$}: If the controller gain $k$ exactly inverses the system's dc gain ($k=g^{-1}$), one root of the characteristic equation \eqref{CharEqn} is at $s=0$, similar to a pure integrator, making the closed-loop marginally stable.

\subsubsection{Tuning NRC Gain}
\label{TuningNMPRCGain}
From the inner closed-loop function \eqref{ICL_Eqn}, it is evident that tuning $\gamma$ is influential for the frequency response characteristics of the inner closed-loop. Motivated by the dual closed-loop shaping guidelines \eqref{Eq_Objective2}, presented in \ref{LoopShapingGuidelines}, to reduce the effect of input reference $r$ to real error $e_r$, $\gamma$ is tuned to be unity such that $|G(s)C_d(s)| = 1$ and $\angle G(s)C_d(s) = \pi$ for $\omega \ll \omega_a$. 

As established in \ref{StabilityAnalysis}, tuning $\gamma=1$ will result in a pole being located at $s=0$, thus resulting in marginal stability. However, it will be shown later in \ref{Tuning_kp} that the tracking controller $C_t(s)$ in the outer loop can be tuned to guarantee sufficient stability margins. 

\begin{figure}[t!]
    \centering
    \includegraphics[width=1\columnwidth]{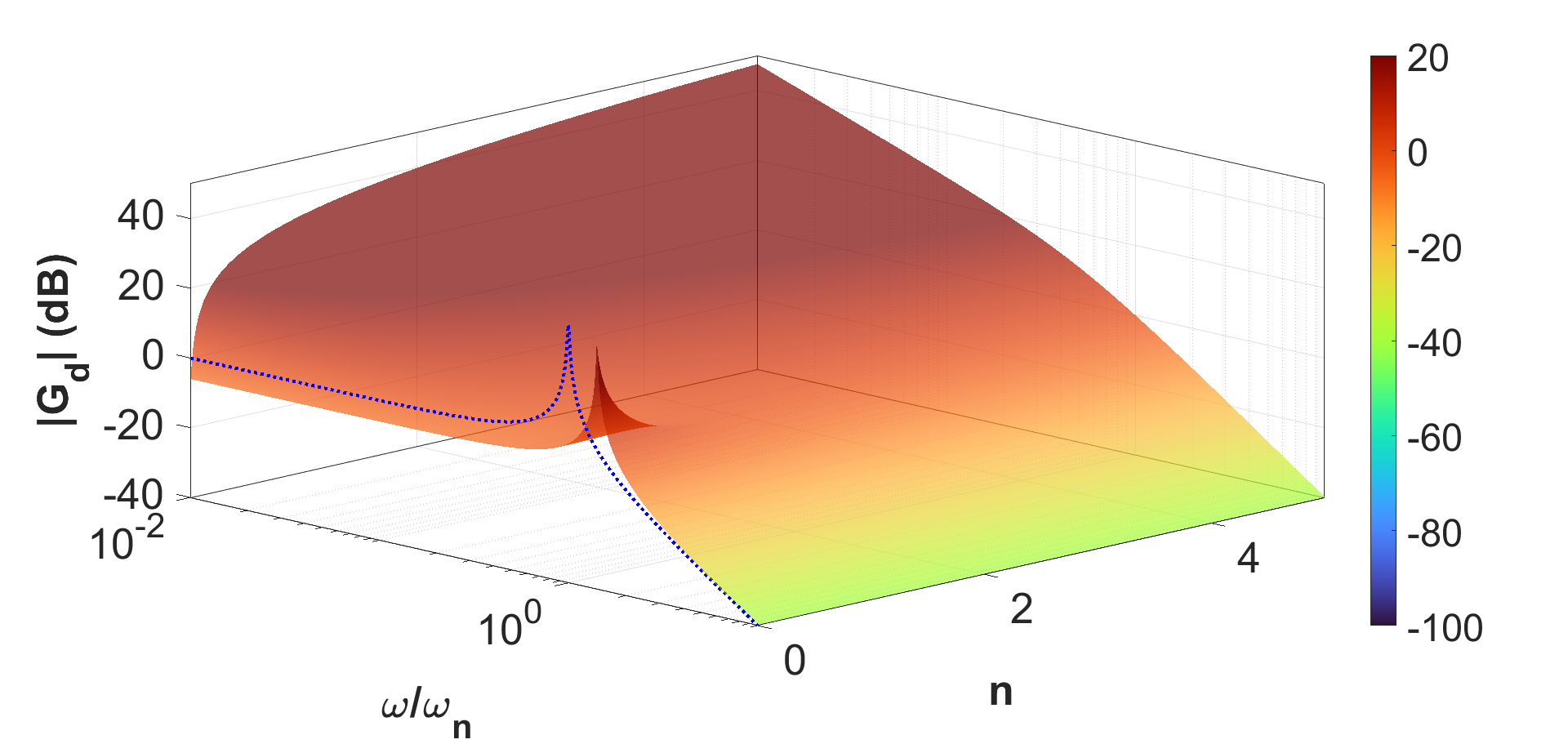}
    \caption{Frequency response magnitude $|G_d(s)|$ vs $n$ for $\gamma=1$. The dashed blue line depicts $|G(s)|$.}
    \label{fig:Damping_ControllerFrequency}
\end{figure}

Fig. \ref{fig:Damping_ControllerFrequency} illustrates the inner closed-loop magnitude response as the normalized controller frequency $n$ varies for the NRC tuned for $\gamma=1$. It can be deduced that for a fixed $\gamma$, the controller frequency $\omega_a$ needs to be further tuned to achieve inner closed-loop damped poles.

\subsubsection{Tuning NRC Corner Frequency}
Here, the achievable closed-loop damping is evaluated as a function of $\omega_a$ for the gain condition established in \ref{TuningNMPRCGain}. For $\gamma=1$, the three poles of $G_d(s)$ are located at: 
\begin{equation}
\label{ICL Poles}
\begin{aligned}
& p_1=0 \\
& p_{2,3}=\resizebox{0.85\hsize}{!}{$\frac{-\left(2 \zeta_n \omega_n+ n\omega_n\right) \pm \sqrt{\left(2 \zeta_n \omega_n+n\omega_n\right)^2-4\left(2 \zeta_n n\omega_n^2+2 \omega_n^2\right)}}{2}$}.
\end{aligned}
\end{equation}

From \eqref{ICL Poles}, it can be deduced that such a specific tuning of the damping controller enables manifesting an integrator effect in the inner closed-loop by placing a pole at zero and the remaining two poles, the location of which depends on $\omega_a$. 

The primary objective of the damping controller in this study is to effectively introduce damping to the lightly damped poles inherent in plant dynamics, which is achieved through strategic manipulation of the inner closed-loop poles, thereby facilitating a shift towards the left-hand plane (LHP) and consequently augmenting the negative real part of these poles. The damping ratio ($\zeta_d$) of the closed-loop double poles is characterized by:
\begin{equation}
\zeta_d = -\cos \left(\angle p_{2,3}\right),
\end{equation}
where $\angle p_{2,3}$ is the angle of the line connecting the pole to the origin in the s-domain measured clockwise from the negative real axis. 
\begin{figure}[t!]
    \centering
    \includegraphics[width=0.5\columnwidth]{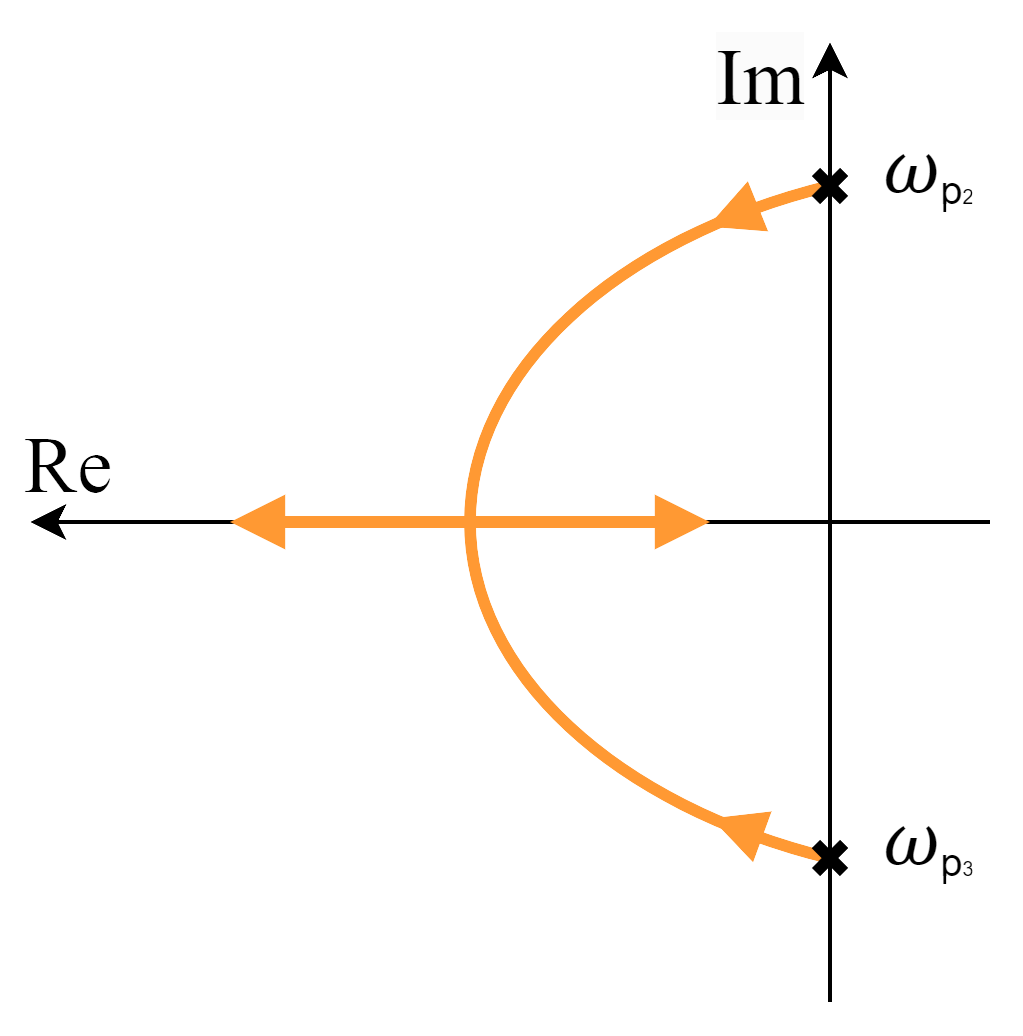}
    \caption{Root locus schematic of inner closed-loop double resonant poles $p_{2,3}$.}
    \label{fig:RootLocusSchematic}
\end{figure}

Fig. \ref{fig:RootLocusSchematic} provides a schematic representation of the typical root locus trajectory associated with the double resonant poles $p_{2,3}$. In a lightly damped system, these poles start as complex conjugates residing in the LHP, with their real parts becoming increasingly negative as the parameter $n$ escalates. However, at a certain $n = f(\zeta_n)$, these double poles coalesce and subsequently bifurcate along the negative real axis, signifying complete damping ($\Im(p_{2,3}) = 0$). Subsequently, they follow an opposite trajectory along the real axis. Thus, the condition for completely damped poles is given as: 
\begin{equation}
\label{Eq_DampingCondition}
    n \geq 2(\sqrt2 + \zeta_n).
\end{equation}

\begin{figure}[t!]
    \centering
    \includegraphics[width=1\linewidth]{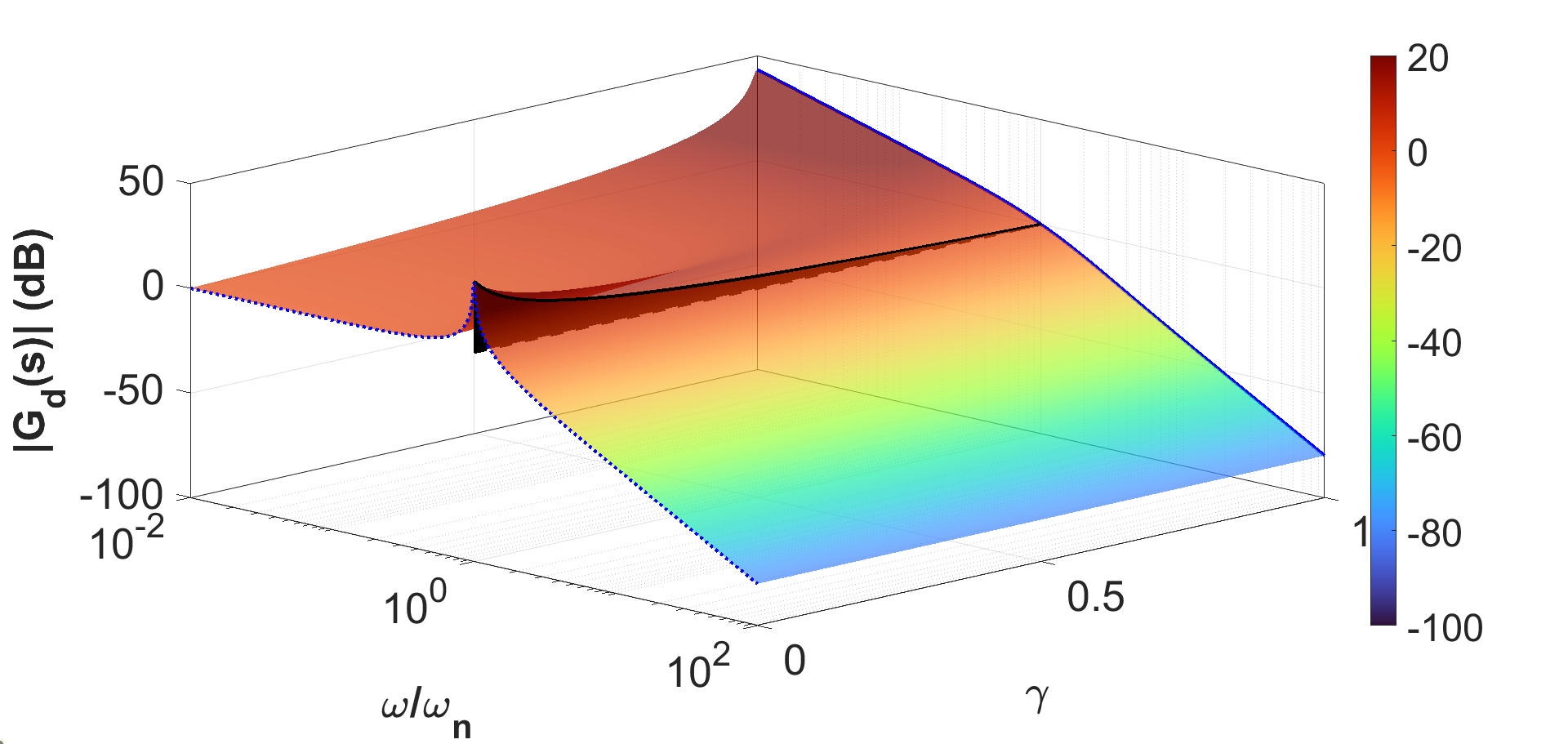}
    \caption{Frequency response magnitude $|G_d(s)|$ vs $\gamma$ for $n=3$. The dashed blue line depicts $|G(s)|$, solid blue depicts $|G_d(s)|$ when $n=3$, black line traces the resonance peak magnitude $|G_d(s)|_{\omega=\omega_n}$.}
    \label{fig:Damping_ControllerGain}
\end{figure}

Thus, for $n \gtrapprox 2\sqrt2$ (for a small damping ratio $\zeta_n \approx 0.01$), the double resonant poles $p_{2,3}$ can achieve complete damping. It should be noted that this condition is valid for the case when $\gamma=1$. Fig. \ref{fig:Damping_ControllerGain} illustrates the variation in the closed-loop frequency response $G_d(s)$ as $\gamma$ varies for a fixed $n=3$. 

\begin{figure}[t!]
  \centering
  {\includegraphics[width=1\columnwidth]{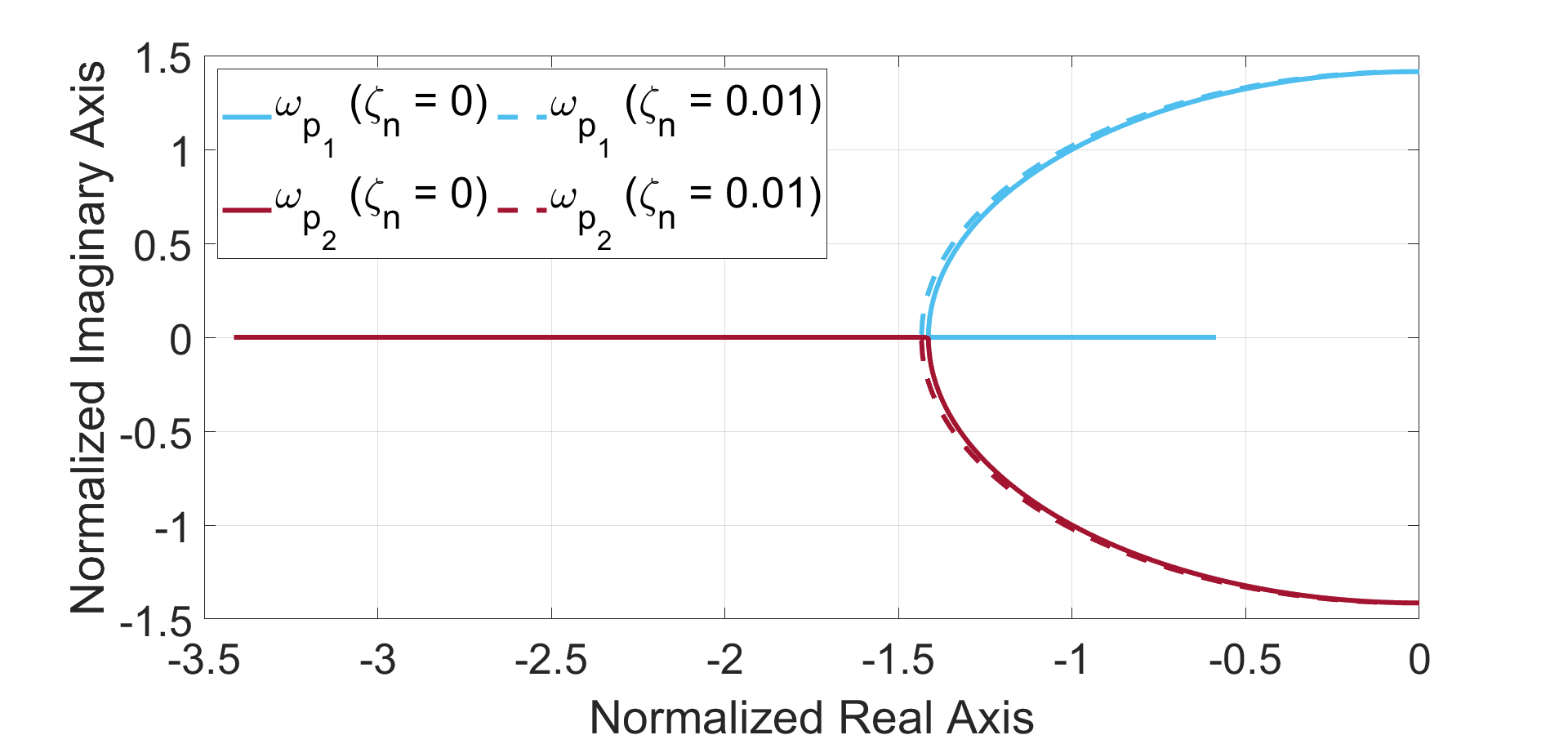}}
  \caption{Influence of plant damping on root locus of resonant poles $p_{2,3}$.}
  \label{fig:PlantDamping}
\end{figure}

\textit{Effect of Damping Ratio:} Despite the dependence of the damping ratio of the damped poles ($\zeta_d$) on the normalized corner frequency $n$, its impact remains negligible for typically lightly damped plants ($\zeta_n \approx 0.01$). Fig. \ref{fig:PlantDamping} underscores the minimal influence exerted by the plant's damping ratio on the root locus of double resonant poles.

\subsubsection{Effect of Closed-Loop Zero}
\begin{figure}[t!]
    \centering
    \includegraphics[width=\columnwidth]{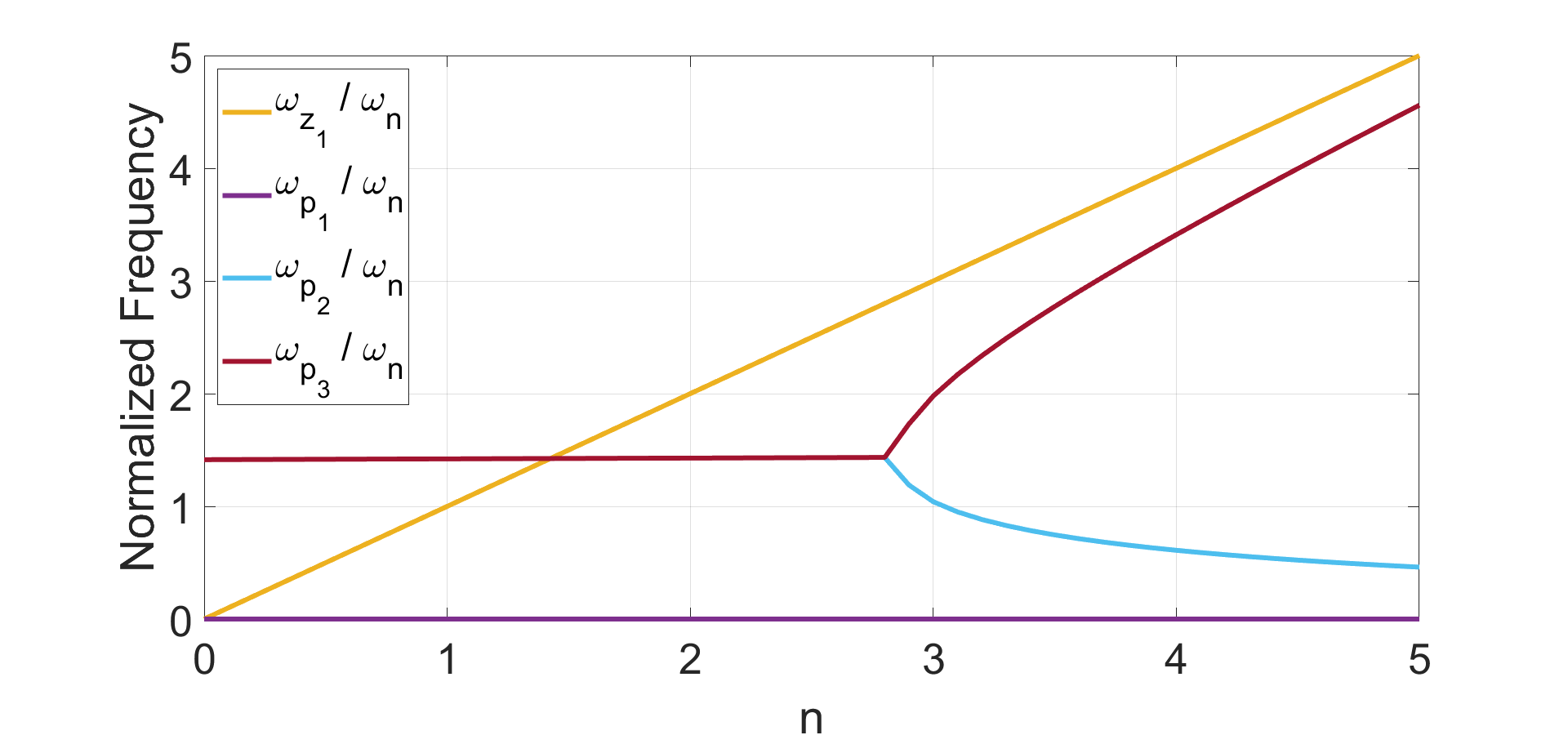}
    \caption{Inner closed-loop poles $p_{1,2,3}$ and zero $z_1$ variation with tuned controller normalized corner frequency $n$.}
    \label{fig:PZVariation}
\end{figure}

From \eqref{ICL_Eqn}, it is evident that the inner closed-loop transfer function consists of three poles ($p_{1,2,3}$) and one zero ($z_1$). The phase response of the inner closed-loop system depends on the positioning of these poles and the zero within the frequency domain. Although closed-loop zero remains fixed for a constant controller corner frequency, it was previously demonstrated that double poles follow a specific trajectory and split as $n$ increases. In scenarios characterized by low plant damping values, the split poles remain consistently lower in frequency than the zero frequency, with one pole situated closer to the zero frequency, as shown in Fig. \ref{fig:PZVariation}. 

The additional $-90^\circ$ phase shift introduced by the pole at a higher frequency is approximately counteracted by the $+90^\circ$ phase shift from the zero near its frequency. Consequently, in the absence of system delay, the inner closed-loop response transitions from $-90^\circ$ to $-180^\circ$ as the frequency increases. Once more, the system's damping ratio has minimal influence on the relationship between the closed-loop pole-zero frequencies. Henceforth, for the sake of mathematical convenience, the plant under consideration will be simplified as an undamped ($\zeta_n= 0$) second-order transfer function.

\subsubsection{Effect of Delay} 
\label{EffectofDelay}
As discussed in \ref{SystemArchitecture}, the dynamics of the nanopositioning system often contains significant delays that can be attributed to the extremely high resolution of the analog-to-digital conversion process and signal filtering. Thus, taking delay into account, the undamped system dynamics can be represented as: 
\begin{equation}
    \label{Eq_SystemDynamicsDelay}
    G(s,\tau) = \frac{g \omega_n^2}{s^2 + \omega_n^2} e^{-\tau s},
\end{equation}
where $\tau$ is the time delay in seconds.

To understand the influence of delay on the closed-loop performance, the analysis utilizes the $1^{\text{st}}$-order Pade approximation of the delay term, represented as follows: 
\begin{equation}
    \label{Eq_PadeApprox}
    e^{-\tau s} \approx \frac{1 - \frac{\tau}{2}s}{1 + \frac{\tau}{2}s} = \frac{\omega_b - s}{\omega_b + s},
\end{equation}
where $\omega_b = 2/\tau$ captures the frequency dependence of time delay to the induced phase lag. Here, we introduce $m=\omega_b/\omega_n$ representing the normalized frequency to parametrize the phase lag due to the delay around the system resonance frequency $\omega_n$. 

Thus, the closed-loop dynamics with delay, for $\gamma=1$, can be computed as: 
\begin{equation}
    \label{Eq_DelayClosedLoop}
    \begin{aligned}
         G_d(s,\tau) &=\frac{-w_n^2\left[s^2+(\omega_a-\omega_b) s-\omega_a \omega_b\right]}{s^4+(\omega_a+\omega_b) s^3+\omega_a \omega_b s^2+2 w_n^2(\omega_a+\omega_b) s} \\
         &= \frac{-w_n^2\left[s^2+(n-m) w_n s-m n w_n^2\right]}{s^4+(n+m) w_n s^3+ n m w_n^2 s^2 + 2 (n+m) w_n^4 s}.
    \end{aligned}
\end{equation}

\begin{figure}[t!]
    \centering
    \includegraphics[width=1\linewidth]{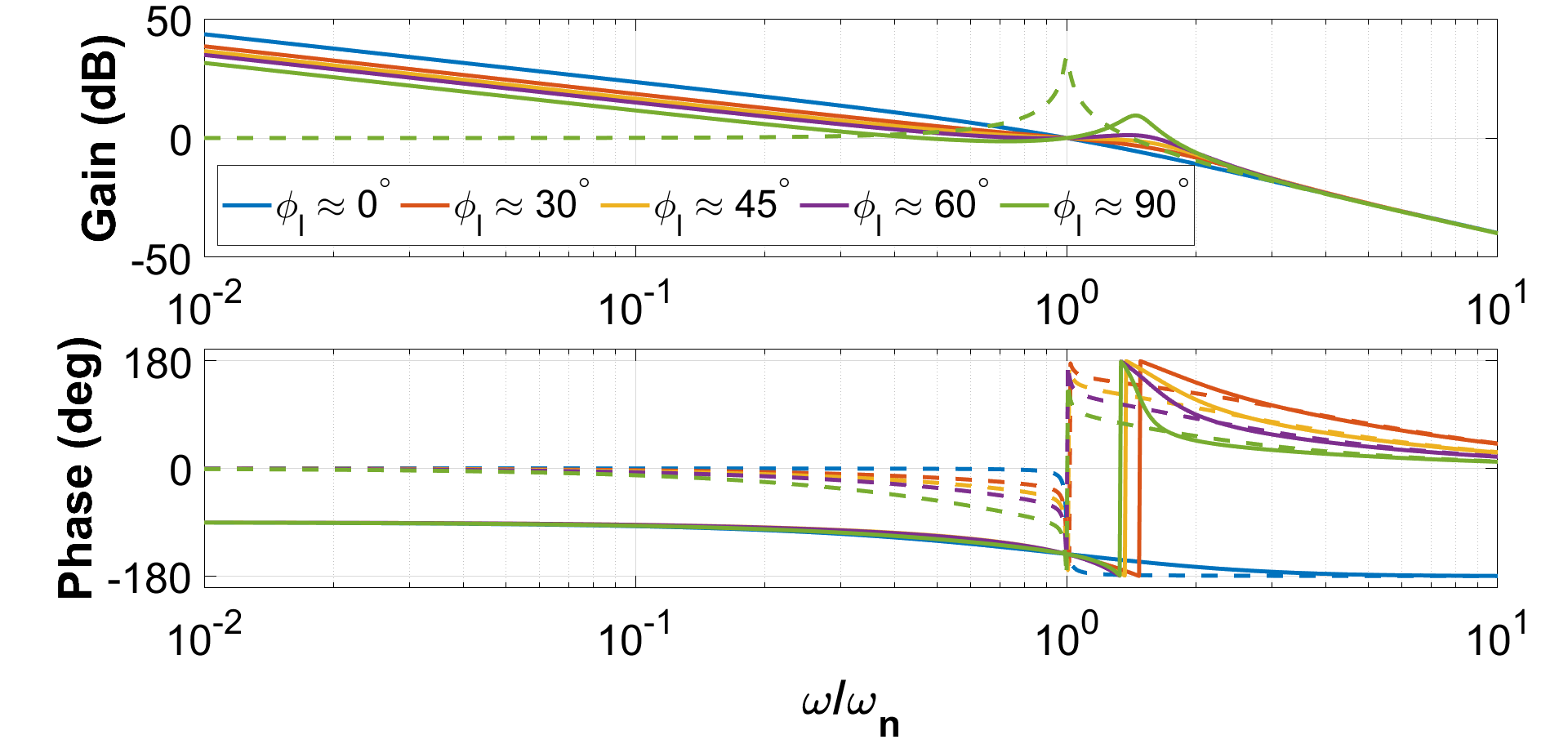}
    \caption{Illustration of the effect of delay on the closed-loop performance. Dashed lines depict $G(s,\tau)$, and solid lines depict $G_d(s,\tau).$}
    \label{fig:RobustnessDelay}
\end{figure}

The denominator of \eqref{Eq_DelayClosedLoop} thereby reduces to a quartic equation with a zero constant term, where the $4^{\text{th}}$-pole ($p_4$) is introduced due to the delay term. For mathematical brevity, a numerical illustration, with the NRC tuned for $n=3$, is shown in Fig. \ref{fig:RobustnessDelay}, to depict the influence of delay on inner closed-loop dynamics. The parameter $m$ is varied decreasingly to correspond to the induced phase lags ($\phi_l$) of $\approx 0-90\deg$ at $\omega_n$, in the system $G(s,\tau)$. Due to additional delay, the inner closed-loop gain in low frequencies ($\omega<\omega_n$) and high frequencies ($\omega>\omega_n$) are affected due to the NRC gain ($k$) and corner frequency ($\omega_a$) respectively. However, since the system delay is known via identification, NRC can be re-tuned to avoid exaggerated effects.

\subsection{Robustness to Resonance Frequency Variations}
\label{Robustness_LoadVariations}
Often, applications require samples/payloads to be positioned using the nanopositioning stages, the mass ($m_p$) of which affects the resonance frequency ($\omega_n \propto \sqrt{\frac{k}{m_s + m_p}}$; $k$ and $m_s$ being the flexural stiffness and stage mass, respectively). Although typical damping controllers are tuned for the identified nominal (unloaded) system, their damping performance can be severely affected in closed-loop with significant variations in system resonance frequency and therefore require more advanced adaptive algorithms for real-time tuning \cite{yong2012design}. 
\begin{figure}[b!]
    \centering
    \includegraphics[width=1\linewidth]{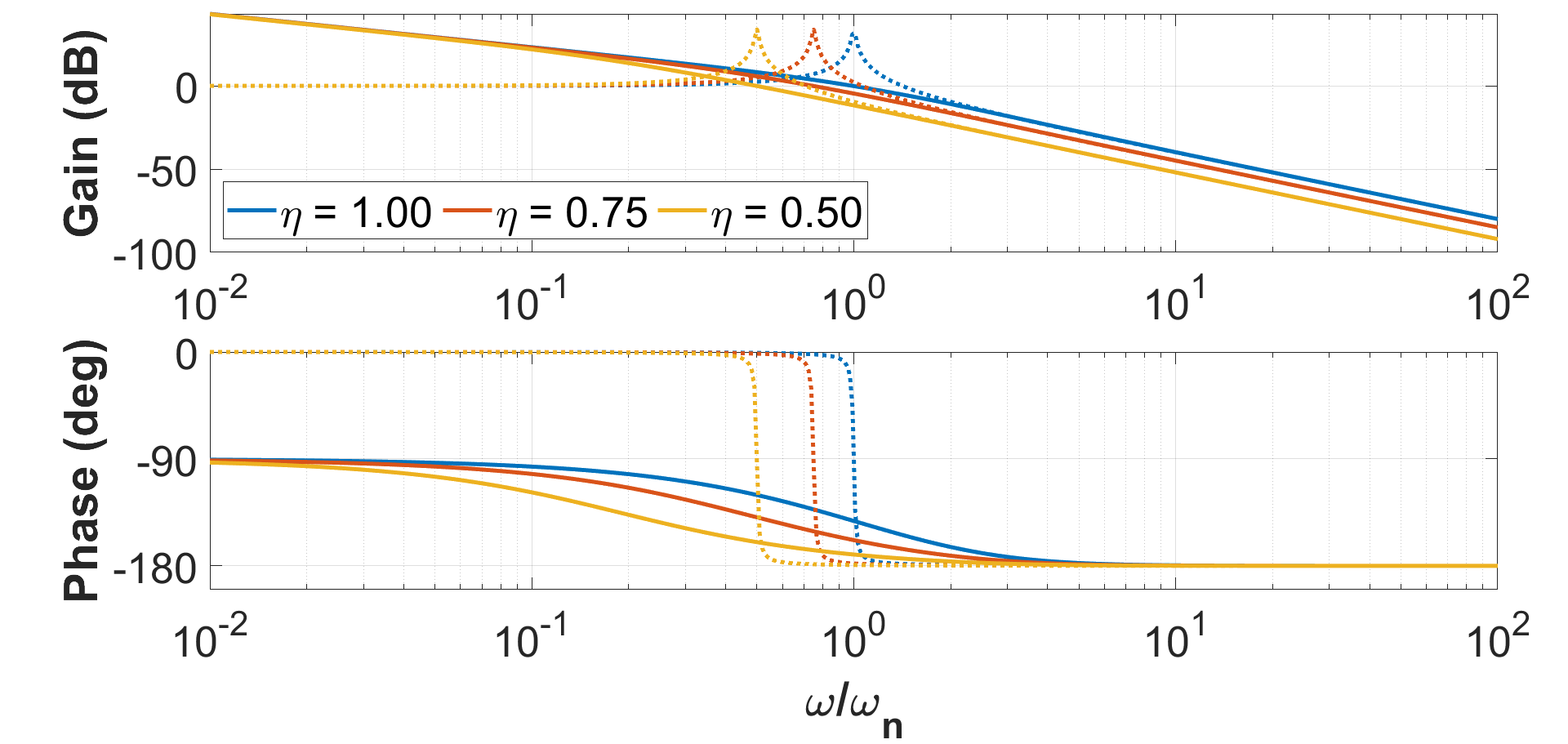}
    \caption{Illustration of the robustness of NRC to variations in resonance frequency. Dotted lines depict $\hat{G}(s)$ and solid lines depict $\hat{G}_d(s)$}.
    \label{fig:Robustness_Load}
\end{figure}
This section examines NRC's damping robustness under varying resonance frequencies. The altered frequency is $\hat{\omega}_n=\eta\omega_n$, with $\eta \in (0,1)$, and the loaded plant is modeled by $\hat{G}(s)=\frac{g\hat{\omega}_n^2}{s^2 + 2\zeta_n\hat{\omega}_n s + \hat{\omega}_n^2}$. For NRC tuned with $\gamma=1$ and $\omega_a = n\omega_n$ in an unloaded state, the inner closed-loop is given by:
\begin{equation}
    \label{Eq_ICL_Load}
    \hat{G}_d(s) = \frac{g \hat{\omega}_n^2\left(s+n \omega_n\right)}{s^3+\left(n \omega_n+2 \zeta_n \hat{\omega}_n\right) s^2+\left(2 \zeta_n n \omega\hat{\omega}_n+2\hat{\omega}_n^2\right) s}.
\end{equation}

The condition for complete damping, as derived from \eqref{ICL Poles} and \eqref{Eq_DampingCondition}, is:
\begin{equation}
    \label{Eq_LoadDampingCondition}
    n \geq 2\eta(\sqrt{2}+\zeta_n).
\end{equation}

Since $\eta<1$, any $n$ that satisfies \eqref{Eq_DampingCondition} also satisfies \eqref{Eq_LoadDampingCondition}, indicating that an NRC set for a nominal system achieves complete damping across loaded systems, showing its robustness to changes in resonance frequency. Fig. \ref{fig:Robustness_Load} demonstrates this, using an NRC tuned for the unloaded case ($\eta=1$) to dampen two loaded conditions ($\eta=0.75,0.5$).

\subsection{Damping Multiple Resonant Modes}
\label{DampingMultipleModes}
The use of NRC to actively dampen the first dominant resonance mode has been extensively discussed in \ref{ADC_NMPC}. However, as briefly outlined in \ref{SystemArchitecture}, the presence of higher-order modes in the vicinity of the dominant resonance mode can also further lead to deterioration of the position accuracy, thereby necessitating the dampening of these modes. As investigated in further detail in this section, the NRC used for the first mode can also induce damping in these significant higher-order modes in the vicinity. 

\begin{figure}[t!]
    \centering
    \includegraphics[width=1\linewidth]{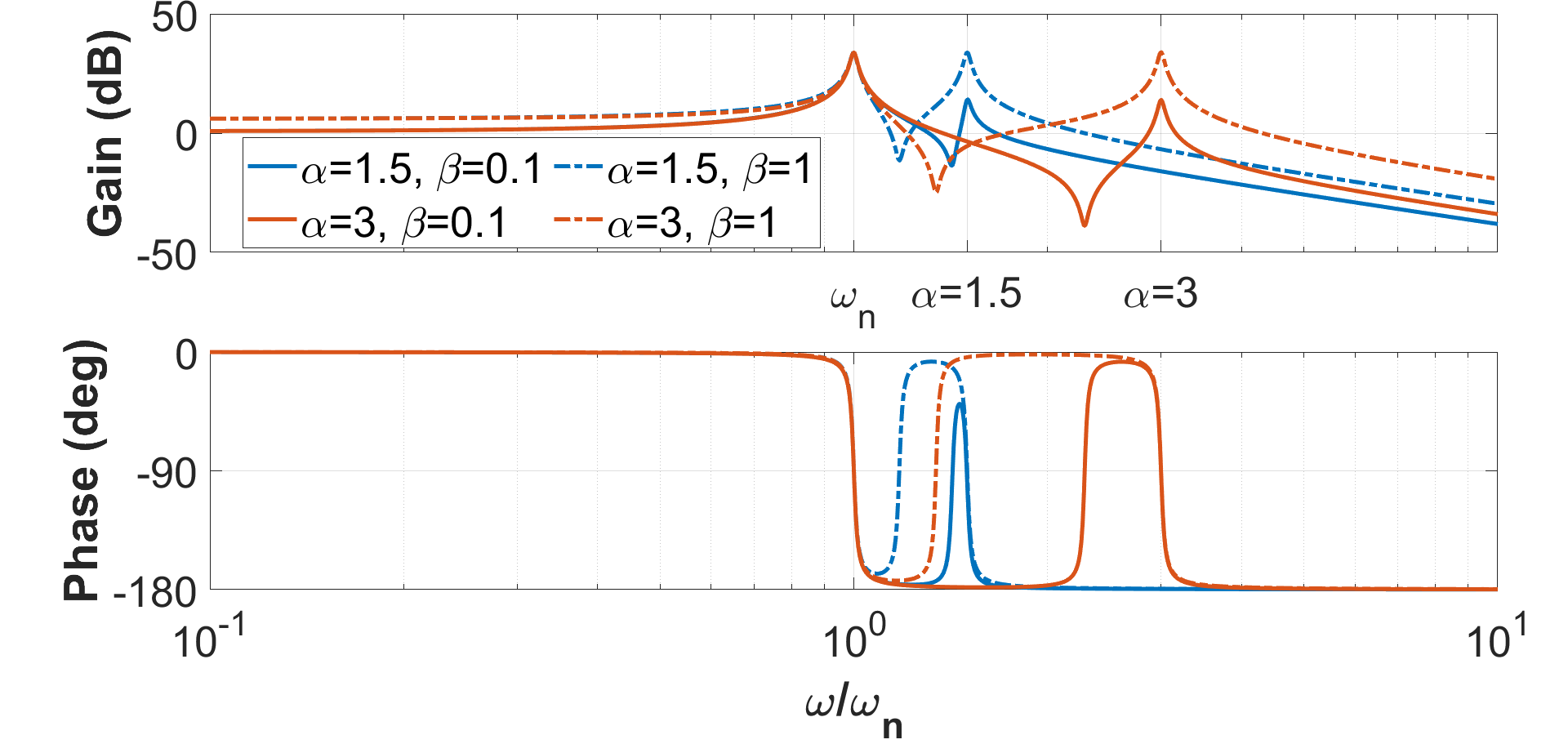}
    \caption{Illustration of Frequency Response of $G_2(s)$ as function of $\alpha$ and $\beta$.}
    \label{fig:TwoModeSystem}
\end{figure}

For simplicity, an undamped system with two modes is considered for the analysis. Based on the theory of modal decomposition, the system $G_2(s)$ can be expressed as: 
\begin{equation}
    \label{Eq_System2}
    \begin{aligned}
        G_2(s) &= \frac{\omega_n^2}{s^2 + \omega_n^2} + \beta\frac{\omega_2^2}{s^2 + \omega_2^2} \\
        &= \frac{(1+\alpha^2\beta)\omega_n^2s^2 + \alpha^2\omega_n^4(1+\beta)}{s^4 + (1+\alpha^2)\omega_n^2s^2 + \alpha^2\omega_n^4},
    \end{aligned}
\end{equation}
where $\omega_2 > \omega_n$ is the resonance frequency of the second mode, and the factor $\beta \in (0,1)$ relates the high-frequency contribution of the second mode to that of the first mode. A parameter $\alpha > 1$ is introduced to normalize $\omega_2$ with respect to $\omega_n$, as $\omega_2 = \alpha\omega_n$. The sum of the two modes leads to a double zero $\omega_z$ being created such that $\omega_n<\omega_z<\omega_2$, the location of which is given by $s=\pm i\alpha\omega_n\sqrt{\frac{1+\beta}{1+\alpha^2\beta}}$. The DC gain is $G_{2}(0) = 1+\beta$. Fig. \ref{fig:TwoModeSystem} illustrates the frequency response of $G_2(s)$ as the parameters $\alpha$ and $\beta$ vary. 

The inner closed-loop frequency response $G_{2d}(s)$ with an NRC tuned to dampen the first resonance mode ($k=\frac{\gamma}{1+\beta}$), is symbolically computed as:
\begin{equation}
    \label{Eq_ICL_TwoMode}
    \begin{aligned}
    G_{2d}(s) &= \frac{((1+\alpha^2\beta)\omega_n^2s^2 + \alpha^2\omega_n^4(1+\beta))(s+n\omega_n)}{c_5s^5 + c_4s^4 + c_3s^3 + c_2s^2 + c_1s + c_0}
    \\ \text{where,}& \\
        c_5 &= 1; \hspace{1mm}
        c_4 = n\omega_n; \\
        c_3 &= ((1+\alpha^2) + k(1+\alpha^2\beta))\omega_n^2; \\
        c_2 &= ((1+\alpha^2)-k(1+\alpha^2\beta))n\omega_n^3; \\
        c_1 &= \alpha^2\omega_n^4(1+k(1+\beta)); \\
        c_0 &= \alpha^2n\omega_n^5(1-k(1+\beta)). 
    \end{aligned}
\end{equation}

The magnitude of $G_{2d}(s)$ at $\omega_2$ is calculated by substituting $s=i\alpha\omega_n$ and simplified as:
\begin{equation}
    \label{Eq_Damping2mode}
    |G_{2d}(s)|_{\omega=\alpha\omega_n} = \frac{1}{k} = \frac{1+\beta}{\gamma}.
\end{equation}

\begin{figure}[t!]
    \centering
    \includegraphics[width=1\linewidth]{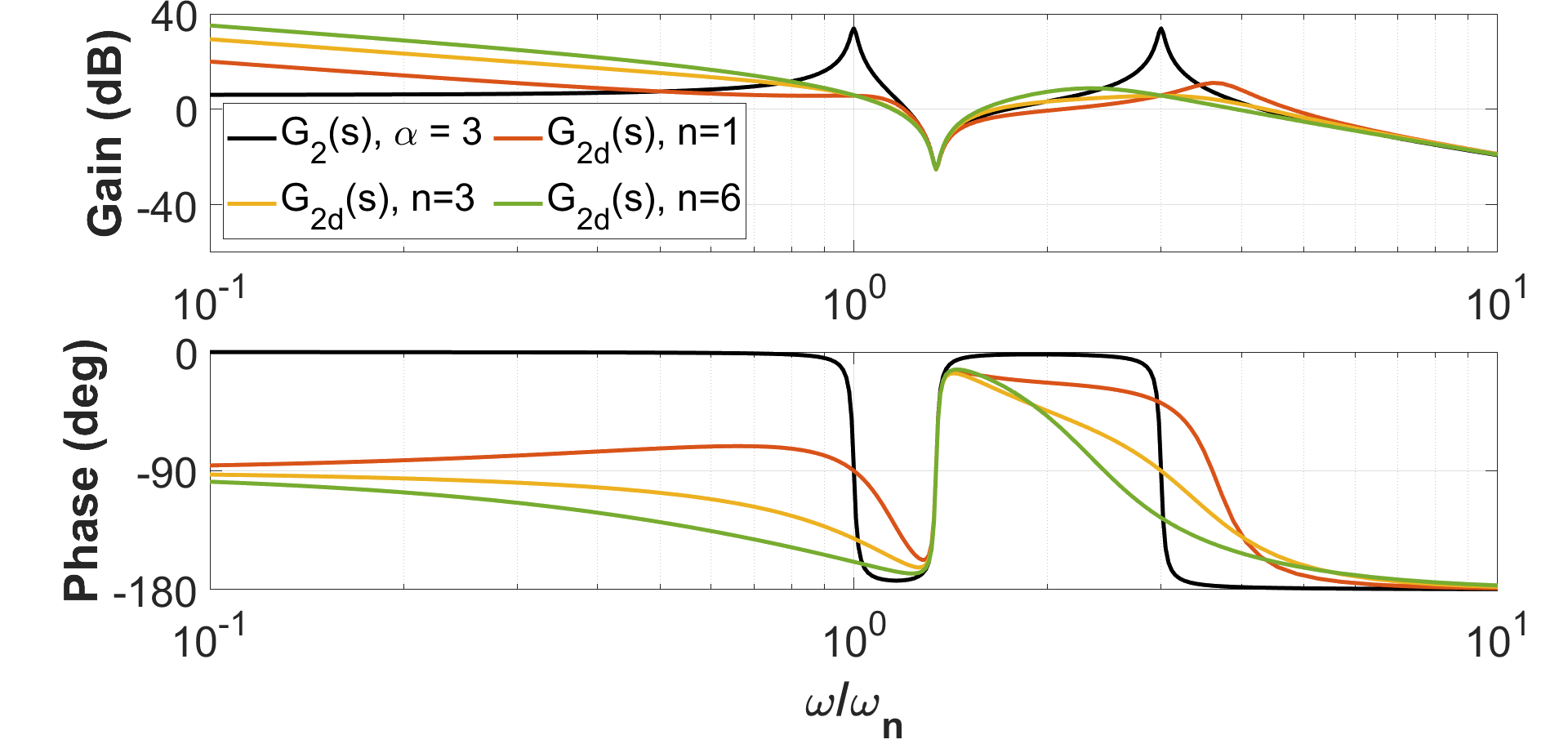}
    \caption{Illustration of frequency response of $G_{2d}(s)$ as $n$ varies relative to $\alpha$.}
    \label{fig:TwoModeSystemDamped}
\end{figure}

When comparing \eqref{Eq_Damping2mode} with $|G_2(s)|_{\omega=\alpha\omega_n} \approx \infty \text{ for } \zeta_n \approx 0$, based on the reduction of magnitude, it can be deduced that damping is achieved in the second resonance mode. However, the second resonance peak frequency, $\hat{\omega}_2$ in closed-loop could vary from $\omega_2$ based on the relation of $n$ with respect to $\alpha$, such as when (a) $n=\alpha$, then $\hat{\omega}_2=\omega_2$, (b) $n<\alpha$, then $\hat{\omega}_2>\omega_2$, (c) $n>\alpha$, then $\hat{\omega}_2<\omega_2$, as illustrated in Fig. \ref{fig:TwoModeSystemDamped}. It is, however, to be noted that delay in the system shall affect the magnitude of damping induced, as highlighted in \ref{EffectofDelay}.

\subsection{Taming NRC}
The NRC lacks roll-off at high frequencies owing to its constant gain characteristic. This leads to the noise being fed back into the closed-loop system when augmented by the amplifier, which can potentially lead to the accumulation of real errors in the system due to the significant higher-order modes present. Mathematically, the NRC output signal $v$ (see Fig. \ref{fig:Inner_CL_Diagram}) can be represented as $v = C_d(s)\cdot y = C_d(s)\cdot (x+n)$. At high frequencies, where noise typically becomes predominant, due to the system roll-off ($x\approx 0$ as $\displaystyle \lim_{s\rightarrow \infty} G(s) = 0$), the noise entering the amplifier through the inner feedback loop can be expressed as: 
\begin{equation}
    v = \lim_{s\rightarrow\infty}C_d(s)\cdot n \approx k\cdot n.
    \label{NoiseApproxEquation}
\end{equation}

\eqref{NoiseApproxEquation} emphasizes the importance of high frequency roll-off in $C_d(s)$ to attenuate the transmitted noise, if significant. Motivated by the dual closed-loop shaping guideline \eqref{Eq_Objective4}, presented in Subsection \ref{LoopShapingGuidelines}, the NRC can be tamed using a simple $1^{\text{st}}$-order low-pass filter to achieve $|C_d(s)| \ll 1 \hspace{1mm} \forall \hspace{1mm}\omega \gg \omega_n$. However, such a taming can affect the damping performance in the closed-loop system, depending on the filter taming frequency $\omega_l$, as further investigated in the following. The taming frequency is parameterized by normalizing with respect to resonance frequency as $\omega_l = l\cdot\omega_n$.

The tamed NRC filter can be expressed as:
\begin{equation}
    \label{Eq_TamedNMPRC}
    C_{d_t}(s) = k \left(\frac{s-\omega_a}{s+\omega_a}\right) \left(\frac{\omega_l}{s+\omega_l}\right).
\end{equation}

Thus, the inner closed-loop equation of the tamed NRC (tuned for $\gamma=1$) in feedback with a $2^{\text{nd}}$-order undamped system can be computed as: 
\begin{equation}
    \label{Eq_ICL_tamed}
    G_{d_t}(s)= \frac{\omega_n^2(s^2 + (n+l)\omega_n s + nl\omega_n^2)}{s(s^3 + (n+l)\omega_n s^2 + (nl+1)\omega_n^2 s + (n+2l)\omega_n^3}.
\end{equation}

\begin{figure}[t!]
    \centering
    \includegraphics[width=1\linewidth]{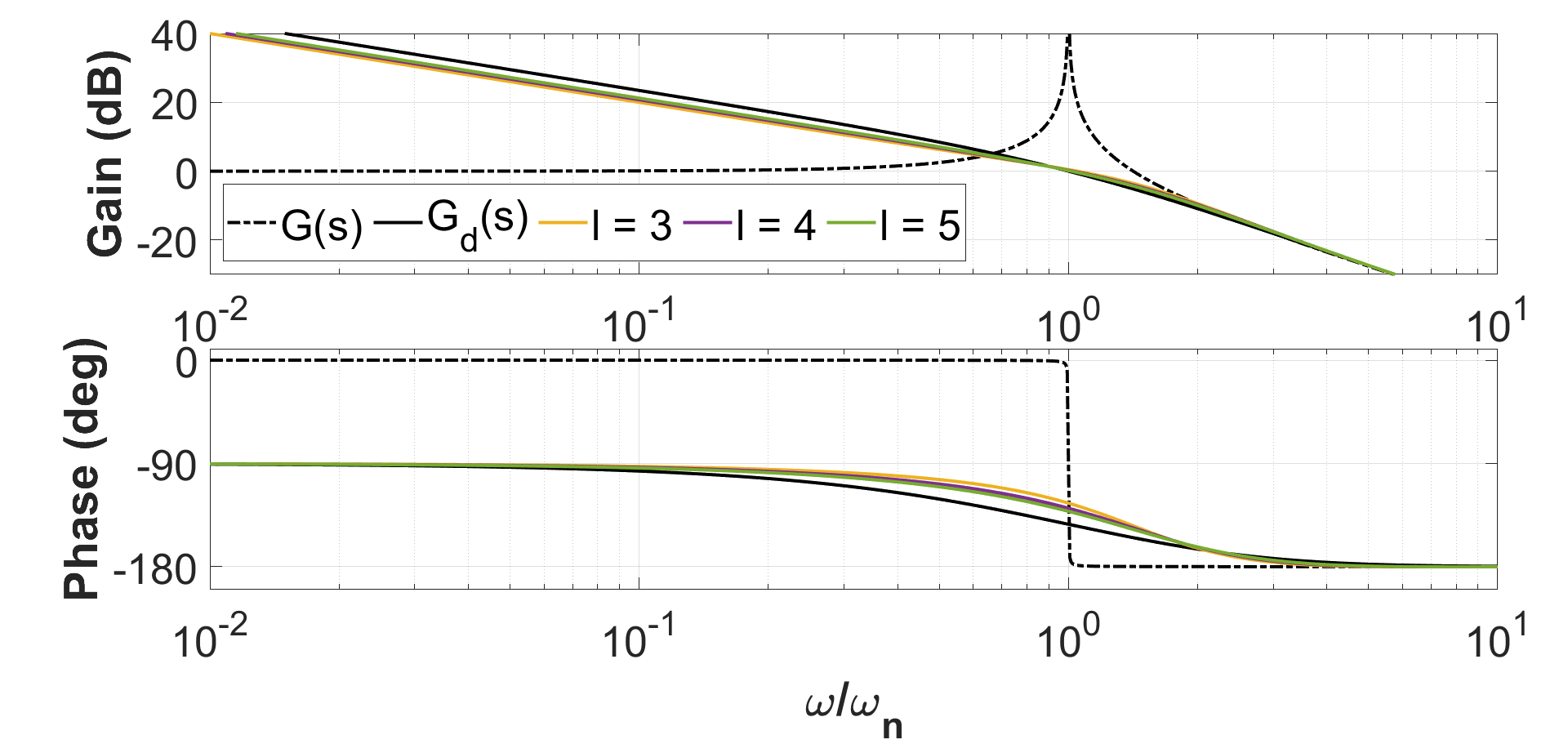}
    \caption{Illustration of variation in $G_{d_t}(s)$ as parameter $l$ varies. $G_d(s)$ is presented in a black solid line for comparison.}
    \label{fig:TamingNMPRC_FreqResp}
\end{figure}

The denominator of \eqref{Eq_ICL_tamed} simplifies to a quartic equation with a zero constant term, introducing the $4^{\text{th}}$-pole ($p_4$) via the taming filter. The damping of the double resonant poles ($p_{2,3}$) is dependent on the parameter $l$. Fig. \ref{fig:TamingNMPRC_FreqResp} shows how the inner closed-loop frequency response $G_{d_t}(s)$ varies when the tamed NRC $C_{d_t}(s)$ is tuned for $n=3$ \eqref{Eq_DampingCondition} as $\omega_l$ changes, highlighting the variation of the damping performance.
\begin{figure}[t!] 
\captionsetup[subfloat]{farskip=0pt,captionskip=2pt}
    \centering
   \subfloat[]{\includegraphics[width=0.5\columnwidth]{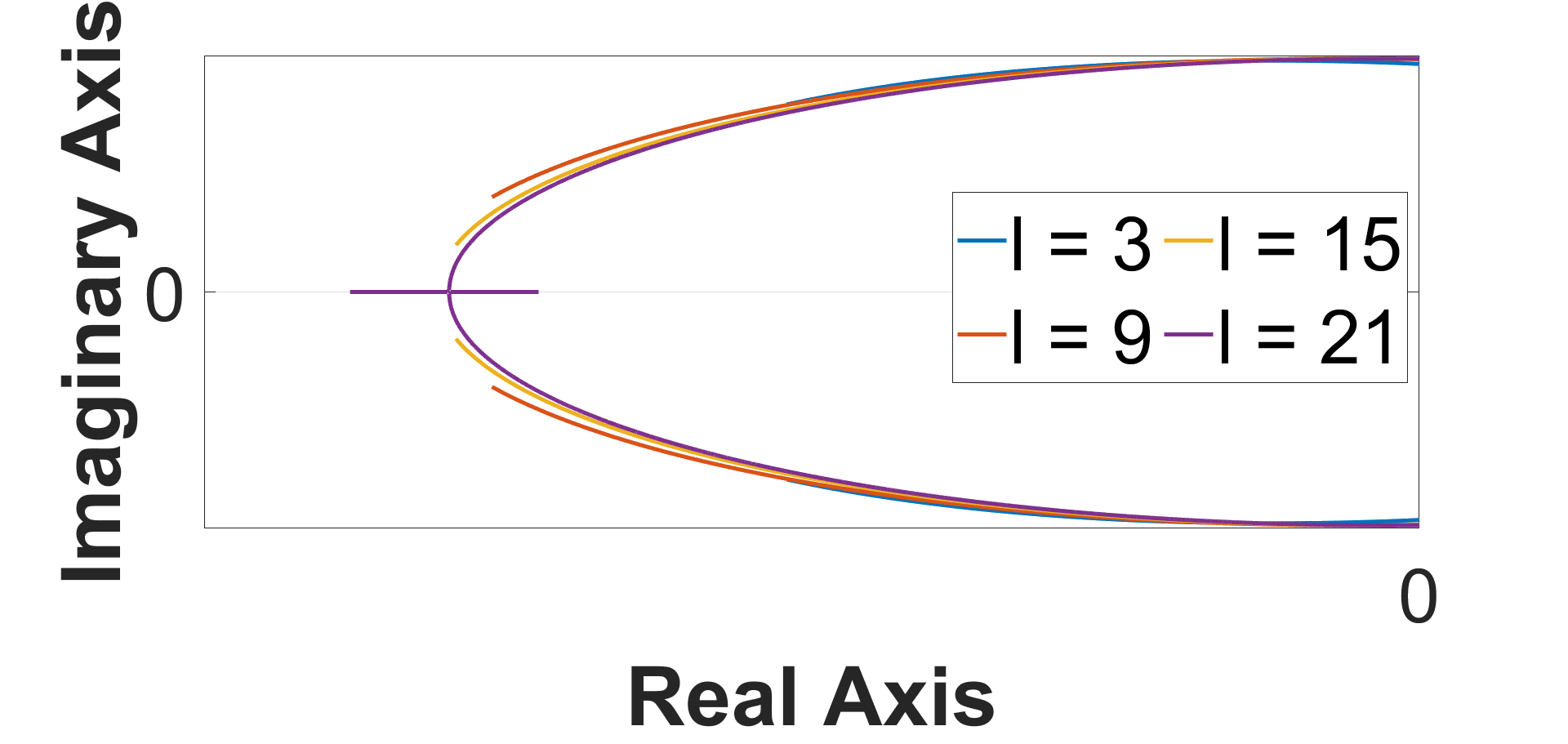} \label{fig:TameHigh}}
    \subfloat[]{\includegraphics[width=0.5\columnwidth]{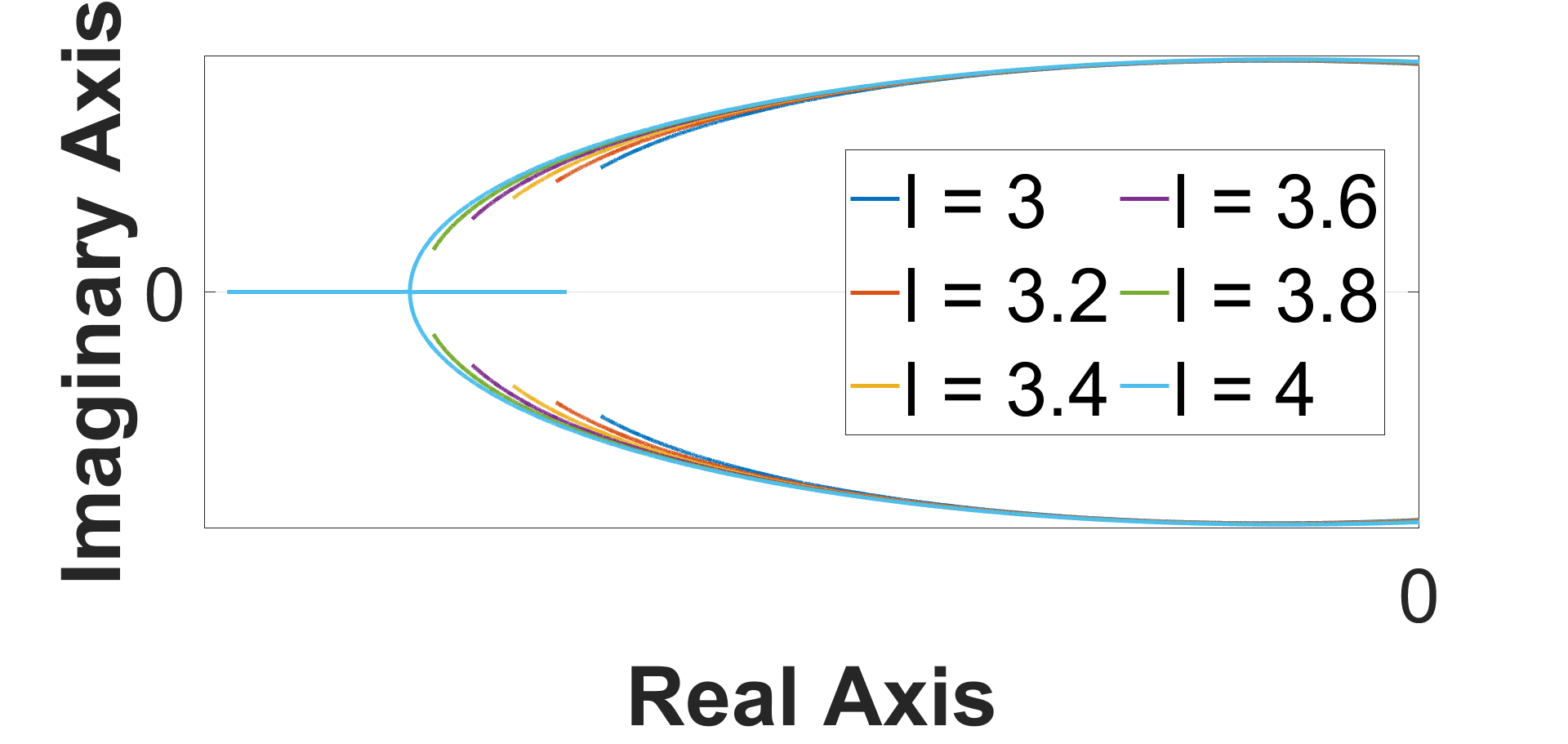}
    \label{fig:TameLow}}
  \caption{Root locus trajectory of double resonant poles $p_{2,3}$ when (a) $0\leq n \leq 3$ and complete damping is achieved for higher $l$, (b) $0\leq n \leq 4.5$ and complete damping is achieved for smaller $l$.}
  \label{fig:TamedRootLocus} 
\end{figure}
Thus, $l$ affects the achievable damping performance. To ensure complete damping, $n$ must be re-tuned based on $l$. Fig. \ref{fig:TamedRootLocus}(a) shows the root locus trajectory of the double resonant poles when $0\leq n \leq 3$ and $l$ are varied. It can be observed that for complete damping to be achieved without re-tuning $n$, $l$ needs to be significantly increased. However, as illustrated in Fig. \ref{fig:TamedRootLocus}(b) (when $0\leq n \leq 4.5$), if $n$ is re-tuned to slightly higher values, complete damping can be achieved for much smaller $l$.

\section{Dual Closed-Loop Control based on NRC}
\label{Dual Loop Control based on NMPRC}
This section presents the combination of NRC with the conventional PI tracking controller and the controller tuning in the outer loop for desired dual closed-loop performance requirements.
\subsection{Motion Tracking Control Loop}
\label{MotionTrackingControlLoop}
In nanopositioning systems, a tracking controller $C_t(s)$ ensures accurate reference tracking, which is designed using a loop-shaping technique in the frequency domain. Typically, a Proportional-Integral (PI) controller is employed to set the desired bandwidth and ensure zero steady-state error, represented in transfer function form as:
\begin{equation}
\label{PI_Controller}
    C_t(s) = k_p \cdot \left( 1 + \frac{w_i}{s} \right),
\end{equation}
where, $k_p$ denotes the proportional gain, while $w_i$ represents the integrator corner frequency.

However, in configurations where an active damping controller is integrated into a dual closed-loop structure, as depicted in Fig. \ref{fig:ControlArchitectures}(b), the design of the tracking controller typically hinges on the dynamics of the inner closed-loop system. The following elucidates the controller design process based on the inner loop featuring the NRC for active damping.

\subsubsection{Proportional Gain} 
\label{Tuning_kp}
Incorporating an integrator in the inner loop simplifies the outer tracking loop with a proportional gain $k_p$. This setup achieves zero steady-state error in the dual closed-loop system, based on a specified open-loop bandwidth $\omega_b$, and calculated as:
\begin{equation}
k_p=\left|\frac{1}{G_d(s)}\right|_{\omega=\omega_b}.
\end{equation}

The selection of the desired bandwidth $\omega_b$ is critical to meeting the specified gain margin (GM) and phase margin (PM) criteria for robustness, as outlined below:
\begin{equation}
    \begin{aligned}
        \text{GM} & \geq \text{6 dB};\\
        \text{PM} & \geq 60^\circ.
    \end{aligned}
\end{equation}

Consequently, the outer open-loop transfer function $L(s)$ (with inner closed-loop), mapping from the reference input $r$ to the output $y$, is expressed as: 
\begin{equation}
\begin{aligned}
L(s) & =k_p \cdot G_d(s) \\ & =k_p \cdot \frac{G(s)}{1+G(s)C_d(s)} \\
& =k_p \cdot \frac{\omega_n^2(s+\omega_a)}{s\left(s^2+\omega_a s+2 \omega_n^2\right)},
\end{aligned}
\end{equation}
where, $k_p=\left|G_d(s)^{-1}\right|_{\omega=\omega_b}$. Thus, at $\omega_b$ (0 dB crossover frequency); $|L(s)|_{\omega=\omega_b} = 1$. The phase of $L(s)$ at $\omega_b$ can then be simplified and expressed as: 
\begin{equation}
\label{OL_Phase}
 \angle L(s)_{\omega=\omega_b}= \left(i\omega_b +\omega_a\right)\left[-\omega_a\omega_b^2-i\left(2 \omega_n^2-\omega_b^2\right) \omega_b \right].
\end{equation}

Using $\nu = \omega_n/\omega_b$ and $n = \omega_a/\omega_n$ to normalize \eqref{OL_Phase} ensures $\text{PM}\geq 60^\circ$, allowing the equation to be rewritten as:
\begin{equation}
\tan 60^{\circ}\leq\frac{n \cdot 2 \nu^3}{n^2 \nu^2-2 \nu^2+1}.
\end{equation}

\begin{figure}[t!]
    \centering
    \includegraphics[width=1\linewidth]{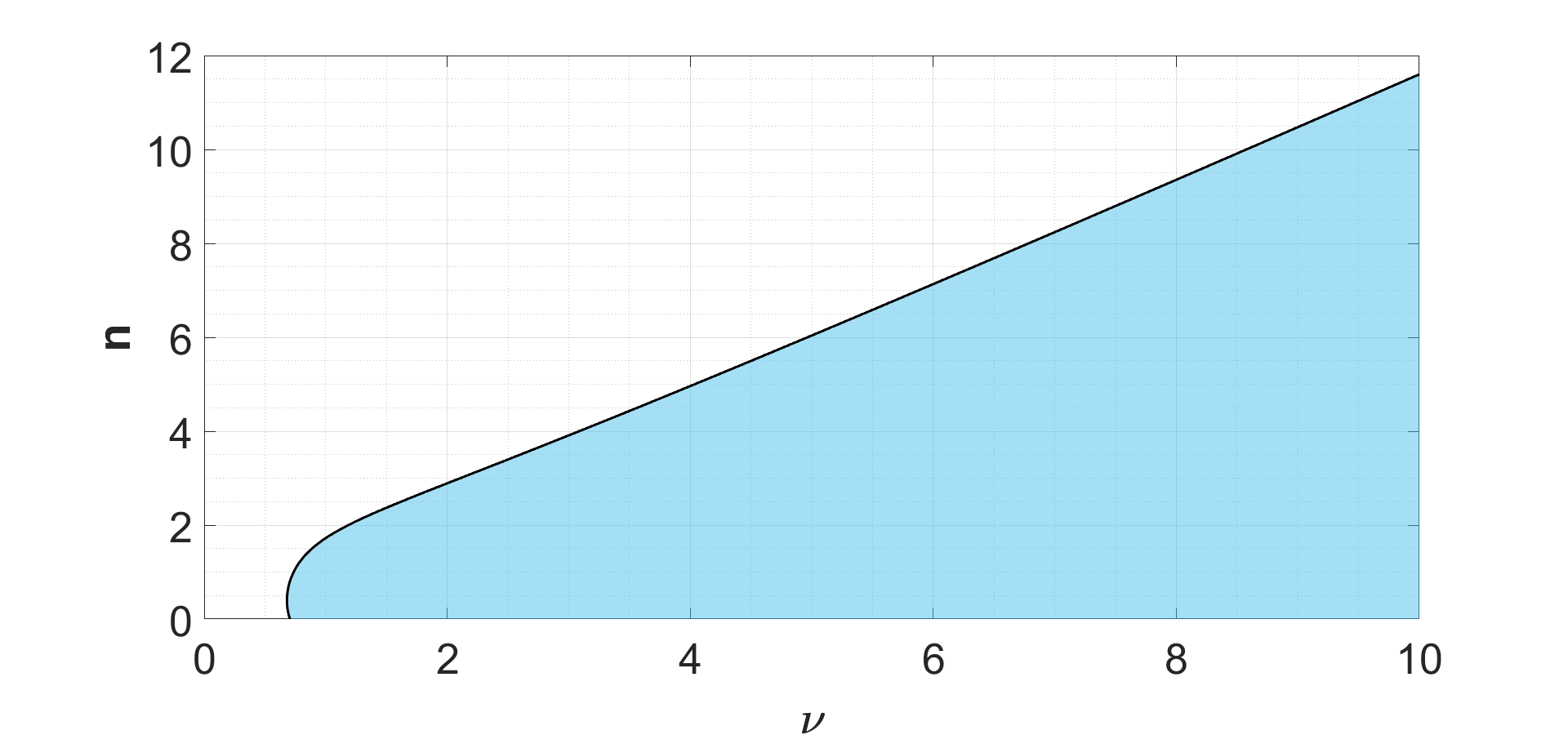}
    \caption{Illustration of solution space (shaded light blue) to tune $n$ for specific $\nu$ to ensure $60^{\circ}$ PM.}
    \label{fig:BandwidthEqSolution}
\end{figure}

The $n$ can be tuned for a selected $\omega_b$ to approximately satisfy the equation, as shown in Fig. \ref{fig:BandwidthEqSolution}. 
\begin{equation}
1.75 \nu^2 n^2-2 \nu^3 n+1.75\left(1-2 \nu^2\right)\leq 0.
\end{equation}

Thus, it can be inferred that to obtain high $\omega_b$ (or low $\nu$), a lower $n$ must be chosen, leading to a trade-off of achievable damping.

\subsubsection{Integrator requirement in Tracking Controller}
\eqref{ICL_Eqn} shows that perfectly inversely matching the controller gain with the plant's DC gain causes an integrator effect by placing a pole at zero in the inner closed-loop ($p_1=0$). However, exact gain matching is impractical due to limited numerical precision in discrete-time controller implementations.

For an undamped second-order system, assume that an NRC is tuned so that the controller gain $k^{\prime}=k+\Delta k$, where $\Delta k < 0$ is infinitesimally small compared to unity, but not zero ($|\Delta k| \approx 0$). Note that for $\gamma=1$ and $\Delta k >0$, $C_d(s)$ becomes unstable, requiring careful tuning of $\gamma\lessapprox1$ in practice. Consequently, $C_{d}^{\prime}(s) = k^{\prime}(s-\omega_a)/(s+\omega_a)$. Then, the inner closed-loop function from $d$ to $y$ can be expressed as:
\begin{equation}
\begin{aligned}
    G_d^{\prime}(s) &=\frac{G(s)}{1+G(s)C_d^{\prime}(s)} \\ &=\frac{w_n^2(s+\omega_a)}{s^3+\omega_a s^2+(2+\Delta k) w_n^2 s-\Delta k w_n^2 \omega_a}.
\end{aligned}
\end{equation}

When $|\Delta k| \neq 0$, the closed-loop characteristic equation has a small finite root instead of $s=0$. Therefore, when $C_t(s) = k_p$, the dual closed-loop system cannot achieve steady-state performance. Mathematically, for a reference input $r$, the Laplace domain representation of the error $e$ can be expressed as:
\begin{equation}
    \begin{aligned}
         E(s) &= R(s) - Y(s) \\
          &= \frac{1 + G(s)C_d(s)}{1+ G(s)(C_t(s) + C_d(s))} R(s).
    \end{aligned}
\end{equation}

The steady-state error $e_{ss}$, for a step reference input ($R(s) = 1/s$), is given by the final value theorem as:
\begin{equation}
e_{ss} = \lim _{s \rightarrow 0} s E(s) = \frac{2}{2 + k_p} \neq 0,
\end{equation}
where, 
\begin{equation}
    k_p =\left|G_d^{-1}(s)\right|_{\omega= \omega_b} \approx \left|\frac{\omega_n^{2} - \omega_b^{2}}{\omega_n^{2}}\right|.
\end{equation}

If the inner closed-loop lacks an integrator, the tracking controller must incorporate one to ensure zero steady-state error in the dual closed-loop system. The tracker $C_t(s)$ is then as shown in \eqref{PI_Controller}, but the integrator can have a very small corner frequency.

\begin{figure}[t!]
    \centering
    \includegraphics[width=1\linewidth]{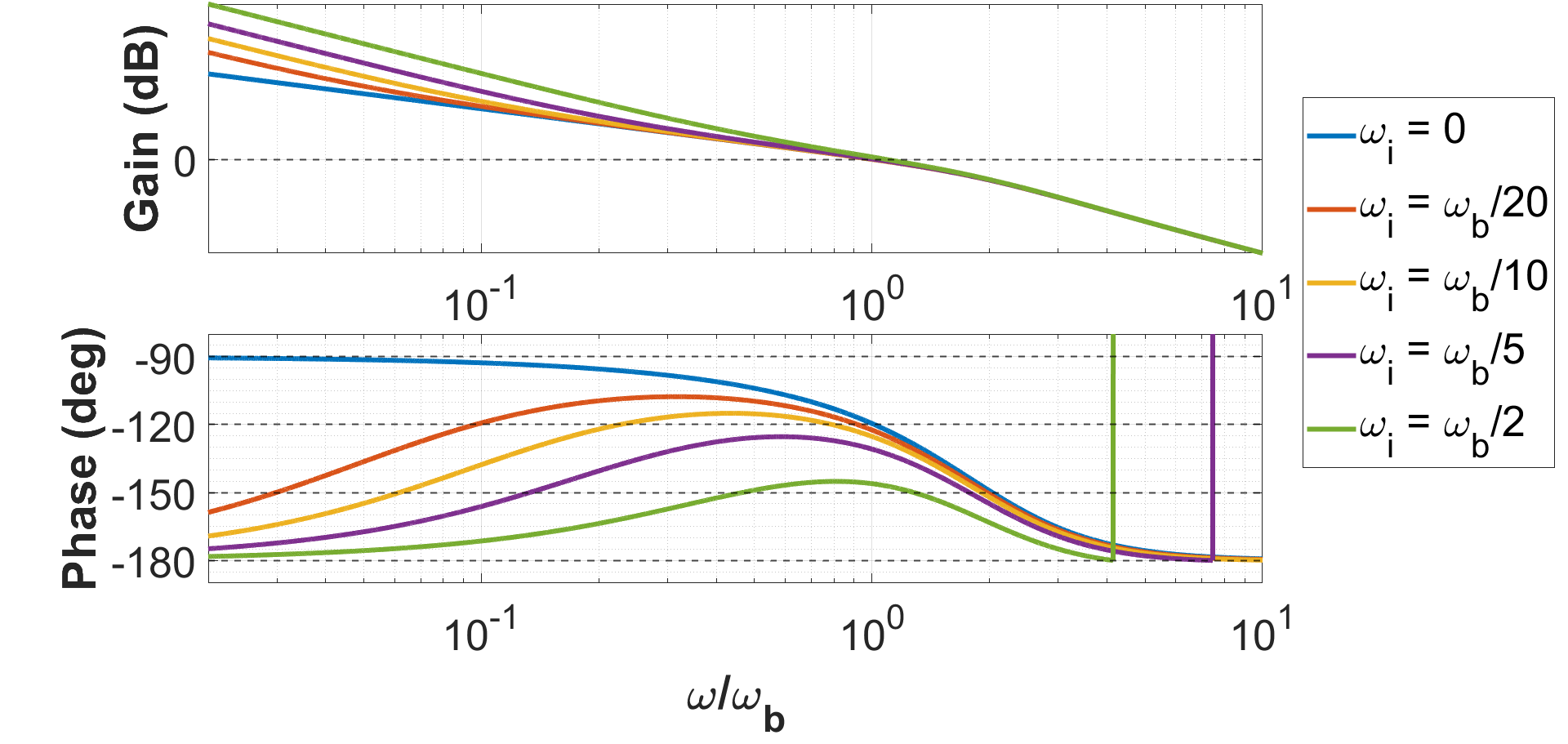}
    \caption{Influence of integrator frequency $\omega_i$ on open-loop $L(s)$ stability margins.}
    \label{fig:IntegratorFrequency}
\end{figure}

Fig. \ref{fig:IntegratorFrequency} illustrates the impact of the integrator corner frequency ($\omega_i$) on the open-loop phase margins, following the tuning guidelines outlined above. As $\omega_i$ increases relative to the tuned open-loop bandwidth ($\omega_b$), the phase lag introduced by the integrator becomes more pronounced, reducing the phase margin evaluated at $\omega_b$. However, this increase in $\omega_i$ also results in higher loop gains, as depicted in the magnitude plot, which further helps to achieve better disturbance rejection ($O_2$ \eqref{Eq_Objective2}). However, a trade-off must be met as this increase leads to a violation of upper bounds for dual closed-loop bandwidths (such as $\pm$1 dB or $\pm$3 dB) due to lower phase margins, as illustrated in the following section. Note for this illustration; the following parameters are chosen: $\nu\approx1.33$ and $n=2.2$ (sufficiently dampened resonance) and will be used for the study in the following section.

\subsection{Dual Closed-Loop Shaped Sensitivities}
\label{ClosedLoopDynamics}
In this section, different dual closed-loop functions, with a conventional PI-Controller ($C_t(s)$) and the NRC ($C_d(s)$) in feedback, will be illustrated based on the tuning guidelines of each, elaborated in \ref{ADC_NMPC} and \ref{MotionTrackingControlLoop}. 

\begin{figure}[t!]
\captionsetup[subfloat]{farskip=0pt,captionskip=3pt}
    \centering
    \subfloat[]{\includegraphics[width=\columnwidth]{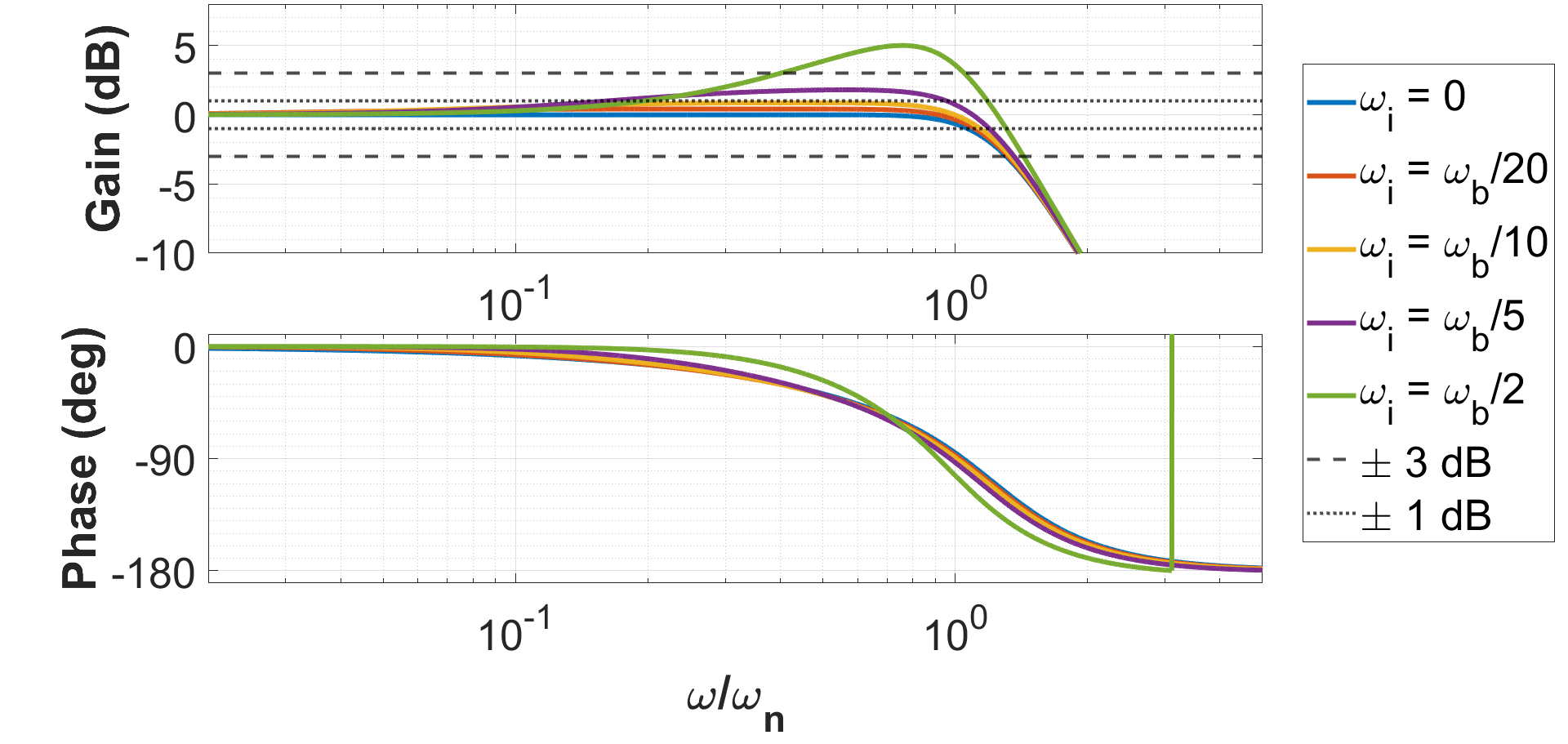}
    \label{fig:T_num}}
    \hfil
    \subfloat[]{\includegraphics[width=\columnwidth]{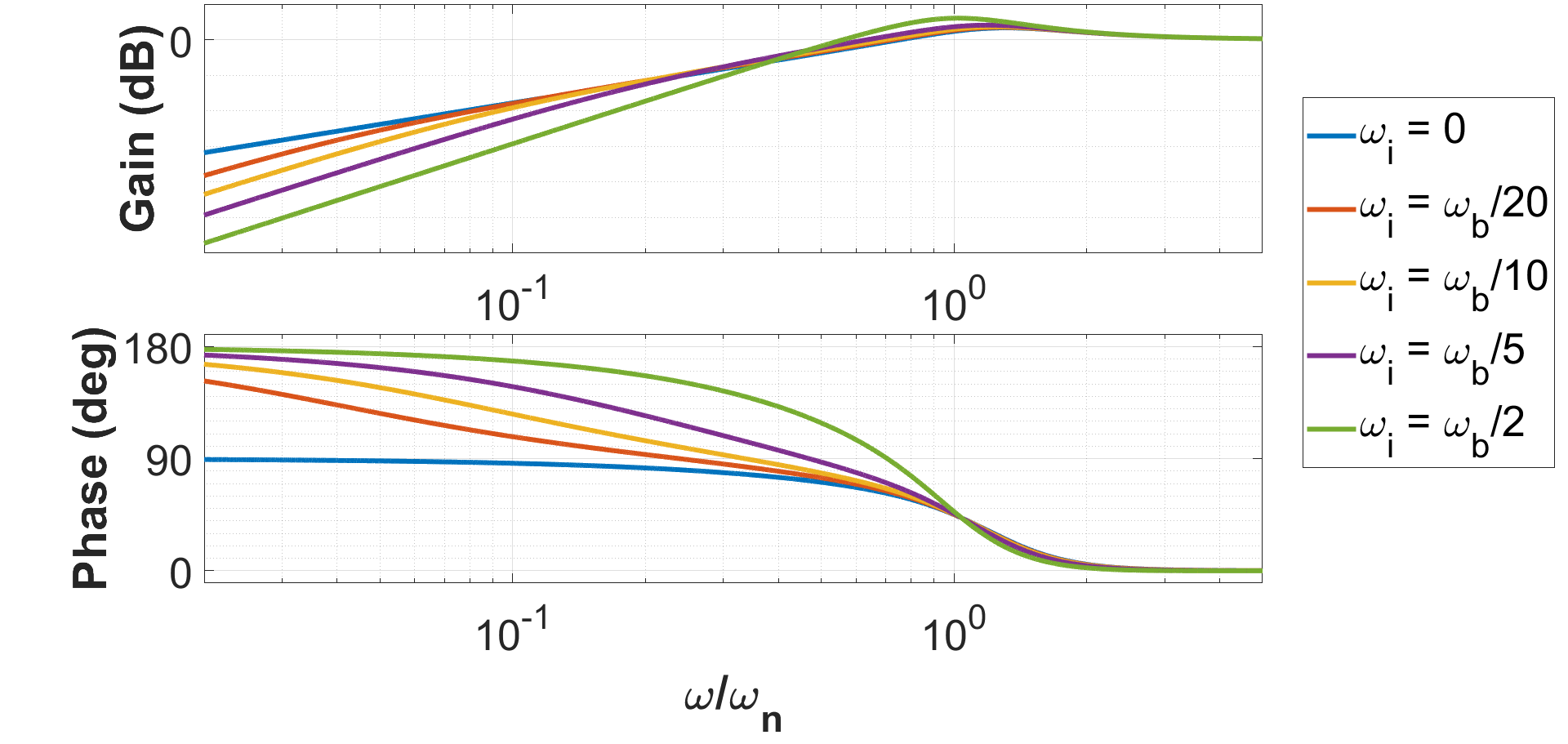}
    \label{fig:ES_num}}
    \caption{Illustration of dual closed-loop frequency response as $\omega_i$ varies (a) $T_{yr}(s): r \mapsto y$, (b) $T'_{xr}(s): r \mapsto e_r$.}
    \label{fig:DCL-TrackingSensitivities}
\end{figure}

The purpose of employing the NRC lies in its ability to completely dampen the resonance peak and enable the dual-closed-loop system to achieve high bandwidths. Fig. \ref{fig:DCL-TrackingSensitivities}(a) visually illustrates the dual closed-loop frequency response $T_{yr}(s)$ for a second-order system incorporating this control architecture and the associated tuning guidelines. The effect of the integrator frequency ($\omega_i$) within $C_t(s)$ is also included for demonstration purposes. As $\omega_i$ increases, it can be observed that the dual closed-loop bandwidth ($\omega_c$) bounds (±1 dB or ±3 dB) may be violated, leading to a reduction in $\omega_c$. Therefore, to meet stringent closed-loop tracking performance requirements, it is advisable to use an integrator with a relatively low $\omega_i$ to ensure zero steady-state error performance, while the high loop gain requirement is managed by the integrator effect provided by the inner closed-loop. It should be emphasized that with the specific tuning of $\nu$ and $n$, it is possible to achieve $\omega_c$, both ±1 dB or ±3 dB bounds, to be higher than the first resonance frequency $\omega_n$ of the system ($\omega_c>\omega_n$). However, this leads to a trade-off of not achieving complete damping of the resonance peak but still sufficiently well-dampened peak. 

In the complementary function $T^{\prime}_{xr}(s)$ (Fig. \ref{fig:DCL-TrackingSensitivities}(b)), it is observed that even without a significant integrator in the outer loop, lower sensitivity gains are achieved at low frequencies. This is primarily due to the tuning of NRC to achieve $|G(s)C_d(s)| = 1$ and $\angle G(s)C_d(s) = \pi$ for $\omega \ll \omega_a$, as previously discussed in \ref{TuningNMPRCGain}. In contrast, for the conventional case (Fig. \ref{fig:ControlArchitectures}(a)), attenuation is limited to a smaller range of frequencies due to lower bandwidths. The proposed method can further enhance this performance by slightly increasing $\omega_i$ of the outer loop integrator, achieving even lower gains at low frequencies. However, it is crucial to properly tune this parameter to balance the trade-offs with other sensitivity function gains at relevant frequencies, as highlighted in \ref{LoopShapingGuidelines}. 

Although tracking performance in the frequency domain is illustrated in Fig. \ref{fig:DCL-TrackingSensitivities}, it is also crucial to understand how loop tuning impacts disturbance rejection and noise attenuation performance in the frequency domain. Typically, noise dominates at high frequencies, while process disturbances can span from low to high frequencies, depending on the system, its environmental conditions, and specific applications. Additionally, internal couplings within the system often introduce disturbances to the motion axis being controlled, particularly at the system's resonance frequencies. Consequently, reducing these sensitivity gains at the relevant frequencies is essential, focusing on the problematic frequency regimes.\\
\begin{figure}[t!]
    \centering
    \includegraphics[width=1\linewidth]{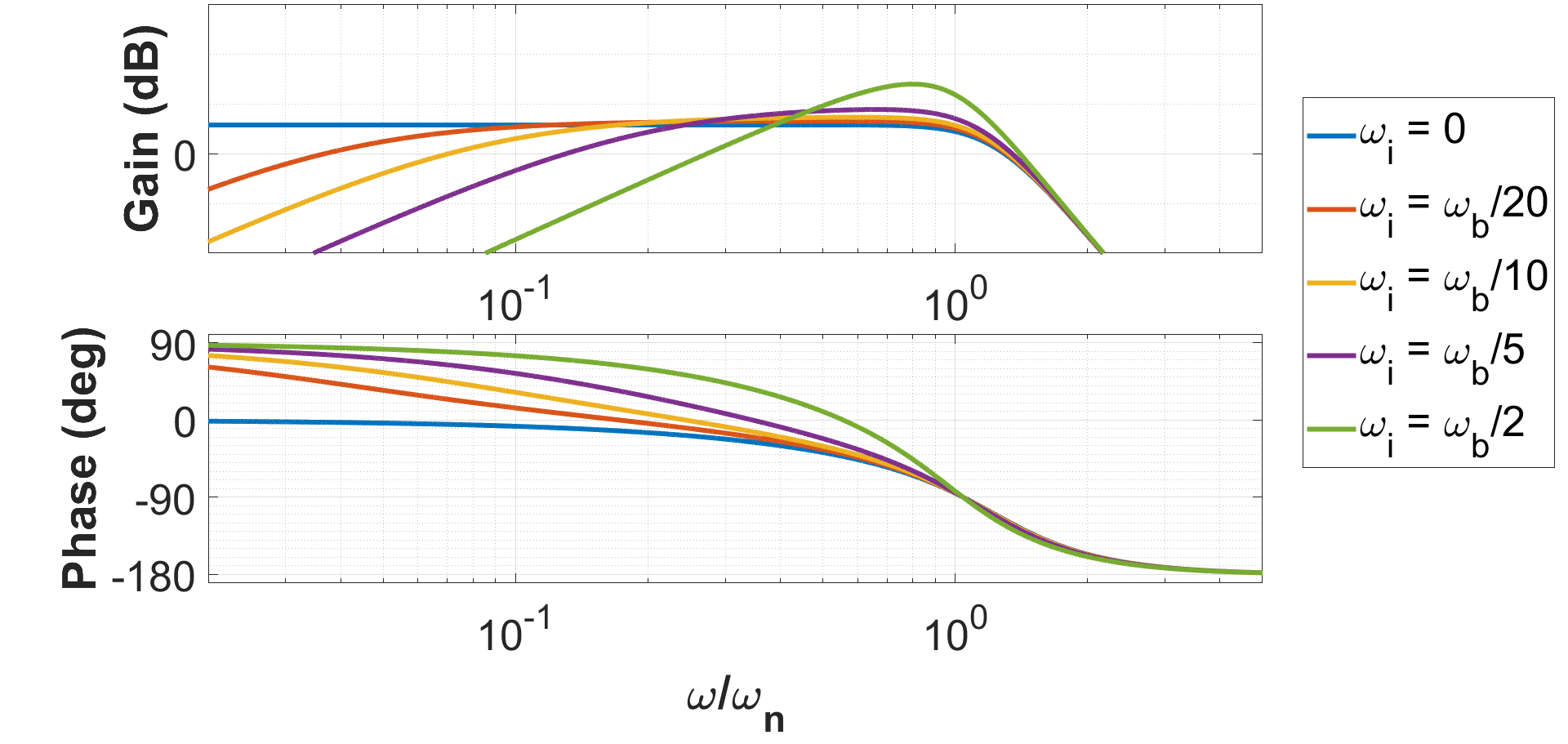}
    \caption{Illustration of dual closed-loop frequency response $PS_{xr}(s): d \mapsto x (\text{or } e_r)$ as $\omega_i$ varies.}
    \label{fig:PS_num}
\end{figure}

The effect of NRC providing a well-dampening effect is evident from the flat gain in the process sensitivity function $PS_{xr}(s)$ around the resonance frequency, as illustrated in Fig. \ref{fig:PS_num}. However, the integrator effect provided by the inner closed loop does not offer any advantage at low frequencies concerning disturbance rejection. This could be a concern in applications where the system is subject to low-frequency disturbances, such as floor vibrations. As shown in Fig. \ref{fig:PS_num}, the process sensitivity gains at low frequencies can be significantly reduced by incorporating a more pronounced integrator in the outer tracking loop. However, as previously discussed, this adjustment comes at the expense of reduced dual closed-loop bandwidth.
\begin{figure}[b!]
\captionsetup[subfloat]{farskip=0pt,captionskip=3pt}
    \centering
    \subfloat[]{\includegraphics[width=\columnwidth]{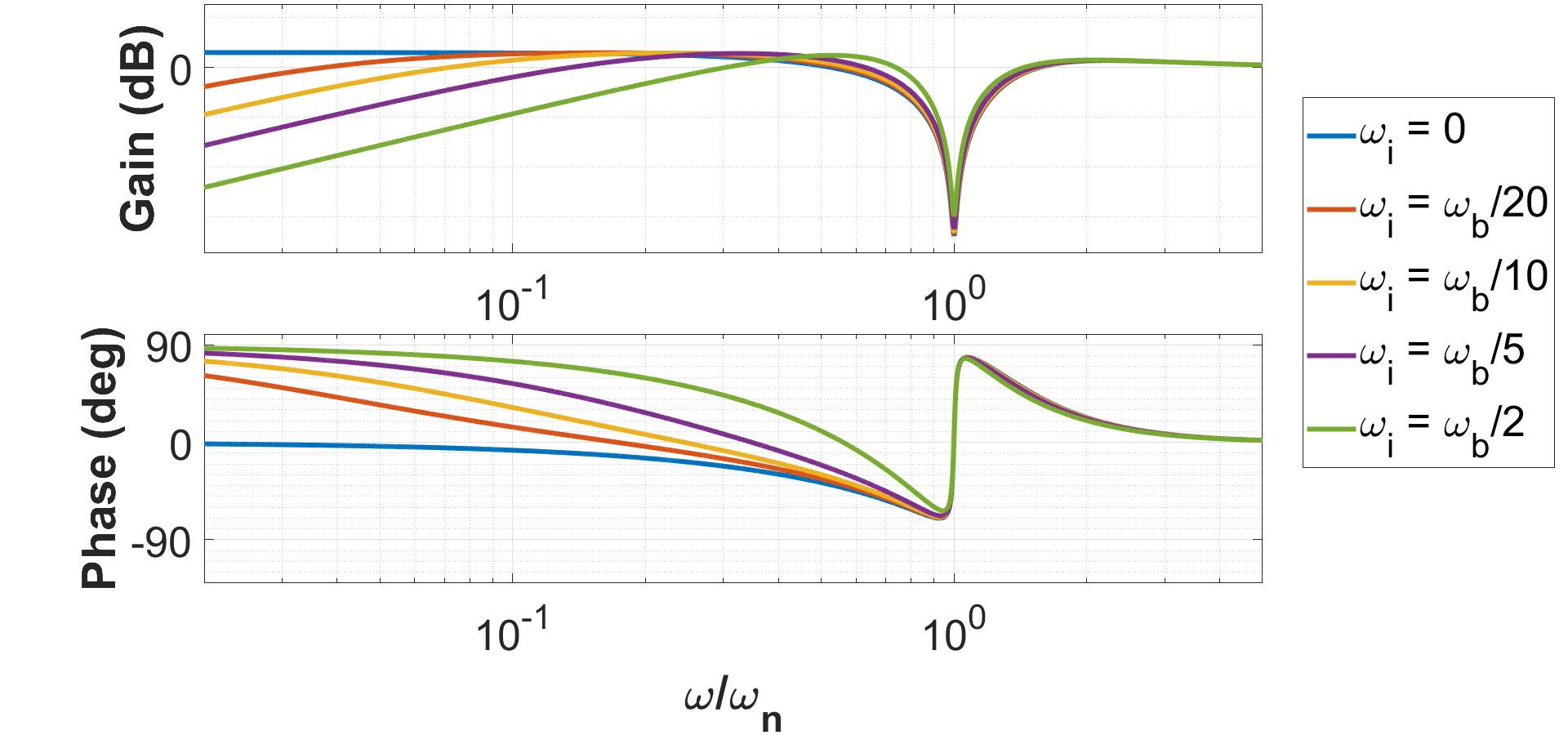}
    \label{fig:S_num}}
    \hfil
    \subfloat[]{\includegraphics[width=\columnwidth]{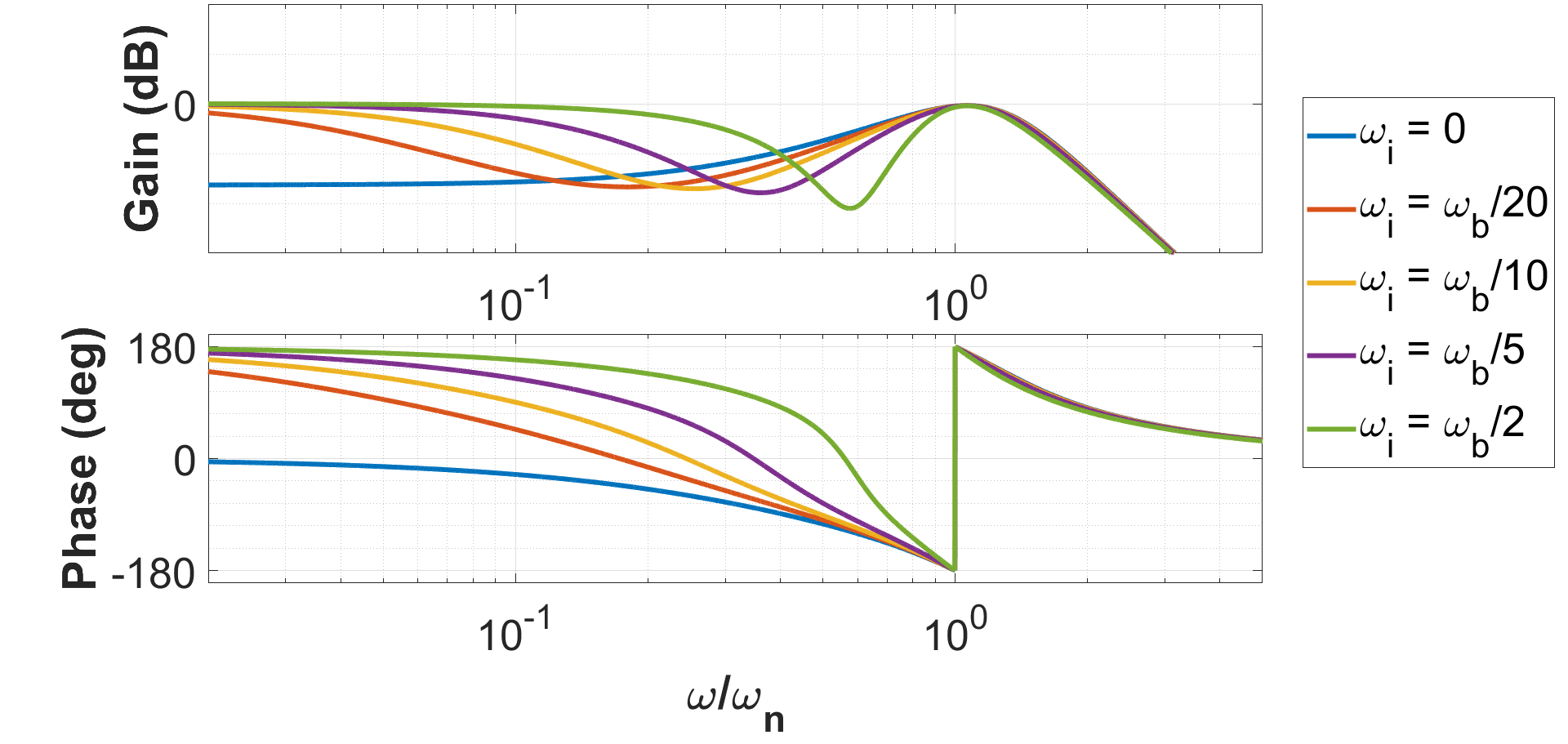}
    \label{fig:CS_num}}
    \caption{Illustration of dual closed-loop frequency response as $\omega_i$ varies (a) $S_{yn}(s): n \mapsto y$, (b) $S_{xr}(s): n \mapsto e_r$.}
    \label{fig:DCL-NoiseSensitivities}
\end{figure}

Similarly, in the noise sensitivity function $S_{yn}(s)$, a deep notch-like behavior indicates substantial noise attenuation near the resonance frequencies, thus implying a reduced influence of noise on the measured position $y$. Again, the low-frequency sensitivity gains can be decreased with a higher integrator corner frequency $\omega_i$, if necessary (Fig. \ref{fig:DCL-NoiseSensitivities}(a)). However, when evaluating the influence of noise $n$ on real error $e_r$ by mapping $S_{xr}(s)$, it can be inferred that even with a pronounced integrator, the low-frequency noise attenuation isn't significant (Fig. \ref{fig:DCL-NoiseSensitivities}(b)). This transpires from the limitation of linear control, as highlighted in \ref{LoopShapingGuidelines}. This is typically not an issue, since noise tends to dominate at high frequencies, where the system gains naturally roll off.

Thus, this section highlights the importance of shaping these dual-closed-loop sensitivity functions to minimize the real error within the controlled system. Combining the advantages of the proposed control method and the limitations of linear control theory, there is considerable flexibility to effectively balance the trade-offs between achievable damping and reference tracking, disturbance rejection, and noise attenuation at relevant frequency regimes.

\section{Experimental Setup and Results}
\label{ExperimentalSection}
This section presents the experimental setup, which incorporates the industrial nanopositioning system, to experimentally demonstrate the damping performance of the proposed NRC and the dual closed-loop performance achieved by shaping different sensitivity functions. 

\begin{figure}[b!]
\captionsetup[subfloat]{farskip=0pt,captionskip=3pt}
    \centering
    \subfloat[]{\includegraphics[width=\columnwidth]{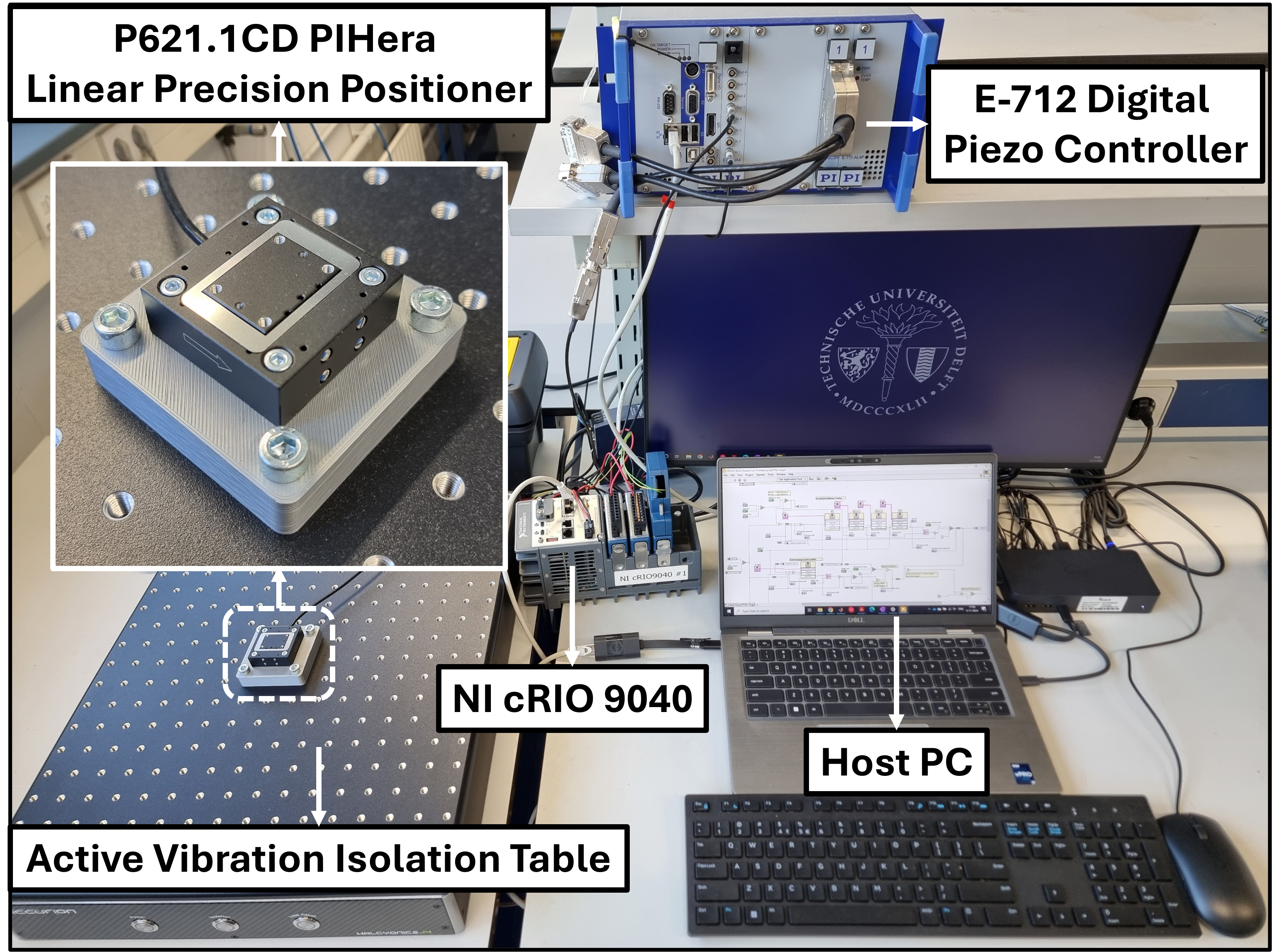}
    \label{fig:ExpPlatform}}
    \hfil
    \subfloat[]{\includegraphics[width=\columnwidth]{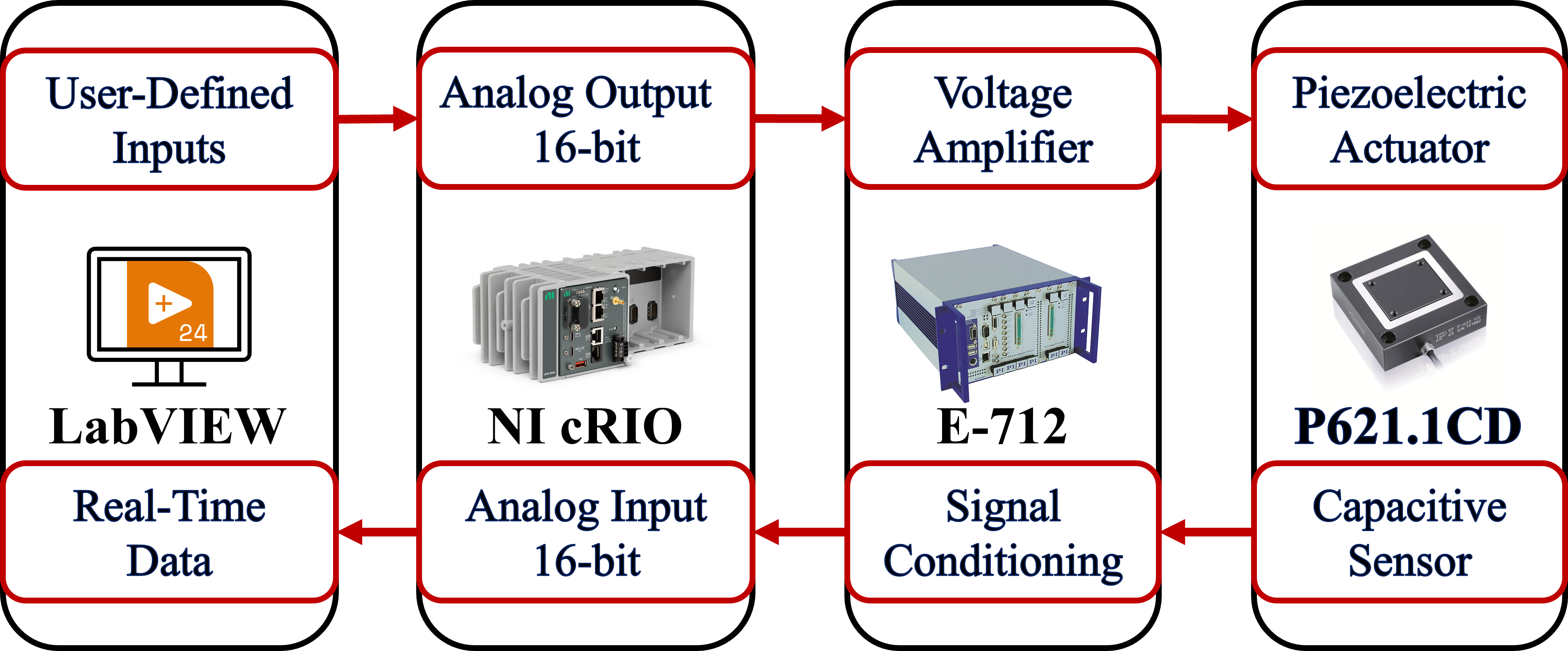}
    \label{fig:SystemWorkflow}}
    \caption{Experimental setup of a piezo-actuated nanopositioning system (a) Experimental platform (b) System workflow.}
    \label{fig:ExpSetup}
\end{figure}
\newpage
\subsection{System Description}
Presented in Fig. \ref{fig:ExpSetup}, the experimental setup utilizes a commercial P-621.1CD PIHera linear precision nanopositioner with a travel range of 100 $\mu$m. The single-axis positioning stage incorporates a ceramic-insulated multilayer piezo-stack actuator, a flexure-based mechanism-guided platform, and a high-resolution capacitive sensor. The stage utilizes a voltage amplifier and sensor signal conditioning modules integrated within the modular E-712 piezo-controller. The commercial hardware is integrated with an NI CompactRio chassis with an embedded FPGA, facilitating actuation signals and control for external control. The chassis includes various analog input-output modules that enable transmission and reception of signals for implementing the control approach. The control scheme is implemented using the NI LabView software, which interfaces the host computer and the nanopositioner. The actuation voltage ranges from 0 to 10 V, and the sampling time $t_s$ of the FPGA-based control loop is set to 30 $\mu$s.

\subsection{System Identification}

A sinusoidal chirp signal (0 to 0.1 V) was generated with LabVIEW and sent to the piezo-actuator for system identification. The capacitive sensor measured the position output, and the input-output signals were imported into MATLAB for analysis. The transfer functions were estimated using MATLAB's signal processing toolbox. A high sampling frequency $F_s$ of 33.3 kHz provided sufficient data for accurate identification.

\begin{figure}[t!]
    \centering
    \includegraphics[width=1\linewidth]{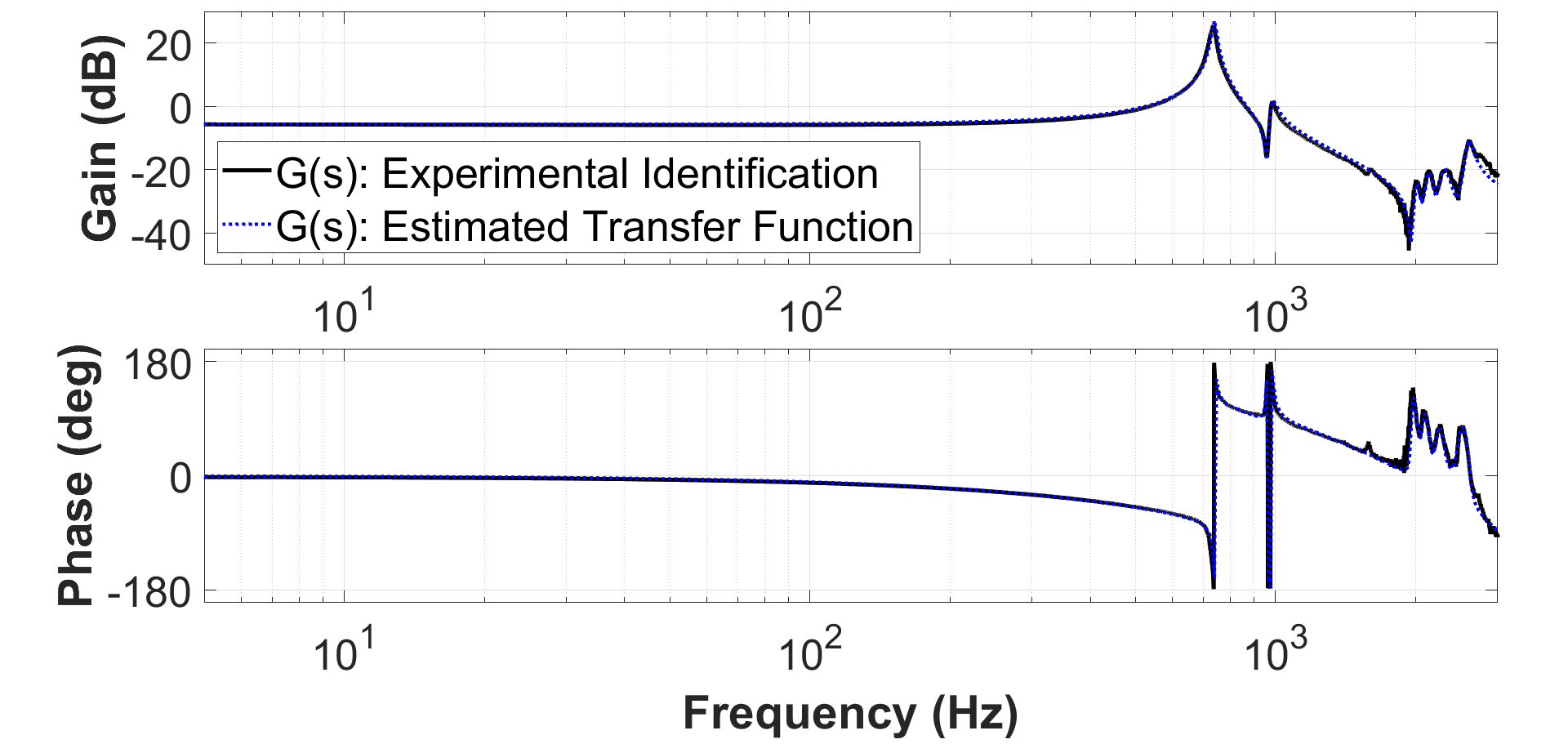}
    \caption{Identified frequency response of the nanopositioning system $G(s)$}
    \label{fig:SystemIdentification}
\end{figure}

The dominant resonance peak \( \omega_n \) is at 739 Hz, with the second mode \( \omega_2 \) nearby at 983 Hz (see Fig. \ref{fig:SystemIdentification}). A notable phase delay occurs due to actuator-amplifier dynamics and system time delay, even at frequencies below \( \omega_n \). Higher modes appear around and beyond 2000 Hz. Pole-zero interlacing indicates the system's collocated nature.

\subsection{Experimental Inner Closed-Loop Frequency Response}
\begin{figure}[t!]
    \centering
    \includegraphics[width=1\linewidth]{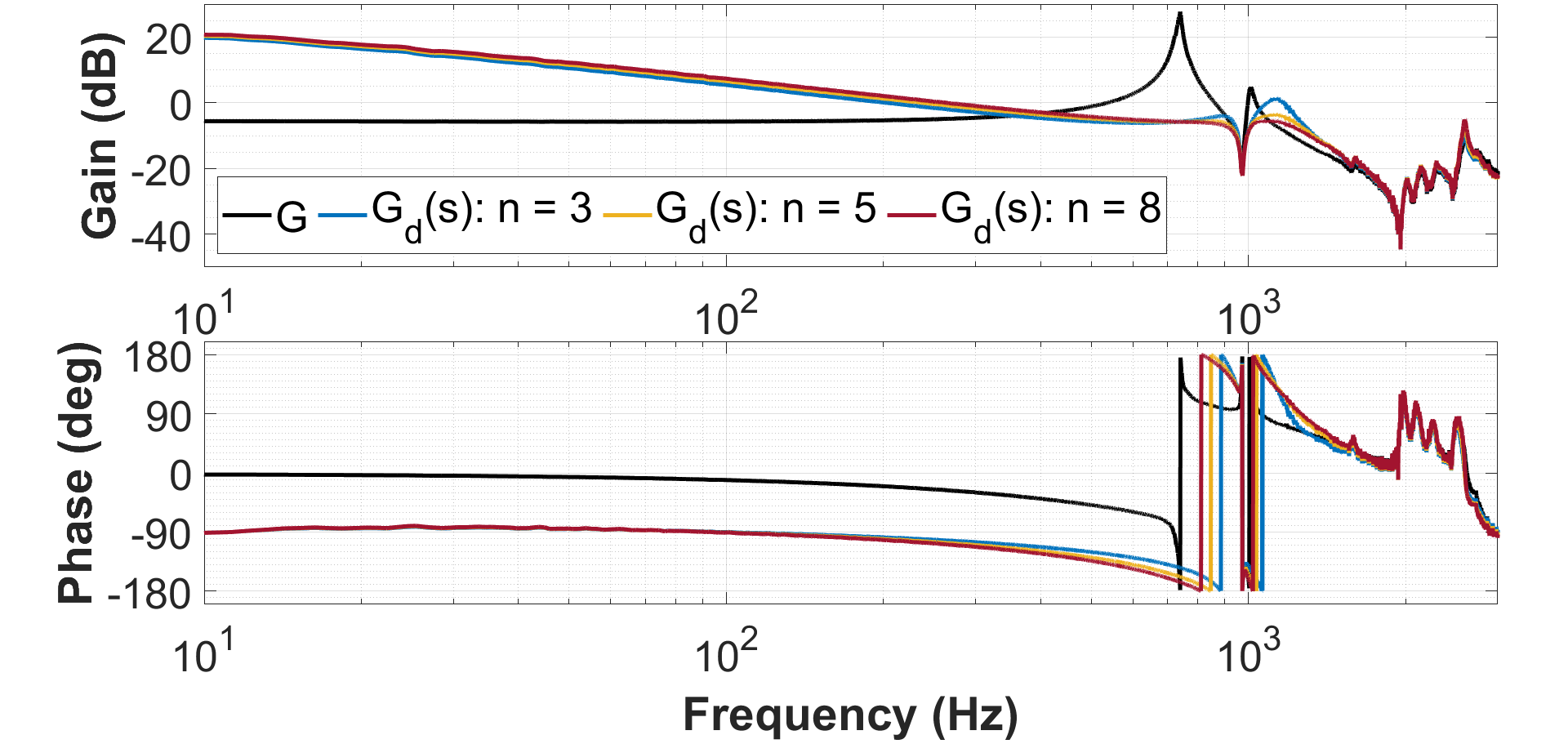}
    \caption{Experimentally identified inner closed-loop frequency response $G_d(s)$.}
    \label{fig:Tuning_NMPC_Exp}
\end{figure}

With the identified system frequency response of the piezoelectric nanopositioning stage, the NRC is designed and implemented to dampen the dominant resonance and extend the system bandwidth. According to the design guidelines presented in \ref{ADC_NMPC}, a normalized corner frequency of $n\approx3$ should ensure the complete damping of the resonance peak. However, due to significant phase lag below $\omega_n$, $n$ is re-tuned to ensure a sufficiently dampened peak (see \ref{EffectofDelay}). Implementing the NRC also induces damping in nearby higher-order modes (see \ref{DampingMultipleModes}). As demonstrated in Fig. \ref{fig:Tuning_NMPC_Exp}, while the dominant resonance is successfully damped, the second mode also experiences reasonable damping, which increases as $n$ increases. It should be noted that this increase in $n$ comes at the cost of an increased phase delay for $\omega\leq\omega_n$. This additional phase lag must be considered during the tuning process to ensure that the tracking controller maintains the required stability margins. Another observation is the second resonance peak $\hat{\omega}_2(s)$ in $G_d(s)$, which experiences a shift toward higher frequencies, contrary to the illustration presented in Fig. \ref{fig:TwoModeSystemDamped}, the effect transpiring due to the delay at those frequencies.

Thus, based on the corner frequency tuning shown in Fig. \ref{fig:Tuning_NMPC_Exp} and the controller gain derived from \eqref{NMPC_Gain_Eq}, the damping controller is implemented with \(n = 8\) and \(k = 1.9095\).

\subsection{Experimental Dual Closed-Loop Frequency Response}
In this dual closed-loop control approach, a feedback tracking controller handles tracking errors. As discussed previously, a proportional-integral (PI) controller is designed based on the inner closed-loop dynamics ($G_d(s)$) to achieve the highest possible dual closed-loop bandwidths. However, in practical systems like this, higher-order modes, if not adequately suppressed, can lead to significant closed-loop gains at their corresponding frequencies. Furthermore, high-frequency noise is a common issue that must be attenuated. To address these challenges, the tracking controller, in this case, is designed as a series combination of a PI controller, two notch filters targeting the two dominant higher-order modes, respectively, and a low-pass filter to mitigate high-frequency noise.

The implemented tracking controller $C_t(s)$ is expressed as follows: 
\begin{equation}
    C_t(s) = \underbrace{k_p \cdot (1 + \frac{\omega_i}{s})}_{\substack{\text{Proportional} \\ \text{Integral Term}}} \cdot \underbrace{N_1(s) \cdot N_2(s)}_{\substack{\text{Notch} \\ \text{Filters}}} \cdot \underbrace{\frac{\omega_l}{s + \omega_l}}_{\substack{\text{Low-Pass}\\ \text{Filter}}},
\end{equation}
where the notch filters $N_1(s)$ and $N_2(s)$ are as follows: 
\begin{equation}
   \begin{aligned}
       N_1(s) & = \frac{(\frac{s}{\omega_{N_{1}}})^2 + \frac{s}{Q_1\omega_{N_{1}}} + 1}{(\frac{s}{\omega_{N_{1}}})^2 + \frac{s}{Q_2\omega_{N_{1}}} + 1} \\
       N_2(s) & = \frac{(\frac{s}{\omega_{N_{2}}})^2 + \frac{s}{Q_3\omega_{N_{2}}} + 1}{(\frac{s}{\omega_{N_{2}}})^2 + \frac{s}{Q_4\omega_{N_{2}}} + 1}.
   \end{aligned} 
\end{equation}  

The tracking controller is tuned to achieve an open-loop bandwidth $\omega_b$ of 280 Hz, with gain and phase margins of 6 dB and \(59^\circ\), respectively, ensuring sufficient robustness. The tuning parameters are as follows: \(k_p = 298.3569\), \(\omega_i = 28\) Hz, \(\omega_{N_{1}} = 1000\) Hz, \(\omega_{N_{2}} = 2600\) Hz, \(\omega_{l} = 5000\) Hz, \(Q_1 = 1.1\), \(Q_2 = 1\), \(Q_3 = 12\), and \(Q_4 = 10\).

\begin{figure}[t!]
    \centering
    \includegraphics[width=1\linewidth]{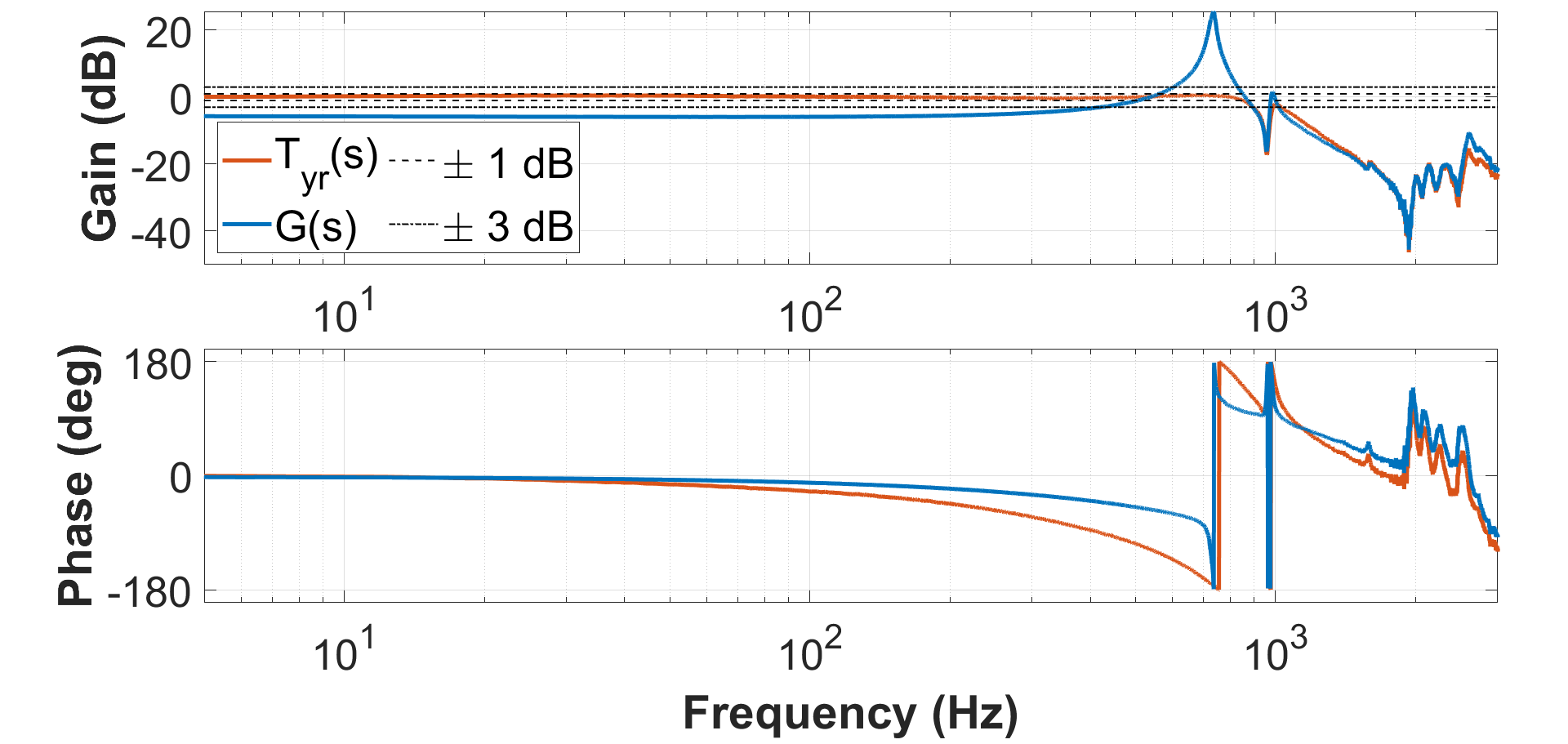}
    \caption{Experimentally identified dual closed-loop frequency response $T_{yr}(s)$.}
    \label{fig:Exp_CL}
\end{figure}

After implementing the tuned controllers, the dual closed-loop frequency response $T_{yr}(s)$ is experimentally estimated through a closed-loop system identification process, as shown in Fig. \ref{fig:Exp_CL}, where it can be observed that both the crossings \(\pm 1\) dB and \(\pm 3\) dB occur beyond the resonance frequency \(\omega_n = 735\) Hz, specifically at 845 Hz and 895 Hz ($\omega_c$), respectively. These experimental results clearly show that effectively tuning the dual closed-loop incorporating the proposed NRC enables the system to achieve dual closed-loop bandwidths that surpass the dominant resonance frequency.

\subsection{Reference Tracking Time-Domain Performance}
The nanopositioning stage aims to accurately follow predefined trajectories, typically periodic ones. Fig. \ref{fig:Exp_CL} shows its ability to track references up to dual closed-loop bandwidths in the frequency domain. This section evaluates time-domain performance, with the system subjected to sinusoidal references, with frequencies spanning from 1 to 900 Hz.
\begin{figure}[!t]
\captionsetup[subfloat]{farskip=0pt,captionskip=2pt}
    \centering
    \subfloat[1 Hz]{\includegraphics[width=0.495\columnwidth]{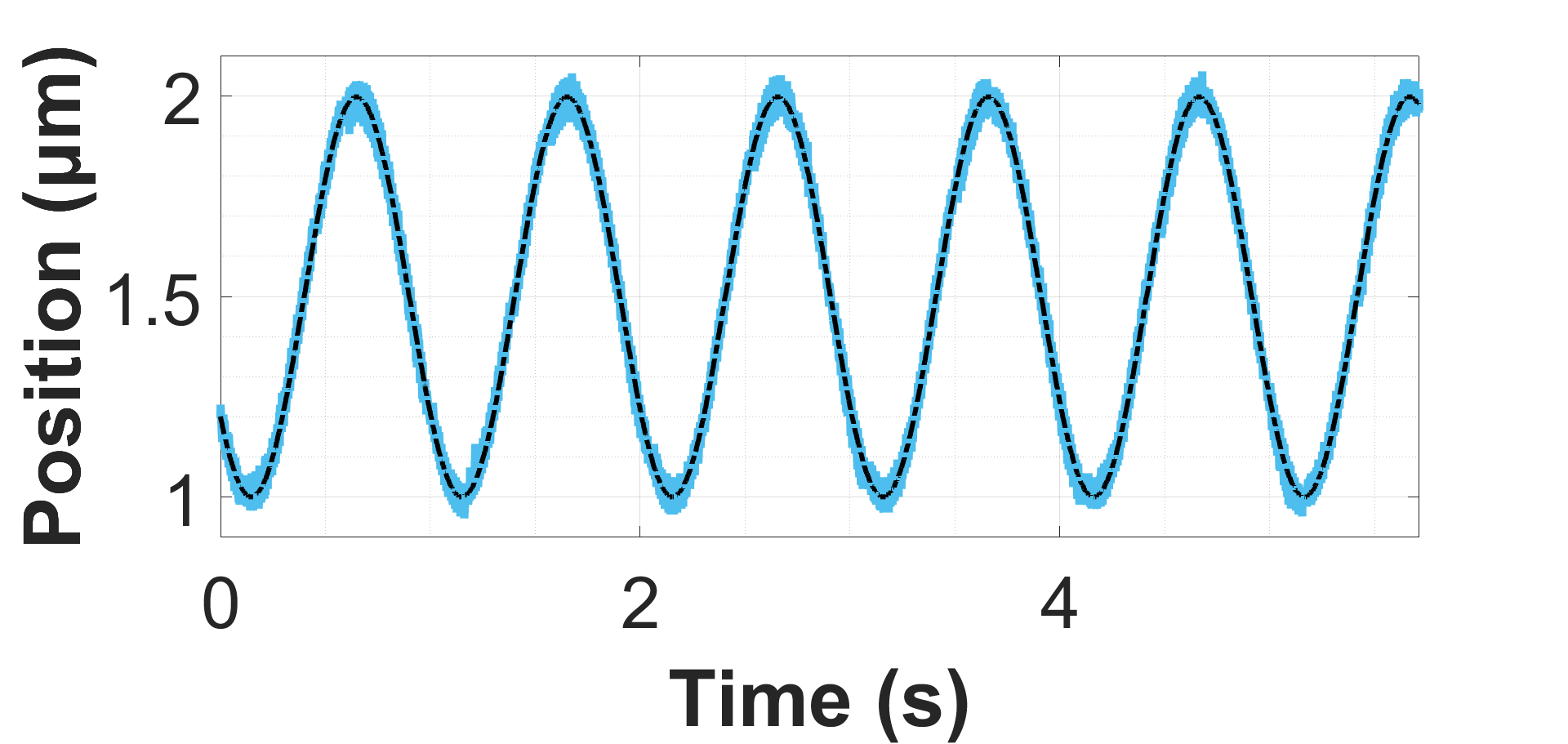}%
    \label{fig:1}}
    \subfloat[10 Hz]{\includegraphics[width=0.495\columnwidth]{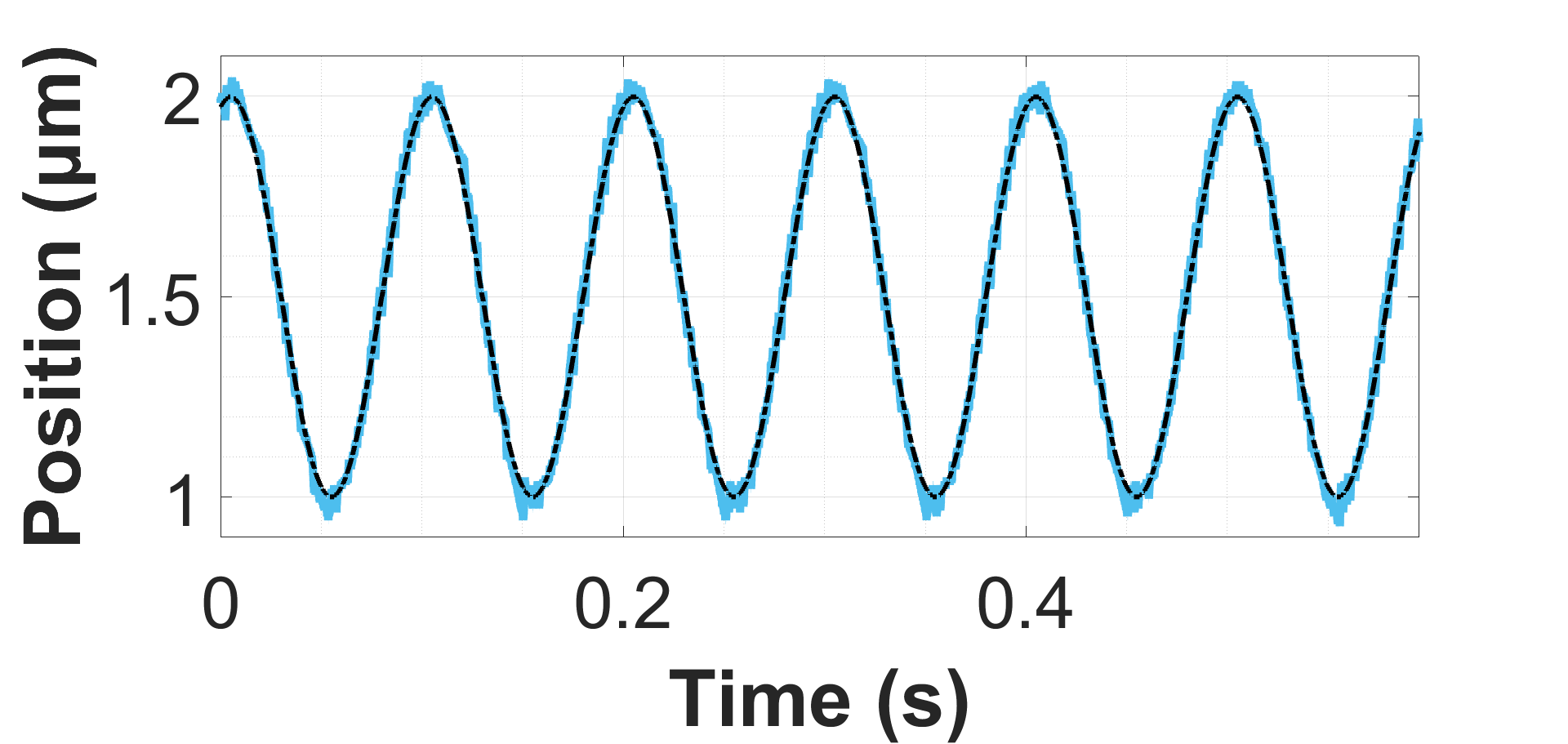}%
    \label{fig:2}}
    \hfil
    \subfloat[20 Hz]{\includegraphics[width=0.495\columnwidth]{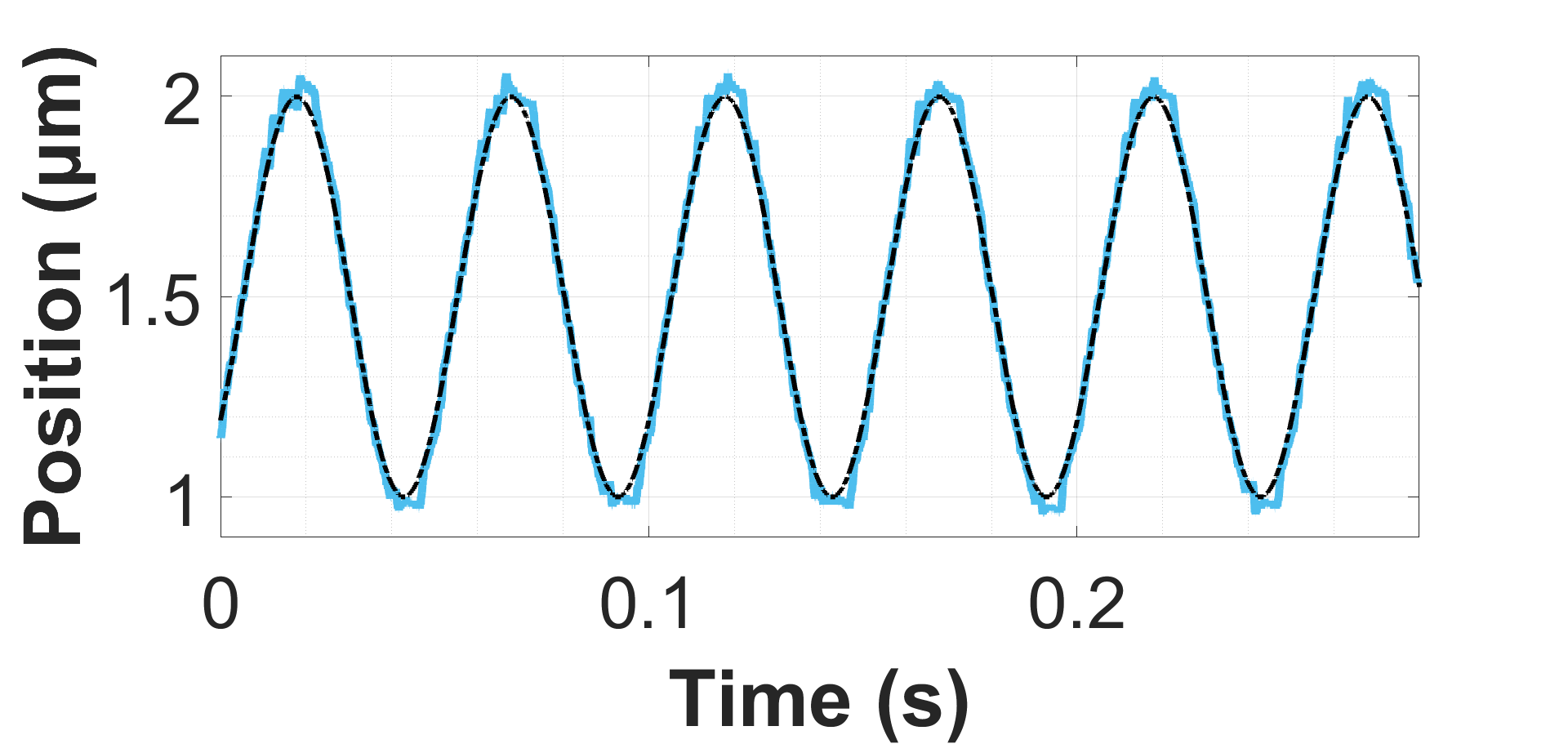}
    \label{fig:3}}
    \subfloat[50 Hz]{\includegraphics[width=0.495\columnwidth]{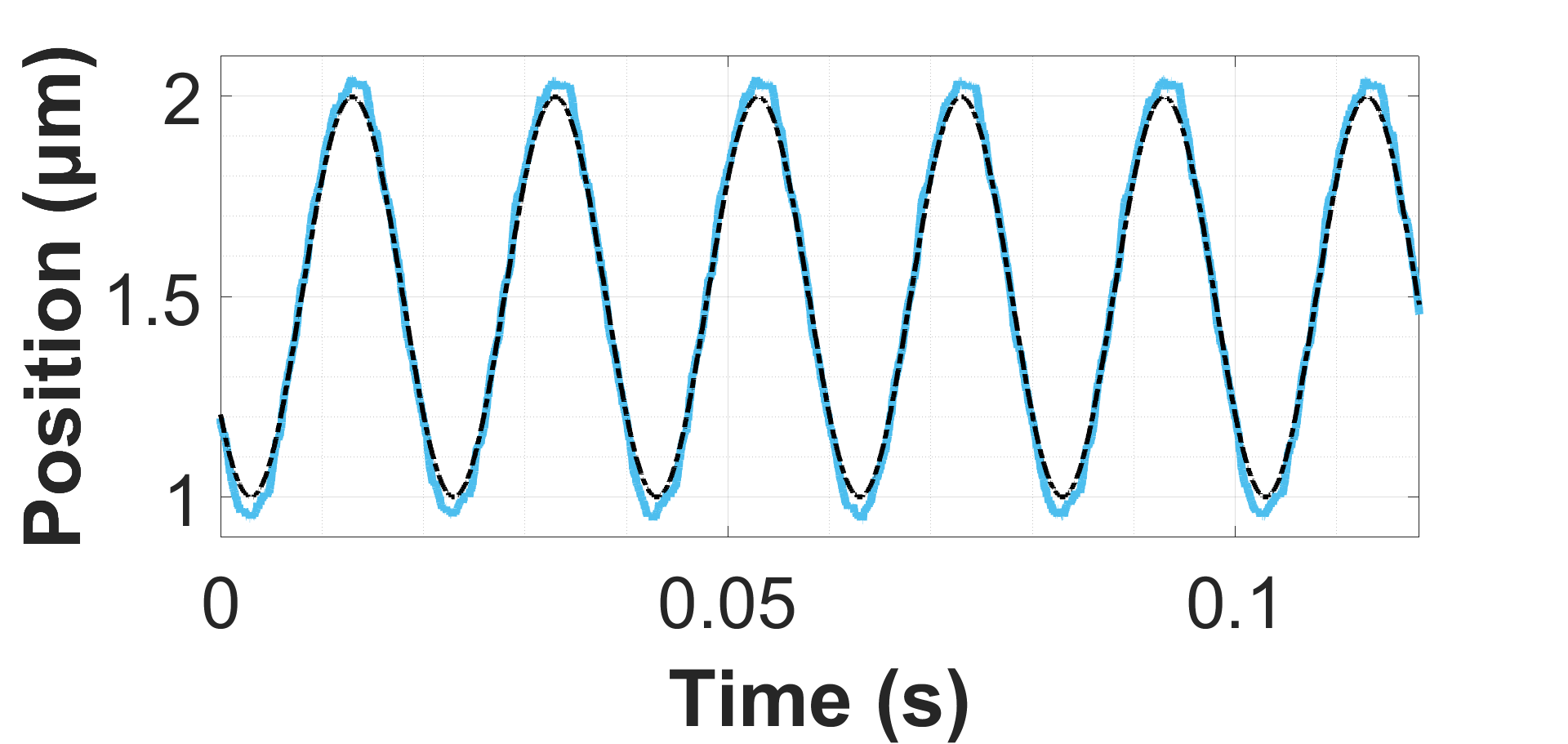}
        \label{fig:4}}
    \hfil
    \subfloat[100 Hz]{\includegraphics[width=0.495\columnwidth]{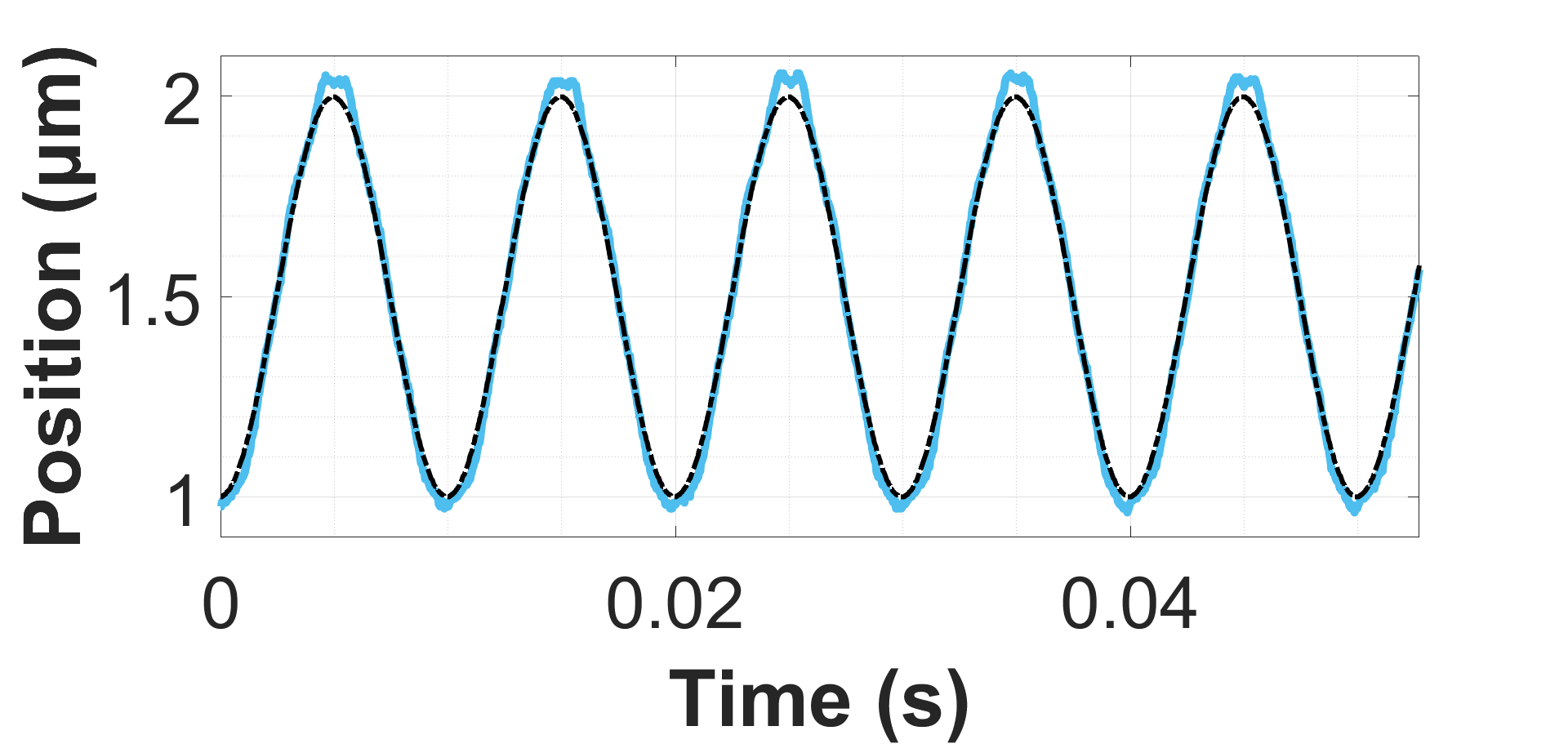}
        \label{fig:5}}
    \subfloat[200 Hz]{\includegraphics[width=0.495\columnwidth]{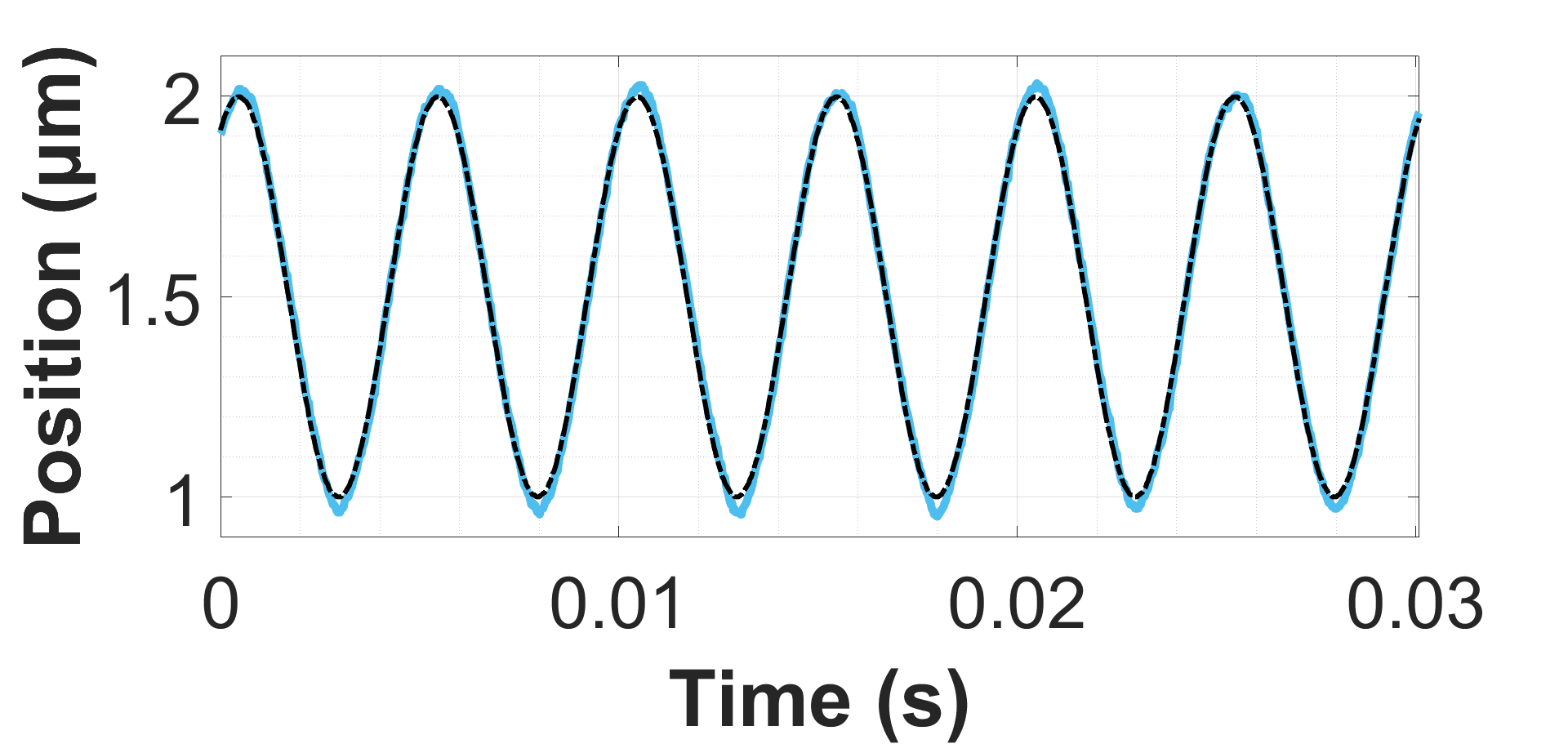}
    \label{fig:6}}
    \hfil
    \subfloat[500 Hz]{\includegraphics[width=0.495\columnwidth]{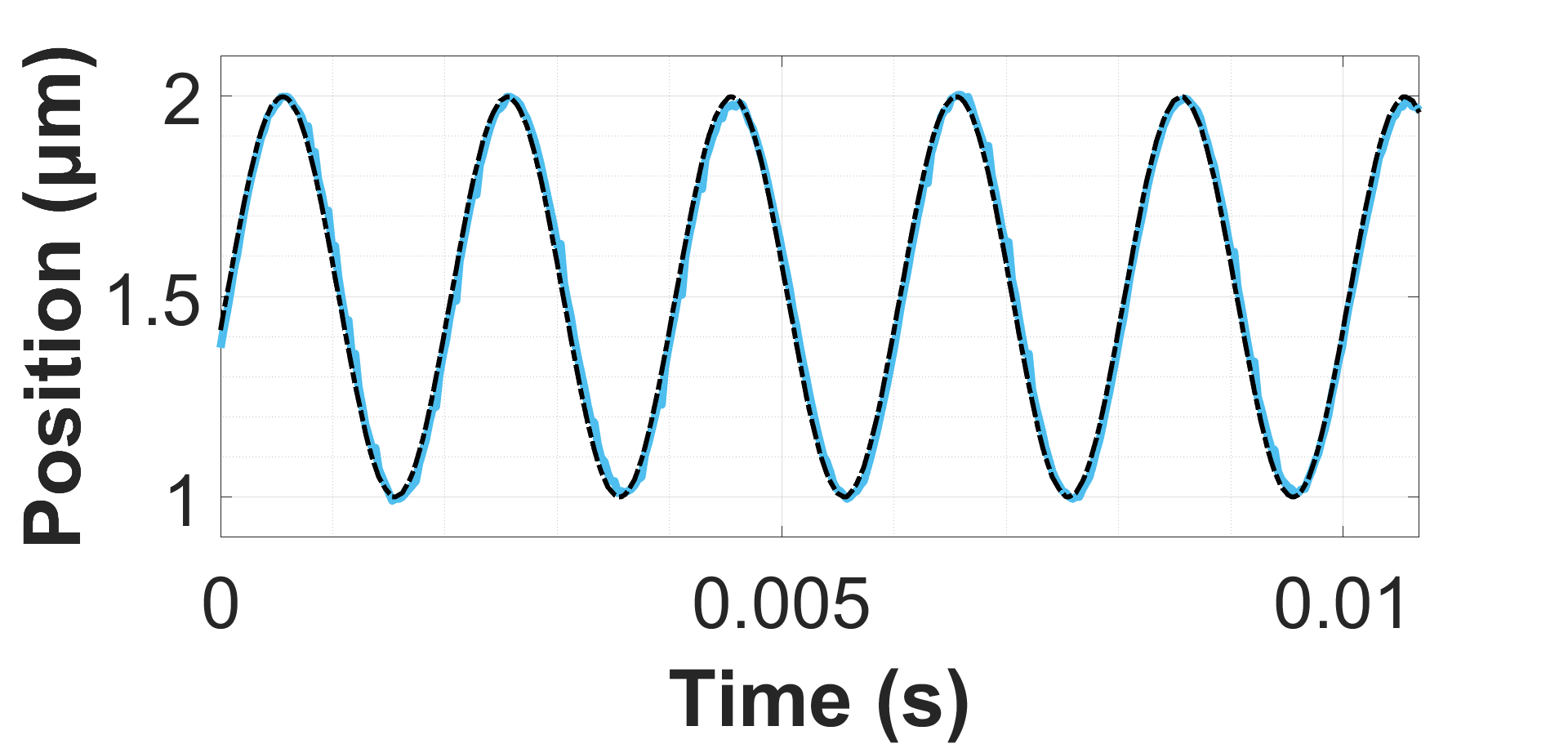}
    \label{fig:7}}
    \subfloat[700 Hz]{\includegraphics[width=0.495\columnwidth]{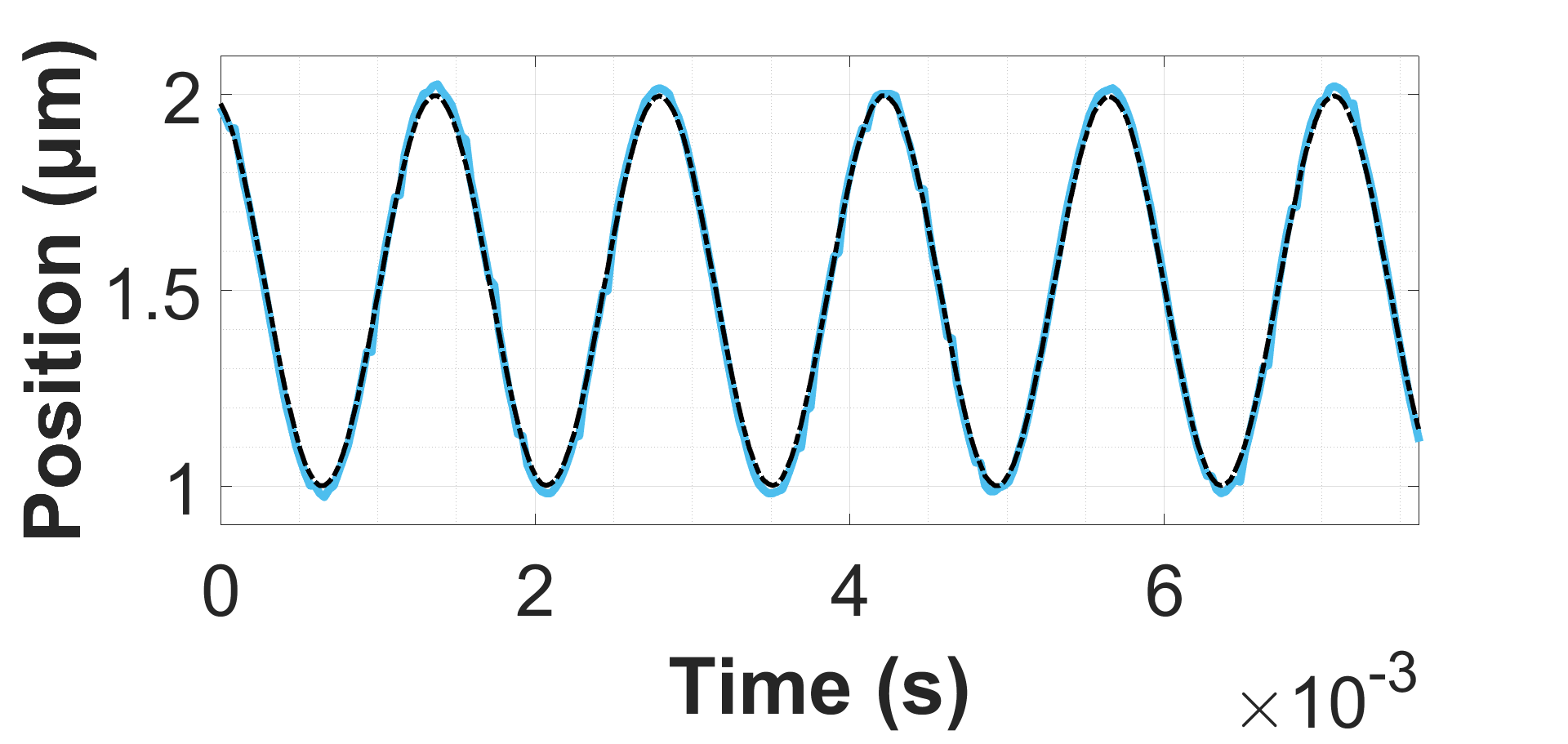}
    \label{fig:8}}
    \hfil
    \subfloat[800 Hz]{\includegraphics[width=0.495\columnwidth]{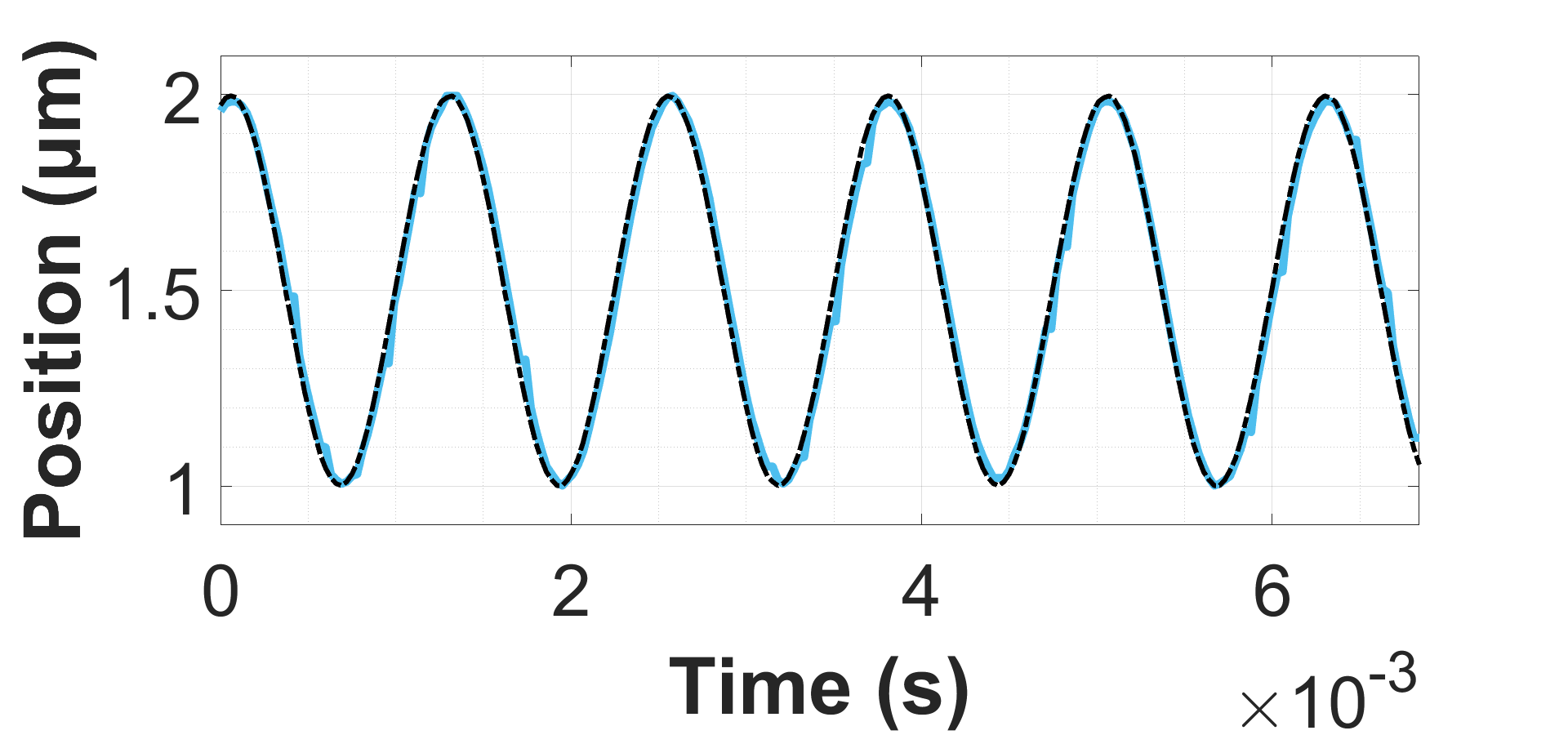}
    \label{fig:9}}
    \subfloat[900 Hz]{\includegraphics[width=0.495\columnwidth]{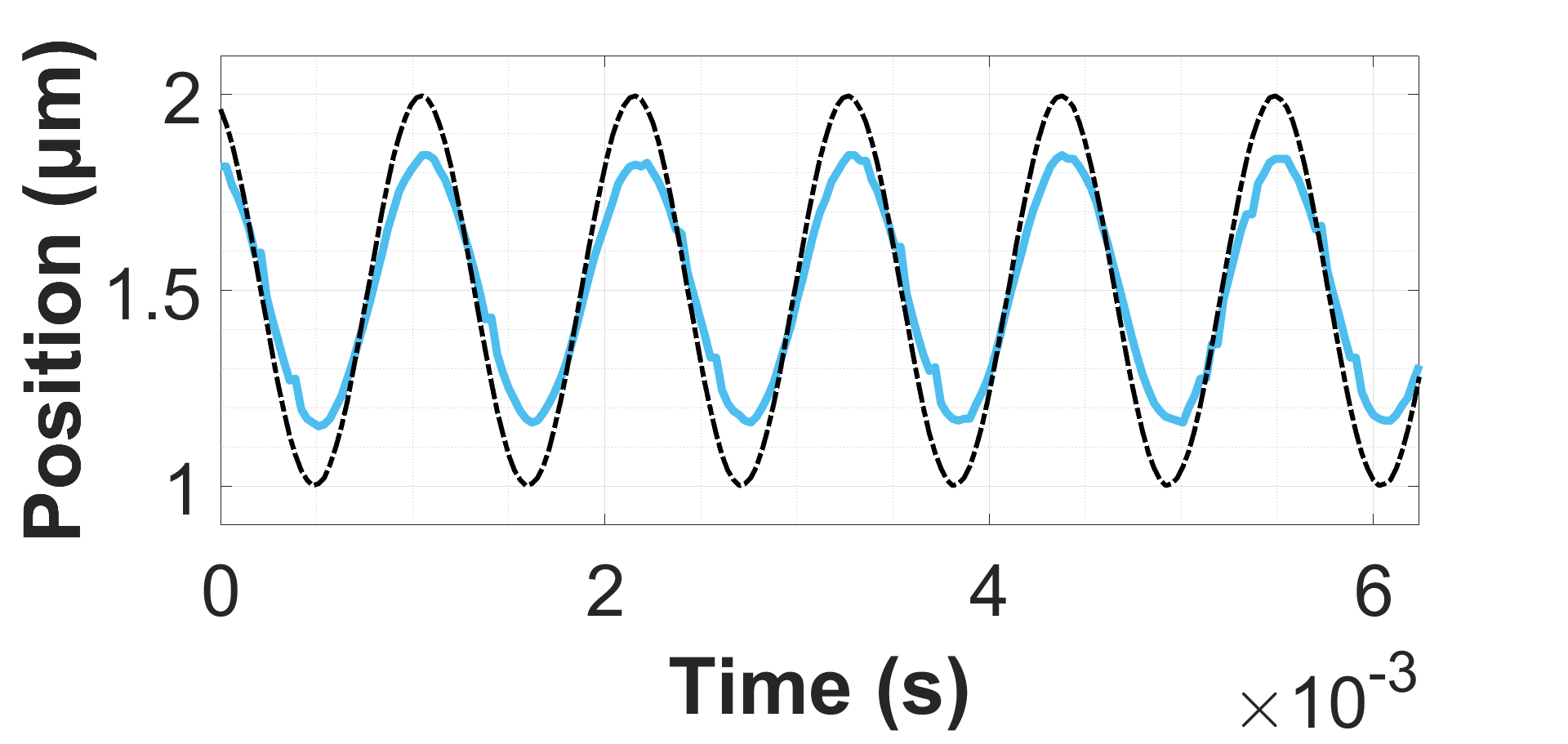}
    \label{fig:10}}
    \caption{Sinusoidal reference tracking with different frequencies.}
    \label{fig:SineRefTracking}
\end{figure}

It is important to note that the phase lag increases with increasing frequencies in the dual closed-loop system. Perfectly delayed tracking is often implemented in typical periodic scanning applications, provided the delay is well known. Thus, phase lags are removed using post-processing techniques to reasonably represent the tracking performance. The known phase lag \(\phi_l\) (in degrees) at the corresponding frequency \(f\) (in Hz) is utilized to compute the resulting time delay \(t_d\), given by:
\begin{equation}
    t_d = \frac{\phi_l}{f\cdot 360}.
\end{equation}

Subsequently, the shifted outputs \(y^*(t)\) can be computed as follows:
\begin{equation}
    y^*(t) = y(t_{i + N_d}:t_N) \text{ for } i=[1,N] \text{ and } i\in \mathbb{Z}.
\end{equation}

In discrete time, phase lags are compensated by shifting $N_d = \lfloor |t_d|/t_s \rceil$ samples, where \(t_s\) is the sampling time, and \(\left\lfloor \cdot \right\rceil\) is the round function. Fig. \ref{fig:SineRefTracking} shows the phase-corrected system tracking response for small amplitude references from $1$ $\mu$m to $2$ $\mu$m.

To assess the tracking performance of the dual closed-loop system with the proposed NRC, we consider two common indices: the maximum tracking error (\(e_{max}\)) and the root mean square tracking error (\(e_{rms}\)).
\begin{equation}
e_{max}=\max \left(\left|y\left(t_i\right)-r\left(t_i\right)\right|\right),
\end{equation}

\begin{equation}
e_{rms}=\sqrt{\frac{1}{N} \sum_{i=1}^N\left(y\left(t_i\right)-r\left(t_i\right)\right)^2},
\end{equation}
where \(r(t_i)\) and \(y(t_i)\) represents the reference signal and output signal at discrete time step \(i\), respectively, and \(N\) is the total number of samples, with \(i = 1, 2, 3,..., N\).

\begin{table}[t!]
    \centering
    \caption{Maximum and RMS Tracking Errors}
    \label{tab:TrackingErrors}
    \begin{tabular}{|c|c|c|}
        \hline Frequency (Hz) & $e_{max}$ ($\mu m$) & $e_{rms}$ ($\mu m$) \\
        \hline 1 & 0.0928 & 0.0218 \\
        \hline 10 & 0.0928 & 0.0247 \\
        \hline 20 & 0.1074 & 0.0289 \\
        \hline 50 & 0.0977 & 0.0344 \\
        \hline 100 & 0.0879 & 0.0282 \\
        \hline 200 & 0.0586 & 0.0168 \\
        \hline 500 & 0.1123 & 0.0347 \\
        \hline 700 & 0.0977 & 0.0221 \\
        \hline 800 & 0.1270 & 0.0302 \\
        \hline 900 & 0.2344 & 0.1226 \\
        \hline
    \end{tabular}
\end{table}

Table \ref{tab:TrackingErrors} presents the computed tracking errors, showing that the system maintains very low errors for reference signals up to 800 Hz. However, performance degrades for frequencies beyond this point, aligning with the drop in sensitivity gains as frequencies exceed the dual closed-loop bandwidth $\omega_c$. It is important to note that these error magnitudes can vary depending on the actual tuning of all controller parameters, which are adjusted based on specific application requirements. The presented values indicate the system's tracking capabilities upto $\omega_c$.

\subsection{Experimental Dual Closed-Loop Sensitivities}
\begin{figure}[t!]
    \centering
    \includegraphics[width=1\linewidth]{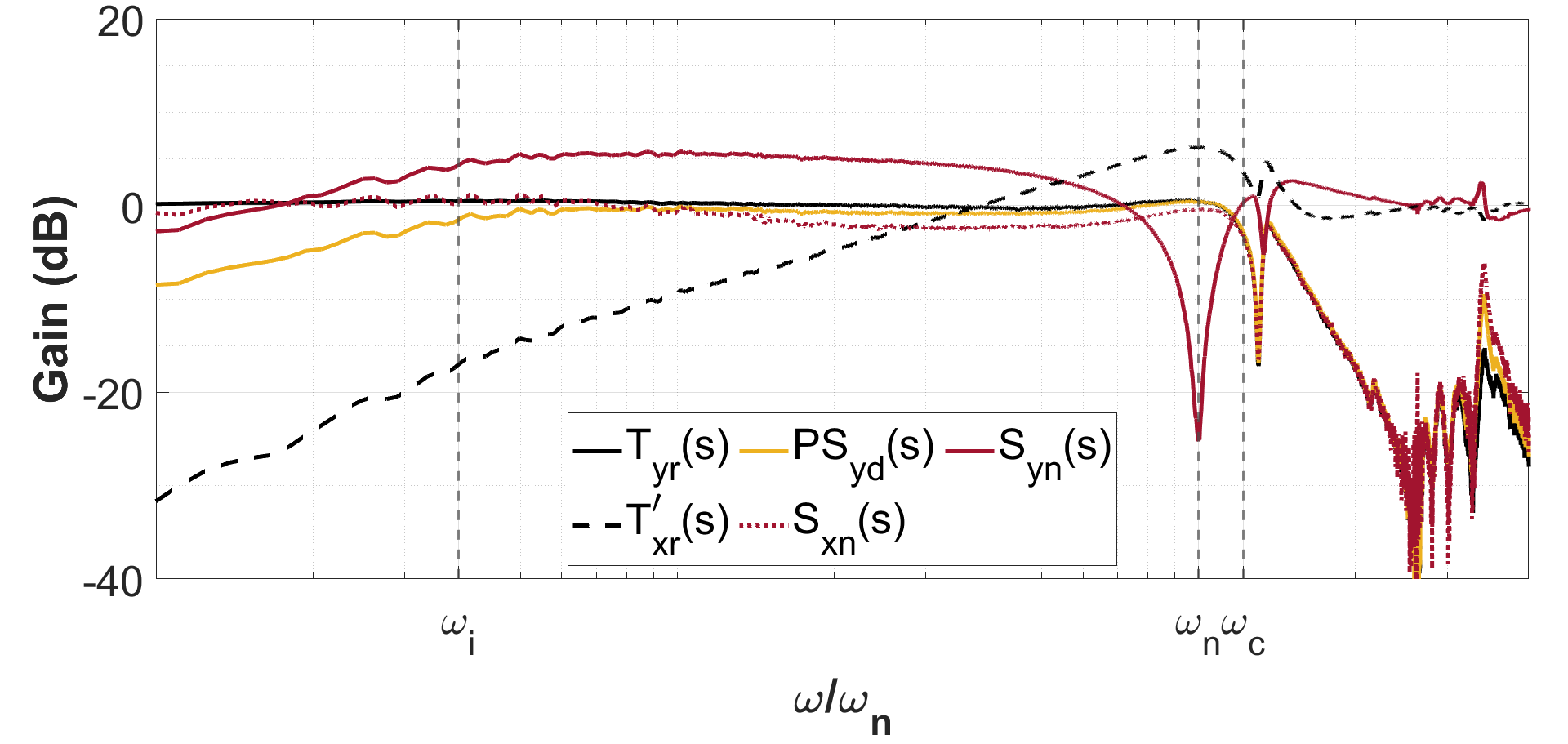}
    \caption{Experimentally identified ($T_{yr}(s), PS_{yd}(s), S_{yn}(s)$) and estimated ($T'_{xr}(s), S_{xn}(s)$) dual closed-loop sensitivities.}
    \label{fig:Exp_AllSens}
\end{figure}

This section reflects on experimentally identified sensitivities to dual closed-loop shaping guidelines (see \ref{LoopShapingGuidelines}) and those shaped using the proposed NRC (see \ref{ClosedLoopDynamics}). The experimental sensitivity magnitudes are shown in Fig. \ref{fig:Exp_AllSens}. It is crucial to assess the impact of input signals $r$, $d$, and $n$ on both the measured position $y$ and the real error $e_r$ in the dual closed-loop system. Sensitivities are shaped to minimize the impact of these signals on $e_r$ in relevant frequency regimes.

As shown in Fig. \ref{fig:Exp_CL}, $\omega_c$ exceeds $\omega_n$, aided by the NRC ($C_d(s)$) tuning, so that $|G(s)C_d(s)|=1$ and $\angle G(s)C_d(s) \approx \pi$ up to $\omega \approx \omega_{C_d}$. The 0 dB crossing of $T'_{xr}(s)$ near $\omega_{C_d}$ implies minimal influence of $r$ on $e_r$ for $\omega < \omega_{C_d}$.

The NRC ensures a well-dampened peak, as shown by flat gains around $\omega_n$ in the process sensitivity function $PS_{yd}(s)$. To reduce the low-frequency gains of $PS_{yd}(s)$ and ensure zero steady-state tracking, the integrator in $C_t(s)$ is tuned according to $\omega_i = \omega_b/10$. Increasing $\omega_i$ decreases $\omega_c$ and slightly amplifies the dynamics around $\omega_n$. The system's high-frequency gain roll-off and low-pass filter in $C_t(s)$ ensures sufficiently low gains for $S_{xn}(s)$. However, limitations in \ref{LoopShapingGuidelines} prevent attenuation of contributions of $n$ to $e_r$ at low frequencies. The experimental findings align with the guidelines and analysis presented in this paper.

\section{Conclusions}
\label{Conclusions}
This paper introduces a novel Non-Minimum-Phase Resonant Controller (NRC) tailored for active damping control in dual-closed-loop architectures, applied to piezo-actuated nanopositioning systems. The NRC leverages non-minimum-phase characteristics to achieve complete damping and the bifurcation of the double resonant poles of the primary resonance peak through a constant-gain design with a tunable phase, ensuring robustness even under varying load conditions. In addition, the paper demonstrates the controller's capability to dampen higher-order flexural modes. In alignment with the dual closed-loop shaping guidelines delineated in this paper, the proposed NRC can provide high gains at low frequencies within the inner loop, which, complemented by a standard PI tracking controller, facilitates the system to achieve high dual closed-loop bandwidths that potentially surpass the primary resonance frequency. Furthermore, the NRC minimizes the impact of low-frequency reference signals on real feedback errors and ensures robust disturbance rejection near the resonance frequency. Experimental results validate NRC performance, demonstrating dual closed-loop bandwidths of 895 Hz and 845 Hz (within $\pm3$ dB and $\pm1$ dB bounds, respectively) that exceed the first resonance frequency at 739 Hz, even amidst significant system delay. These outcomes underscore the potential of the NRC for high-performance, precise nanopositioning applications.

\section*{Acknowledgments}
The authors express their sincere gratitude to both Mathias Winter, Head of Piezo System \& Drive Technology, and Dr.-Ing. Simon Kapelke, Head of Piezo Fundamental Technology, from Physik Instrumente (PI) SE \& Co. KG, for their invaluable collaboration in providing technical insights concerning the system and its applications.

\bibliographystyle{IEEEtran}
\bibliography{References}



\section{Biography Section}
\begin{IEEEbiography}[{\includegraphics[width=1in,height=1.25in,clip,keepaspectratio]{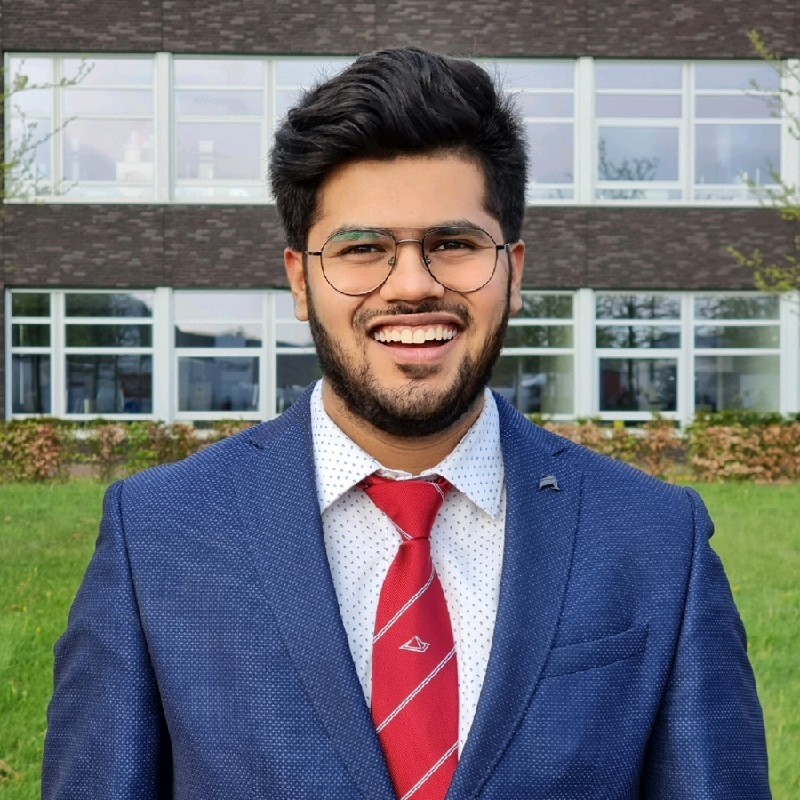}}]{Aditya Natu}
received the MSc in Mechanical Engineering (\textit{cum laude}) with a specialization in Mechatronics System Design and is currently employed as a PhD Candidate at the Department of Precision and Microsystems Engineering, Delft University of Technology, Delft, The Netherlands. His research interests include active damping control, distributed actuation and sensing, precision mechatronics, and nanopositioning systems.
\end{IEEEbiography}
\begin{IEEEbiography}[{\includegraphics[width=1in,height=1.25in,clip,keepaspectratio]{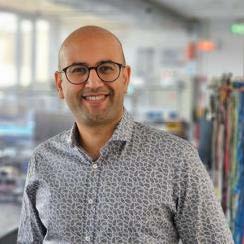}}]{S.Hassan HosseinNia}
(Senior Member, IEEE) received the Ph.D. degree (Hons.) (\textit{cum laude}) in electrical engineering specializing in automatic control: application in mechatronics from the University of Extremadura, Badajoz, Spain, in 2013. He has an industrial background, having worked with ABB, Sweden. Since October 2014, he has been appointed as a Faculty Member with the Department of Precision and Microsystems Engineering, Delft University of Technology, Delft, The Netherlands. He has co-authored numerous articles in respected journals, conference proceedings, and book chapters. His main research interests include precision mechatronic system design, precision motion control, and mechatronic systems with distributed actuation and sensing.

Dr. HosseinNia served as the General Chair of the 7th IEEE International Conference on Control, Mechatronics, and Automation (ICCMA
2019). Currently, he is an editorial board member of “Fractional Calculus
and Applied Analysis,” “Frontiers in Control Engineering,” and “International Journal of Advanced Robotic Systems (SAGE).”

\end{IEEEbiography}

\vfill

\end{document}